\documentclass[12pt]{article}
\usepackage{a4wide,epsfig,psfrag,amsmath,amssymb,cite,scalefnt}
\usepackage[dvipsnames]{xcolor}
\usepackage{comment,braket}
\usepackage{hyperref}
\usepackage{subfig}
\usepackage{cleveref}
\usepackage{slashed}
\usepackage{booktabs}
\usepackage{siunitx}

\graphicspath{ {figs/} }

\newcommand{\lt}{\left}
\newcommand{\rt}{\right}

\newcommand{\ov}{\overline}
\newcommand{\eq}[1]{Eq.~(\ref{#1})}
\newcommand{\eqsand}[2]{Eqs.~(\ref{#1}) and (\ref{#2})}
\newcommand{\eqsto}[2]{Eqs.~(\ref{#1}) to (\ref{#2})}
\newcommand{\gev}{\,\mbox{GeV}}

\newcommand{\Bbar}{\bar{B}}
\newcommand{\bbd}{\ensuremath{B_d\!-\!\Bbar{}_d\,}}
\newcommand{\bb}{\ensuremath{B\!-\!\Bbar\,}}

\newcommand{\bbq}{\ensuremath{B_q\!-\!\Bbar{}_q\,}}
\newcommand{\bbm}{\bb\ mixing}

\newcommand{\bbmd}{\bbd\ mixing}
\newcommand{\bbmq}{\bbq\ mixing}

\newcommand{\fig}[1]{Fig.~\ref{#1}}

\newcommand{\lqcd}{\Lambda_{\rm QCD}} 
\newcommand{\dm}{\ensuremath{\Delta M}}
\newcommand{\dg}{\ensuremath{\Delta \Gamma}}
\newcommand{\gqtf}{\ensuremath{\Gamma (B_q(t) \rightarrow f )}}
\newcommand{\gqbtf}{\ensuremath{\Gamma (\Bbar{}_q(t) \rightarrow f )}}

\newcommand{\gdtfb}{\ensuremath{\Gamma (B_q(t) \rightarrow \ov{f} )}}

\newcommand{\guntf}{\ensuremath{\Gamma  [f,t] }}
\newcommand{\guntfb}{\ensuremath{\Gamma  [\ov{f},t] }}

\newcommand*{\ie}{i.e.\ }

\newcommand{\centerhfill}[1][\quad]{\hspace{\stretch{0.5}}#1\hspace{\stretch{0.5}}}

\newcommand\numberthis{\addtocounter{equation}{1}\tag{\theequation}}

\allowdisplaybreaks

\begin{document}

\title{\vskip-3cm{\baselineskip14pt
    \begin{flushleft}
      \normalsize P3H-25-034, SI-HEP-2025-11, TTP25-016 
    \end{flushleft}} \vskip1.5cm 
  Current-current operator contribution to the decay matrix in $B$-meson mixing at 
  next-to-next-to-leading order of QCD
}
\author{
  Marvin Gerlach$^a$, 
  Ulrich Nierste$^a$, 
  Pascal Reeck$^a$, 
  \\
  Vladyslav Shtabovenko$^b$,
  and Matthias Steinhauser$^a$
  \\[1em]
  {\it $^a$\small\it Institut f{\"u}r Theoretische Teilchenphysik,}\\
  {\small\it Wolfgang-Gaede Stra\ss{}e 1, Karlsruhe Institute of Technology (KIT)}\\
  {\small\it 76131 Karlsruhe, Germany}  
  \\[.5em]
  {\it $^b$\small Theoretische Physik 1, Center for Particle Physics Siegen,} \\
\small\it Universit\"at Siegen, 57068 Siegen, Germany}

\date{}

\maketitle

\thispagestyle{empty}

\begin{abstract}
  \noindent
  We compute next-to-next-to-leading order perturbative corrections to the
  decay width difference of mass eigenstates and the charge-parity asymmetry
  $a_{\rm fs}$ in flavour-specific decays of neutral $B$ mesons.  In our calculation we
  take into account the full dependence on the charm and bottom quark masses
  for the current-current operator contributions up to three-loop
  order. Special emphasis is put on the proper construction of the so-called
  $|\Delta B|=2$ theory such that Fierz symmetry is preserved. We provide
  updated phenomenological predictions, for $\Delta\Gamma$,
  $\Delta\Gamma/\Delta M$ and $a_{\rm fs}$ for the $B_d$ and $B_s$ system,
  including a detailed analysis of the uncertainties of our predictions.  
  The calculated NNLO correction reduce the perturbative uncertainty of the 
  leading term of the $1/m_b$ expansion of the width difference $\dg_s$ in the $B_s$ system 
  to the level of the current experimental error. The uncertainty of our prediction 
  $\dg_s=({0.077}\pm 0.016)\,\mbox{ps}^{-1}$ is dominated by the sub-leading term of this expansion. 
  We further illustrate how better future measurements of $\dg_d$ and $a_{\rm fs}^d$ will help to gain a better understanding of $B_d$-$\bar B_d$ mixing.
\end{abstract}

%- }}}

\thispagestyle{empty}

\newpage

%- {{{ Introduction:

\section{Introduction}

\begin{figure}[t]
  {\centering \mbox~~~~~
    \includegraphics[width=0.3\textwidth]{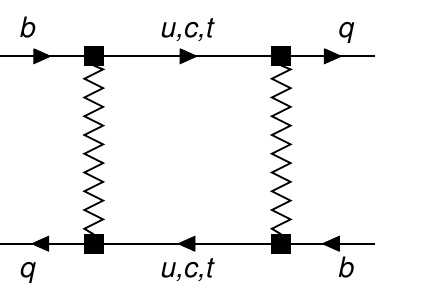} \hspace{5mm}
    \includegraphics[width=0.6\textwidth]{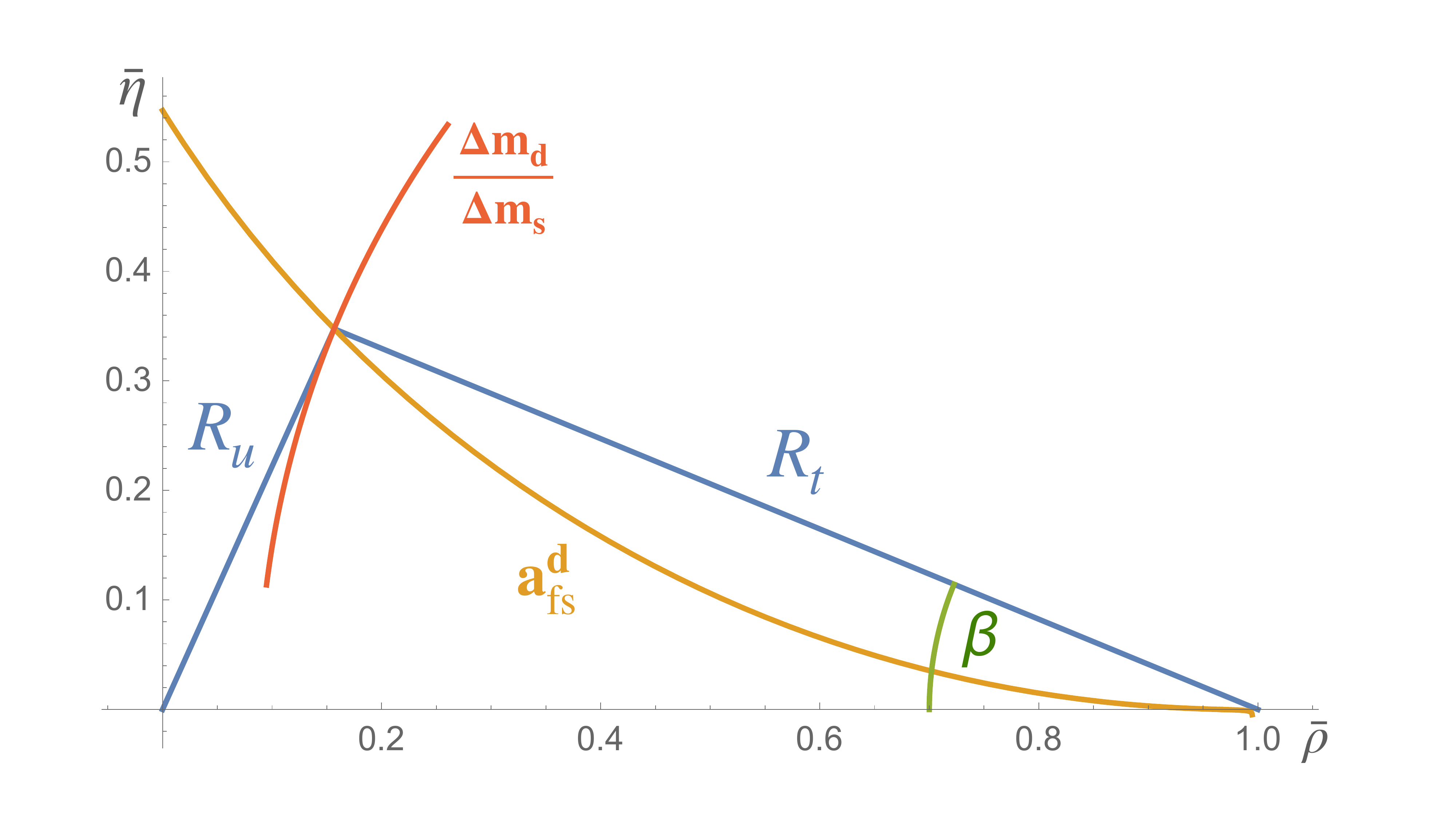}
    }
  \caption{Left: box diagram describing \bbmq\ with $q=d$ or $s$ to leading
    order in QCD. Right: constraints from the three \bbm\ observables
    $\dm_d/\dm_s\propto R_t^2$, $a_{CP} (B_d (t) \to J/\psi K_S) \propto
    \sin(2\beta)$, and $a_{\rm fs}^d\propto (\sin\beta)/R_t$ on the
    apex $(\bar\rho,\bar \eta)$ of the CKM unitarity triangle.\label{fig:box}}
%~\\[-5mm]
%\hrule  
\end{figure}

Neutral $B_q$ mesons mix with their antiparticles because the weak interaction 
permits transitions which change the beauty quantum number $B$ by two
units.  \bbmq\ requires the exchange of two virtual $W$ bosons and is
therefore a loop-induced process, mediated by the box diagram in \fig{fig:box}. 
The valence quark $q$ can be $d$ or $s$; since their masses are negligible, the corresponding box diagrams
only differ by the elements of the Cabibbo-Kobayashi-Maskawa (CKM)
matrix accompanying the $W$ couplings.  The time evolution of the
two-state system $(\ket{B_q},\ket{\bar B_q})$ is governed by the
$2\times 2$ matrix $M^q-i\Gamma^q/2$ built from the Hermitian mass and
decay matrices $M^q$ and $\Gamma^q$. To leading order (LO) in the strong
coupling constant $\alpha_s$, the off-diagonal elements of both matrices
are calculated from the box diagram in \fig{fig:box}. While $M_{12}^q$
is dominated by the diagrams with one or two internal top quarks, 
only box diagrams with light quarks $u,c$ contribute to
$\Gamma_{12}^q$. The latter diagrams describe the interference between
the decay amplitudes of $B_q\to f$ and $\Bbar_q\to f$, which is related
to the absorptive part of the $B_q\to \Bbar_q$ box diagram.

The state $\ket{B_q(t)}$ of a meson tagged as $B_q$ at time $t=0$
evolves into a superposition of $\ket{B_q}$ and $\ket{\Bbar_q}$ and
experimental studies of time dependent decay rates $\Gamma(B_q(t)\to f)$
permit the determination of the three theoretical quantities $|M_{12}^q|$,
$|\Gamma_{12}^q|$, and $\arg (-\Gamma_{12}^q/M_{12}^q)$ of 
the \bbq\ complex. One defines
\begin{align*}
   \Delta M_q &= M_H^q - M_L^q\,,\\
  \Delta \Gamma_q &= \Gamma_L^q - \Gamma_H^q\,,\numberthis\label{eq:dg}
\end{align*}
where $M_{L,H}$ and $\Gamma_{L,H}$ denote the masses and widths of the
lighter and heavier eigenstates
\begin{align*}
  \ket{B_{q,L}} &=  p \ket{B_q} + q \ket{\Bbar_q}\,, &
  \ket{B_{q,H}} &=  p \ket{B_q} - q \ket{\Bbar_q}\,.\numberthis\label{eq:defpq}
\end{align*}
The coefficients $p$ and $q$ are found by diagonalising $M^q-i\Gamma^q/2$.
Then  
theoretical and experimental quantities are related as
\begin{align}
  \dm_q & =  2|M_{12}^q|  {\, ,} & \Delta \Gamma_{{q}} &= 2 |\Gamma_{12}^q| \cos \phi_q
  %\left(\arg(-\Gamma_{12}^q / M_{12}^q)\right) 
  % + \mathcal{O}\left(\frac{|\Gamma^q_{12}|^2}{|M^q_{12}|^2}\right)
  \nonumber\\%   \label{eq:dm}\\ 
  \frac{\dg_q}{\dm_q} &=
                 - \mbox{Re}\frac{\Gamma_{12}^q}{M_{12}^q}\, , &
  a_{\rm fs}^q &= \mbox{Im} \frac{\Gamma_{12}^q}{M_{12}^q}\,, 
                 \label{eq:dgdmafsq}
\end{align}
where $\phi_q\equiv \arg(-\Gamma_{12}^q / M_{12}^q)$ 
is the small fundamental CP phase quantifying CP violation in mixing.
$ a_{\rm fs}^q$ denotes the charge-parity (CP) asymmetry in
flavour-specific decays and quantifies CP violation in \bbmq.
Eqs.~\eqref{eq:dgdmafsq} receive negligible relative corrections of order
$|\Gamma_{12}^q/M_{12}^q|^2 \sim 10^{-5}$. 
Furthermore, in
$a_{\rm fs}^q$ we omit the quadratic terms in
$ \mbox{Im} [\Gamma_{12}^q/M_{12}^q] \lesssim  10^{-3}$.

The experimental situation is as follows:
\begin{align}
    \dm^{\rm exp}_s &=  (17.7656 \pm 0.0057)\; \mbox{ps$^{-1}$}\,, && \mbox{\cite{LHCb:2021moh}}
            \label{eq:expdms}\\
              \dg^{\rm exp}_s  &=  
(0.082 \pm 0.005)\;\mbox{ps}^{-1}\,, && \mbox{\cite{HFLAV:2022esi}}
                       \label{eq:expdgs} \\
  a_{\rm fs}^{s, \rm exp} &= -0.0006 \pm 0.0028\,, && 
\mbox{\cite{HFLAV:2022esi}}   
                          \label{eq:expafss}   \\
  \dm^{\rm exp}_d &= (0.5065 \pm 0.0019)  \;  \mbox{ps}^{-1}\,,
 && \mbox{\cite{HFLAV:2022esi}}
            \label{eq:expdmd} \\
  \dg^{\rm exp}_d  &=  (0.7 \pm 6.6)\cdot 10^{-3} \;
                       \mbox{ps}^{-1}\,, && 
\mbox{\cite{HFLAV:2022esi}}
                       \label{eq:expdgd} \\
  a_{\rm fs}^{d, \rm exp} &= -0.0021 \pm 0.0017\,. && 
\mbox{\cite{HFLAV:2022esi}}
                          \label{eq:expafsd}
\end{align}  
The last five numbers are averages from all available data, see
Ref.~\cite{HFLAV:2022esi} for the original references.
$ \dg^{\rm exp}_d$ in \eq{eq:expdgd} is calculated from
$\dg_d/\Gamma_d=0.001\pm 0.010$ quoted in Ref.~\cite{HFLAV:2022esi} and
the $B_d$ lifetime
$1/\Gamma_d^{\rm exp}=\tau(B_d)=(1.519\pm 0.004)\,\mbox{ps}^{-1}$. 

In order to fully determine the three fundamental quantities
$|M_{12}^q|$, $|\Gamma_{12}^q|$, and
$\phi_q = \arg (-M_{12}^q/\Gamma_{12}^q) $ for both the $B_d$ and $B_s$
systems, one must measure all six quantities listed in
\eqsto{eq:expdms}{eq:expafsd}. Ideally, these measurements will be
confronted with precise Standard Model (SM) predictions. While
$\dm^{\rm exp}_q$ accurately constrains $|M_{12}^q|$, the only precise
information on a decay matrix element is on
$\mbox{Re}(\Gamma_{12}^s/M_{12}^s)$ inferred from
$\dg_s^{\rm exp}/\dm_s^{\rm exp}$.

The calculation of $\Gamma_{12}^q$ involves two operator product
expansions (OPE) which separate the strong interactions associated with
the different energy scales $m_t,M_W \gg m_b \gg \lqcd$, where $m_t$,
$m_b$, and $M_W$ are the masses of top and bottom quark and $W$ boson
and $\lqcd \sim 400\,$MeV is the hadronic scale. The first OPE amounts
to the construction of the effective weak $|\Delta B|=1$ Hamiltonian
describing $b$ decays mediated by virtual $W$ bosons or top quarks in
terms of dimension-six operators. The Wilson coefficients multiplying
these operators can be calculated in perturbation theory and are known
to next-to-leading order (NLO)
\cite{Buras:1989xd,Buras:1991jm,Buras:1992tc} and
next-to-next-to-leading order (NNLO)
\cite{Gambino:2003zm,Gorbahn:2004my,Gorbahn:2005sa} of quantum
chromodynamics (QCD). The numerically most important coefficients in the
$|\Delta B|=1$ Hamiltonian are those of the current-current operators
describing the effect of $W$-mediated tree decays. Penguin operators
related to $W$-top loops in the SM have a smaller, yet non-negligible
effect on $\Gamma_{12}^q$.  The second OPE is the Heavy Quark
Expansion (HQE) which matches the \bbmq\ amplitude with two $\Delta B=1$
operators onto local $\Delta B=2$ operators, which amounts to an
expansion in powers of $\lqcd/m_b$. The coefficients multiplying the
$\Delta B=2$ operators of the leading power are fully known at LO and
NLO, including the contributions with one or two penguin operators
\cite{Buras:1984pq,Beneke:1996gn,Beneke:1998sy,Beneke:2003az,
  Ciuchini:2001vx,Lenz:2006hd,Gerlach:2021xtb, Gerlach:2022wgb}. NNLO
corrections to the contribution from two current-current operators have
been calculated first in the approximation of a large number of flavours
\cite{Asatrian:2017qaz,Asatrian:2020zxa,Hovhannisyan:2022miy}, but a
theoretical prediction matching the error of $\dg^{\rm exp}_s$ in
\eq{eq:expdgs} calls for a full NNLO calculation of the current-current
contribution.  The numerical NNLO result for $\dg_s$ presented
in Ref.~\cite{Gerlach:2022hoj} is based on an expansion of the
three-loop integrals to second order in
$m_c/m_b$, where $m_c$ denotes the charm quark mass.  % Our
% numerical result for $\dg^{\rm exp}_s$ at NNLO has been presented in
% Ref.~\cite{Gerlach:2022hoj}.
This paper is devoted to a detailed
description of the calculation including semi-analytic results. We go beyond
Ref.~\cite{Gerlach:2022hoj} by evaluating the three-loop integrals up to
order $z^{50}$ where $z=m_c^2/m_b^2$.\footnote{In Ref.~\cite{Reeck:2024iwk}  expansions around $\sqrt{z}=1/20, 1/5$ and $3/10$ have been constructed, too. For the phenomenological application it is sufficient to use the expansion for $z\to0$.} Contributions from penguin operators are considered up to NLO and up to order $z^1$.

Another focus of this paper is the application of our result to a
phenomenological analysis of $\dg_d$ and $ a_\text{fs}^d$, which can
constrain models of new physics predicting sizeable contributions to
$\Gamma_{12}^d$ (see, e.g.\
Ref.~\cite{Alonso-Alvarez:2021qfd}). Furthermore, $ a_\text{fs}^d$ can play
an important role to constrain new physics in $M_{12}^d$: While we know the
magnitude and phase of $M_{12}^d$ well from $\dm_d$ and the
mixing-induced CP asymmetry $a_{CP} (B_d (t) \to J/\psi K_S)$, the
predictions of both quantities involve the parameters
$\bar\rho,\bar\eta$ characterising the apex of the CKM unitarity
triangle (UT). In fact, in the absence of new physics, the ratio
$\dm_d/\dm_s$ pins down the side $R_t=\sqrt{\bar\rho{}^2+\bar\eta{}^2}$
of the UT precisely, and
$a_{CP} (B_d (t) \to J/\psi K_S)\propto \sin(2\beta)$ accurately
determines the UT angle $\beta$. However, to \emph{test}\ the SM
description of \bbmd\ one must determine $\bar\rho,\bar\eta$ from
\emph{other}\ observables. For example, $\beta$ is well constrained by
the opposing side $R_u\propto |V_{ub}/V_{cb}|$ of the
UT, which is unfortunately affected by the ``exclusive vs.\ inclusive''
controversies on the measurements of $|V_{ub}|$ and $|V_{cb}|$ from
semileptonic decays. Now \cite{Beneke:2003az}
\begin{align}
  a_{\rm fs}^d &\propto \;
             \frac{\sin\beta}{R_t} \, =\,
             \frac{\bar\eta}{\sqrt{ (1 - \bar\rho{} )^2+\bar\eta{}^2}} 
 \label{eq:afdre}
\end{align}  
up to small and calculable corrections. \eq{eq:afdre} means that a
precise measurement of $a_\text{fs}^d$ defines a circle in the
$(\bar\rho,\bar\eta)$ plane. The requirement that the apex of the UT
determined from $\dm_d/\dm_s$ and $a_{CP} (B_d (t) \to J/\psi K_S)$ lies
on this circle probes the SM from \bbmq\ observables \emph{alone}, see
\fig{fig:box}.  Moreover, $a_\text{fs}^d={\cal O}(m_c^2/m_b^2)$, because
$\Gamma_{12}^d$ and $M_{12}^d$ have the same phase for $m_c=0$,
so that $\mbox{Im} (\Gamma_{12}^d/M_{12}^d)$ vanishes in this limit. A small
new-physics contribution $\phi_{12}^{d\,\Delta}$ to $\arg M_{12}^d$ will change
$\beta$ inferred from
$a_{CP} (B_d (t) \to J/\psi K_S)\propto \sin(2\beta +\phi_{12}^{d\,\Delta}) $
slightly, but significantly enhance $a_\text{fs}^d$ because $a_\text{fs}^d$
receives a contribution proportional to $\sin \phi_{12}^{d\,\Delta} $
unsuppressed by $m_c^2/m_b^2$ \cite{Laplace:2002ik,Beneke:2003az}.

The paper is organised as follows: In the next Section we discuss in detail
the effective $\Delta B=1$ and $\Delta B=2$ theories. In the latter case we elaborate in detail on the construction of the evanescent operators such that Fierz symmetry is
preserved in $D\not=4$ dimensions and discuss the mixing among the operators. 
Section~\ref{sec::technicalities} is dedicated to technical details concerning the computation of the amplitudes, and in Section~\ref{sec::analytic} we show results for the matching coefficients for $m_c=0$.
Section~\ref{sec::phen} contains the phenomenological analysis of 
$\Delta\Gamma_q$, $\Delta\Gamma_q/\Delta M_q$ and $a_{\rm fs}^q$.
We conclude in Section~\ref{sec::concl}. In Appendix~\ref{app::Z} we 
discuss the 
renormalisation constants in the $\Delta B=2$ theory and
describe how to obtain them from the computer-readable file. In Appendix~\ref{app::VIA} 
we  provide additional information on the Fierz transformation.
In Appendix~\ref{app::masters} we provide analytic results for the master integrals which are needed in the limit
of massless charm quarks. Appendix~\ref{app:full_ev_consts} contains the coefficients which appear in the definition of the evanescent operators in the $\Delta B=2$ theory keeping the dependence on the number of colours.

%- }}}
%- {{{ Effective theories:

\section{\label{sec::eff}Effective theories}

For the calculation of $\Gamma_{12}^q$ we have to consider two effective
theories which are successively constructed from the Standard Model.  In a
first step one integrates 
out the degrees of freedom with masses of the order of
the electroweak scale to arrive at the  $|\Delta B|=1$ Hamiltonian
$ {\cal H}_{\rm eff}^{|\Delta B|=1} =   {\cal H}_{\rm eff}^{\Delta B=1}+ \mbox{h.c.}$
mentioned in the Introduction; the corresponding Wilson coefficients
are known to NNLO. For our calculation we need the coefficients
of the current-current operators \cite{Gorbahn:2004my}, which emerge
from tree-level $W$ exchange.
In a first step we  write $\Gamma_{12}^q$ from Eq.~(\ref{eq:dgdmafsq}) as
\begin{equation}
  \Gamma_{12}^q = \frac{1}{2 M_{B_s}} \,\mbox{Abs}
                    \langle B_q|i\int{\rm d}^4 x \,\, T\,\,
                    {\cal H}_{\rm eff}^{\Delta B=1}(x)
                    {\cal H}_{\rm eff}^{\Delta B=1}(0)
                    |\bar{B}_q\rangle\,, 
                    \label{eq:ot}
\end{equation}
where ``Abs'' denotes the absorptive part of the matrix element.  
In a second step we perform a HQE which leads to an expansion of
$\Gamma_{12}^q$ in powers of $1/m_b$ in the so-called $\Delta B=2$
theory. This step amounts to the expression of the bi-local matrix
element in \eq{eq:ot} as a sum of terms involving matrix elements of
local $\Delta B=2$ operators multiplied by new $\Delta B=2$ matching
coefficients, which depend on the $\Delta B =1$ coefficients, the strong
coupling constant $\alpha_s$, and the quark masses $m_c$ and $m_b$.  Higher
dimensions of the $\Delta B=2$ operators correspond to higher powers of
$1/m_b$ in the HQE.  The $\Delta B =2$ matching coefficients are
calculated in perturbative QCD as an expansion in $\alpha_s$.  The main
purpose of this work is the calculation of the leading-power
coefficients to NNLO, i.e.\ three-loop accuracy.  The
essentials of the matching calculation between the $\Delta B=1$ and
$\Delta B=2$ theories is described in
Refs.~\cite{Beneke:1998sy,Beneke:2003az,Ciuchini:2001vx,Lenz:2006hd,Asatrian:2017qaz,Asatrian:2020zxa,Gerlach:2021xtb,Gerlach:2022wgb}. We
will therefore focus on the new features occurring first at NNLO, which
are related to evanescent operators and the mixing with an operator
related formally to power-suppressed terms.

%- {{{ $\Delta B=1$ theory:

\subsection{{$\mathbf{\Delta B =1}$} theory}
For definiteness we consider $\Gamma_{12}^s$ from now on, the
generalisation to  $\Gamma_{12}^d$ is straightforward. 
On the $|\Delta B|=1$ side of the matching we work with the operator basis
\cite{Chetyrkin:1997gb}
\begin{align*}
	\mathcal{H}_{\textrm{eff}}^{|\Delta B|=1} 
	=&  \frac{4G_F}{\sqrt{2}}  \left[
	-\, \lambda^s_t \Big( \sum_{i=1}^6 C_i Q_i + C_8 Q_8 \Big) 
	- \lambda^s_u \sum_{i=1}^2 C_i (Q_i - Q_i^u) \right. \\
	& \phantom{\frac{4G_F}{\sqrt{2}} \Big[}
	\left.
	+\, V_{us}^\ast V_{cb} \, \sum_{i=1}^2 C_i Q_i^{cu} 
	+ V_{cs}^\ast V_{ub} \, \sum_{i=1}^2 C_i Q_i^{uc} 
	\right]
	+ \mbox{h.c.}\,,\numberthis
	\label{eq::HamDB1}
\end{align*}
with
\begin{equation}
  \lambda^s_a = V_{as}^\ast V_{ab}\,, 
\end{equation}
where $a=u,c,t$ and $\lambda_t=-\lambda_c-\lambda_u$.  $G_F$ stands for the
Fermi constant. We need the current-current operators
\begin{align*}
	Q_1   &= \bar{s}_L \gamma_{\mu} T^a c_L\;\bar{c}_L     \gamma^{\mu} T^a b_L\,, &
	Q_2   &= \bar{s}_L \gamma_{\mu}     c_L\;\bar{c}_L     \gamma^{\mu}     b_L\,,\\
	% Q_3   &=& \bar{s}_L \gamma_{\mu}     b_L \sum_q \bar{q}\gamma^{\mu}     q\,,\nonumber\\
	% Q_4   &=& \bar{s}_L \gamma_{\mu} T^a b_L \sum_q \bar{q}\gamma^{\mu} T^a q\,,\nonumber\\
	% Q_5   &=& \bar{s}_L \gamma_{\mu_1}\gamma_{\mu_2}\gamma_{\mu_3} b_L
        %           \sum_q \bar{q} \gamma^{\mu_1} \gamma^{\mu_2}\gamma^{\mu_3}     q\,,\\
	% Q_6   &=& \bar{s}_L \gamma_{\mu_1}\gamma_{\mu_2}\gamma_{\mu_3} T^a b_L
        %           \sum_q \bar{q} \gamma^{\mu_1}\gamma^{\mu_2}\gamma^{\mu_3} T^a q\,,\nonumber\\
	% Q_8   &=& \frac{g_s}{16\pi^2} m_b \, \bar{s}_L \sigma^{\mu \nu} T^a b_R \, G_{\mu\nu}^a\,, \nonumber\\
	Q^u_1 &= \bar{s}_L \gamma_{\mu} T^a u_L\;\bar{u}_L     \gamma^{\mu} T^a b_L\,,&
	Q^u_2 &=  \bar{s}_L \gamma_{\mu}     u_L\;\bar{u}_L     \gamma^{\mu}     b_L\,,\\
	Q^{cu}_1 &= \bar{s}_L \gamma_{\mu} T^a u_L\;\bar{c}_L \gamma^{\mu} T^a b_L\,, &
	Q^{cu}_2 &=  \bar{s}_L \gamma_{\mu}     u_L\;\bar{c}_L     \gamma^{\mu}     b_L\,,\\
	Q^{uc}_1 &= \bar{s}_L \gamma_{\mu} T^a c_L\;\bar{u}_L     \gamma^{\mu} T^a b_L\,,&
	Q^{uc}_2  &= \bar{s}_L \gamma_{\mu}     c_L\;\bar{u}_L     \gamma^{\mu}     b_L\,.\numberthis
	\label{operators}
\end{align*}
Here we introduced the left-chiral projector $P_L=(1-\gamma_5)/2$ so that
$q_L=P_Lq$. The  four-quark penguin operators $Q_3,\ldots,Q_6$ and
the chromomagnetic penguin operator $Q_8$ are not needed for the
calculation presented in this paper; their definition can be found, e.g.\ in Refs.~\cite{Chetyrkin:1997gb,Gerlach:2021xtb,Gerlach:2022wgb}.

We employ dimensional regularisation in $D= 4-2\epsilon $ dimensions.
The bare matrix elements  of the operators in \eq{operators} feature ultraviolet
(UV) divergent expressions with more than one Dirac matrix on each fermion
line, which can be reduced to matrix elements of the operators in
$\mathcal{H}_{\textrm{eff}}^{|\Delta B|=1} $ in four dimensions but not
for $D\neq 4$. 
The renormalisation programme therefore involves evanescent operators;
at NLO these are
\begin{align*}
	E_1[Q_1] &= 
	\bar{s}_L \gamma^{\mu_1} \gamma^{\mu_2} \gamma^{\mu_3}
	T^a c \;
	\bar{c} \gamma_{\mu_1} \gamma_{\mu_2} \gamma_{\mu_3}  T^a
	b_L - 16 Q_1\,,\\ 
	E_1[Q_2] &= 
	\bar{s}_L \gamma^{\mu_1} \gamma^{\mu_2} \gamma^{\mu_3}
	c\;\bar{c} \gamma_{\mu_1} \gamma_{\mu_2}
                     \gamma_{\mu_3} b_L - 16 Q_2\, .
	\numberthis\label{evan_operators_lo}
\end{align*}
Similar evanescent operators are needed for $Q_{i}^{u}$, $Q_{i}^{uc}$ and
$Q_{i}^{cu}$ with $i=1,2$; they are obtained by replacing one or both $c$
fields with $u$ fields in Eq.~(\ref{evan_operators_lo}). We refer to $E_1[Q_{1,2}]$
as evanescent operators of the first generation and define the $n$-th generation operators as 
those with $2n$ more Dirac matrices  in each fermion current than the corresponding physical operator.
At NNLO one has to introduce the second-generation operators
%\cite{Chetyrkin:2017gb}
\begin{align*}
	E_2[Q_1] &= 
	\bar{s}_L \gamma^{\mu_1} \gamma^{\mu_2} \gamma^{\mu_3} \gamma^{\mu_4} \gamma^{\mu_5}
	T^a c \;
	\bar{c} \gamma_{\mu_1} \gamma_{\mu_2} \gamma_{\mu_3} { \gamma_{\mu_4} \gamma_{\mu_5}}  T^a
	b_L - 20 E_1 [Q_2] - 256 Q_1\,,\\ 
	E_2[Q_2] &= 
	\bar{s}_L \gamma^{\mu_1} \gamma^{\mu_2} \gamma^{\mu_3} \gamma^{\mu_4} \gamma^{\mu_5}
	c\;\bar{c} \gamma_{\mu_1} \gamma_{\mu_2}
	\gamma_{\mu_3} 
    { \gamma_{\mu_4} \gamma_{\mu_5}} b_L - 20 E_1 [Q_2]
                     - 256 Q_2\,
           ,
	\numberthis\label{evan_operators_nlo}
\end{align*}
which occur in two-loop matrix elements of $Q_{1,2}$ and one-loop matrix
elements of $E_1[Q_{1,2}]$. 
The renormalisation of the physical and evanescent $\Delta B=1$ 
operators has extensively been discussed in the
literature, see, e.g.\ Refs.~\cite{Buras:1989xd,Buras:1991jm,Buras:1992tc,Herrlich:1994kh,Herrlich:1996vf,
  Gorbahn:2004my}. 

%- }}}
%- {{{ $\Delta B=2$ theory:

\subsection{Physical operators in the $\mathbf{\Delta B=2}$ theory}
\label{sec:physical_db2}

In the following we present the basis of physical operators used in our calculation. At leading power in $1/m_b$, there are three physical operators which are
chosen as 
\begin{align*}
	Q &= 4\,(\bar{s}^c \gamma^\mu P_L b^c) \; (\bar{s}^d \gamma_\mu
	P_L b^d)\,, \\
	\widetilde{Q}_S &= 4\,(\bar{s}^c P_R b^d)\; (\bar{s}^d P_R
	b^c)\,, \\
    Q_S &= 4\,(\bar{s}^c P_R b^c) \;(\bar{s}^d P_R
	b^d)\,,\numberthis \label{eq::opDB2}
\end{align*}
where $P_R=(1+\gamma_5)/2$ and $c,d$ are the colour indices attached to quark fields. As in the
$\Delta B=1$ theory, we need evanescent operators with several Dirac
matrices on each line, in analogy to Eqs.~\eqref{evan_operators_lo} and \eqref{evan_operators_nlo}. These will be discussed in Section \ref{sec:eva}.

There is a special relation between the three physical operators,
\begin{equation}
  R_0   = \frac{1}{2}  Q + Q_S      + \widetilde{Q}_S,
      \label{eq::R0}
\end{equation}
which is an operator satisfying $\langle B_s | R_0 | \bar B_s\rangle =
\mathcal{O}(1/m_b)$ \cite{Beneke:1996gn}. Equation~\eqref{eq::R0} can be used to trade $Q_S$ for $R_0$. In perturbation theory,
$\langle  B_s | R_0 | \bar B_s\rangle^{(n)} = \mathcal{O}(1/m_b)$
only holds beyond $n=0$ (LO) if $R_0$ receives a special finite renormalisation, which is discussed in Section~\ref{sec:R0_ren}. The calculation of these finite renormalisation constants requires a subtle treatment of infrared (IR) singularities, to be discussed in Section~\ref{sec:alpha12_finite}. 

In summary, we can trade the operators in \eq{eq::opDB2} for linear
combinations which are either evanescent or have power-suppressed
matrix elements such that our basis of physical operators is $\{Q,\widetilde Q_S, R_0\}$. The evanescent operators, which will be discussed in the following section, are unphysical, meaning that they vanish in $D \rightarrow 4$ dimensions and appear at intermediates steps, while their Wilson coefficients do not enter physical observables. 

\subsection{Evanescent operators in the {$\mathbf{\Delta B=2}$} theory\label{sec:eva}}

In order to obtain results consistent with the low-energy matrix elements, we choose an operator basis that respects Fierz symmetry for the matrix elements of the physical operators. This is because our perturbative Wilson coefficients are combined with non-perturbative hadronic matrix elements, which are calculated in four dimensions, e.g.\ in lattice gauge theory. In such four-dimensional calculations an operator is trivially identical to its Fierz transform. 

The relevance of Fierz symmetry is rooted in the fact that the $\Delta B=2$ matrix elements can be calculated exactly in the limit $N_c \to\infty$, where $N_c=3$ is the number of colours, in terms of meson masses and decay constants. In the case of $\widetilde Q_S$ this 
calculation involves a four-dimensional Fierz transform and only gives the correct result for physical observables if the Wilson coefficients of $\widetilde Q_S$ and its Fierz transform coincide for $N_c \to\infty$. 
It is further desirable to implement Fierz symmetry at all orders of $1/N_c$, so that in any calculation of 
the hadronic matrix element one can easily choose the Fierz arrangement at wish.   
For further explanations see Appendix~\ref{app::VIA}.

To achieve Fierz symmetry in perturbation theory, one must define the evanescent operators appropriately \cite{Buras:1989xd,Herrlich:1994kh}, and this step is part of the definition of the renormalisation scheme. Here ``Fierz symmetry" means that the Wilson coefficient of a given operator
$O$ is identical to that of its Fierz transform $O^F$.

In the following, we will explicitly construct bases of evanescent operators obeying Fierz symmetry. Specifically, we impose the following conditions:
\begin{enumerate}
\item Renormalised matrix elements of physical operators are identical to those of their Fierz-transformed counterparts: $\lim_{d\rightarrow 4} \langle O \rangle^\text{ren} = \lim_{d\rightarrow 4} \langle O^F \rangle^\text{ren}$ at one-loop and two-loop level. 
\item The operator definitions of the evanescent operators should not depend on the number of quark flavours $N_f$.
\item The large-$N_c$ limit of the renormalised physical operator matrix elements fixes the leading term to be of order $N_c^2$,  see Appendix~\ref{app::VIA}.
This constraint must be respected by the definition of the evanescent operators. Hence, the $\mathcal{O}(\epsilon)$ contributions of physical operators to the evanescent operators must be at most $\mathcal{O}(N_c^0)$ as $N_c\rightarrow\infty$. 
\end{enumerate}
One can check that enforcing these conditions leaves no ambiguity in the leading $\mathcal{O}(N_c^2)$ term of the renormalised physical operators in the large-$N_c$ limit and ensures that they evaluate to the correct factorised results. Note that condition~1 must hold for any choice of IR regulator and is most easily implemented by choosing off-shell kinematics,
for which no IR singularities occur.

To enforce the conditions~1 to~3 above, both the regular as well as the Fierz basis have to be defined. We first define the Fierz-transformed counterparts of the physical operators in Eq.~\eqref{eq::opDB2}, namely
\begin{align*}
\widetilde{Q} =  Q^F &= 4\, (\overline{s}^c \gamma_\mu P_L b^d)\,(\overline{s}^d \gamma^\mu P_L b^c)\,,\\
Q_S^F &= \frac{1}{12} Q_T - \frac16 \widetilde{Q}_T\,,\\
\widetilde{Q}_S^F &= \frac{1}{12} \widetilde{Q}_T - \frac16 Q_T\,. \numberthis \label{eq:QSFt_def}
\end{align*}
Note that applying a Fierz transformation to $Q$ leads directly to the operator $\widetilde{Q}$ which has the colour indices $c,d$ contracted across the two spin lines. The operator $Q$ and its evanescent operators renormalise separately from the operators $Q_S$ and $\widetilde{Q}_S$ because even and odd numbers of gamma matrices do not mix. This is also true for the respective Fierz-transformed operators.

The operators $Q_T$ and $\widetilde{Q}_T$ used in the definitions in Eq.~\eqref{eq:QSFt_def} are 
\begin{align*}
Q_T &= 4 (\overline{s}^c \sigma^{\mu\nu} P_R b^c)\,(\overline{s}^d \sigma_{\mu\nu} P_R b^d)\,,\\
\widetilde{Q}_T &= 4 (\overline{s}^c \sigma^{\mu\nu} P_R b^d)\,(\overline{s}^d \sigma_{\mu\nu} P_R b^c)\,, \numberthis
\end{align*}
where $\sigma^{\mu\nu} = \frac{{\rm i}}{2} [ \gamma^\mu, \gamma^\nu ]$. This is a result of applying the Fierz identity
\begin{equation}
(P_R)_{\alpha\beta} (P_R)_{\gamma\delta} = \frac18 (\sigma^{\mu\nu}P_R)_{\alpha\delta} (\sigma_{\mu\nu}P_R)_{\gamma\beta} + \frac12 (P_R)_{\alpha\delta} (P_R)_{\gamma\beta}\,. \label{eq:chiral_Fierz_sigma}
\end{equation}

There is an infinite number of evanescent operators, but it is sufficient to include only a finite number to renormalise physical operators to a given order in perturbation theory.  To renormalise $n$th generation evanescent operators to order $\mathcal{O}(\alpha_s^i)$, where physical operators are counted as $0$th generation, we need to consider evanescent operators up to generation $k=n+i$ at tree-level. At loop order $l$, we need to include operators up to generation $k_l = k - l$. For illustration purposes, we list the necessary operators to renormalise a second generation evanescent operator to $\mathcal{O}(\alpha_s^2)$ in Tab.~\ref{tab:ren_example}.

A small caveat here is that while the procedure above is sufficient to renormalise physical operators to a given order in $\alpha_s$, it is necessary to renormalise evanescent operators to carry out the matching calculation described in Section \ref{sub::match} if those evanescent matrix elements appear at lower orders on the $\Delta B = 1$ side.

\begin{table}[t]
    \centering
    \begin{tabular}{l l}
         Loop order & Inserted operators  \\ \midrule
         Tree-level & $Q, E^{(1)}, E^{(2)}, E^{(3)}, E^{(4)}$ \\
         One-loop & $Q, E^{(1)}, E^{(2)}, E^{(3)}$ \\
         Two-loop & $Q, E^{(1)}, E^{(2)}$
    \end{tabular}
    \caption{An example of which operators need to be inserted at which loop order to renormalise a second generation evanescent operator at NNLO, i.e.~two-loop level. The physical operators are schematically denote by $Q$ while $E^{(i)}$ stands for the $i$th generation evanescent operators.}
    \label{tab:ren_example}
%    ~\\[-5mm]
%\hrule
\end{table}

The evanescent operators of the first generation have the generic definitions
\begin{align*}
  E^{(1)}_1[Q] &= Q- \widetilde{Q}\, , \\
  E^{(1)}_2[Q] &= 4 (\overline{s}^c \gamma_{\mu_1} \gamma_{\mu_2} \gamma_{\mu_3} P_L b^d)\,(\overline{s}^d \gamma^{\mu_3} \gamma^{\mu_2} \gamma^{\mu_1} P_L b^c) - \left(4 + \tilde{f}\epsilon + \tilde{g}\epsilon^2\right) \widetilde{Q}\,,\\
  E^{(1)}_3 [Q] &= 4 (\overline{s}^c \gamma_{\mu_1} \gamma_{\mu_2}
                  \gamma_{\mu_3} P_L b^c)\,(\overline{s}^d \gamma^{\mu_3}
                  \gamma^{\mu_2} \gamma^{\mu_1} P_L b^d) - \left(4 +
                  {f}\epsilon + {g}\epsilon^2\right)
                  Q\,.\numberthis\label{eq:E1_def}
\end{align*}
For the operators $\{Q_S,\widetilde{Q}_S\}$, which mix under renormalisation, the evanescent operators of the first generation are
\begin{align*}
  E^{(1)}[Q_S]
               &= 4 (\overline{s}^c \gamma_{\mu_1} \gamma_{\mu_2}
                 P_R b^c)\,(\overline{s}^d \gamma^{\mu_2} \gamma^{\mu_1}
                 P_R b^d) + \left(8 + \tilde{a}\epsilon +
                 \tilde{b}\epsilon^2\right) \widetilde{Q}_S\ \\
                 &\phantom{=}- \left(a\epsilon + b \epsilon^2\right) Q_S\,, \\
  E^{(1)}[\widetilde{Q}_S]
               &= 4 (\overline{s}^c \gamma_{\mu_1} \gamma_{\mu_2}
                 P_R b^d)\,(\overline{s}^d \gamma^{\mu_2} \gamma^{\mu_1}
                 P_R b^c) + \left(8 + \tilde{a}_2\epsilon +
                 \tilde{b}_2\epsilon^2\right) Q_S\nonumber\\
                 &\phantom{=} -
                 \left(a_2\epsilon + b_2 \epsilon^2\right)
                 \widetilde{Q}_S\,.\numberthis \label{eq:E1QS_def}
\end{align*} 
They appear in one-loop diagrams; for a NLO calculation one only needs
the coefficients of the $\epsilon$ and not the $\epsilon^2$ terms.

In the Fierz-transformed basis we can similarly define the first generation of evanescent operators as
\begin{align*}
      E^{(1)}_1[Q^F] &= E^{(1)}_1[Q]\,, \\
      E^{(1)}_2[Q^F] &= E^{(1)}_2[Q]\,, \\
      E^{(1)}_3[Q^F] &= E^{(1)}_3[Q]\,, \numberthis\label{eq:e1qF}
\end{align*}
i.e.~we choose the $\epsilon$ and $\epsilon^2$ coefficients to be identical for $Q$ and $Q^F$. For the operators with an even number of $\gamma$ matrices we allow for different coefficients in the evanescent operators
\begin{align*}
E^{(1)}[Q_S^F] =& - \frac23 (\overline{s}^c \gamma_{\mu_1} \gamma_{\mu_2} \sigma^{\mu\nu} \gamma^{\mu_2}\gamma^{\mu_1}  P_R b^d)\,(\overline{s}^d \sigma_{\mu\nu} P_R b^c) \\
&+ \frac13 (\overline{s}^c \gamma_{\mu_1} \gamma_{\mu_2} \sigma^{\mu\nu} P_R b^c)\,(\overline{s}^d \sigma_{\mu\nu} P_R \gamma^{\mu_2}\gamma^{\mu_1} b^d)  \\
&+ \left(8 + \tilde{a}^F \epsilon + \widetilde{b}^F \epsilon^2\right) \widetilde{Q}_S^F - \left(a^F\epsilon + b^F\epsilon^2\right) Q_S^F\,, \\
E^{(1)}[\widetilde{Q}_S^F] =& - \frac23 (\overline{s}^c \gamma_{\mu_1} \gamma_{\mu_2} \sigma^{\mu\nu} \gamma^{\mu_2}\gamma^{\mu_1} P_R b^c)\,(\overline{s}^d \sigma_{\mu\nu} P_R b^d)  \\
&+ \frac13 (\overline{s}^c \gamma_{\mu_1} \gamma_{\mu_2} \sigma^{\mu\nu} P_R b^d)\,(\overline{s}^d \sigma_{\mu\nu} P_R \gamma^{\mu_2}\gamma^{\mu_1} b^c)  \\
&+ \left(8 + \tilde{a}_2^F \epsilon + \tilde{b}_2^F \epsilon^2\right) Q_S^F - \left(a_2^F\epsilon + b_2^F\epsilon^2\right) \widetilde{Q}_S^F\,. \numberthis
\label{eq:e1qstF}
\end{align*}
Renormalising the $\langle Q \rangle^{(1), \text{ren}}$, $\langle Q_S \rangle^{(1), \text{ren}}$ and $\langle \widetilde{Q}_S \rangle^{(1), \text{ren}}$ amplitudes in the standard basis and the $\langle \widetilde{Q} \rangle^{(1), \text{ren}}$, $\langle Q_S^F \rangle^{(1), \text{ren}}$ and $\langle \widetilde{Q}_S^F \rangle^{(1), \text{ren}}$ amplitudes in the Fierz-transformed basis, we can then impose condition~1. For the renormalised physical matrix elements in the standard basis we insert the operators in Eqs.~\eqref{eq::opDB2}, \eqref{eq:E1_def} and \eqref{eq:E1QS_def} in tree-level and one-loop diagrams with the corresponding renormalisation constants. Analogously, for the Fierz-transformed basis the calculation is carried out with tree-level and one-loop diagrams of the operators in Eqs.~\eqref{eq:QSFt_def}, \eqref{eq:e1qF} and \eqref{eq:e1qstF}, using the renormalisation constants of that basis. Note that the renormalisation programme must be done off-shell, and we choose vanishing external momenta with $m_b = m_s \neq 0$ on internal lines.

In order to compare the amplitudes, the results need to be mapped onto the same basis of operator matrix elements, which can be either those of the Fierz-transformed or the regular basis. However, since the renormalised amplitudes are finite, all evanescent pieces can be discarded and tree-level matrix elements of the Fierz-transformed operators identified with their regular counterparts. For example, we demand the coefficient of $\langle Q \rangle$ in the renormalised matrix element $\langle Q \rangle^{(1), \text{ren}}$ to be equal to the coefficient of $\langle \widetilde Q \rangle$ in the renormalised matrix element $\langle \widetilde{Q} \rangle^{(1), \text{ren}}$. We arrive at the following result:
\begin{align*}
a=a_2&=0\,, & \tilde{a}=\tilde{a}_2&=-8\,, \\
a^F=a^F_2&=0\,, & \tilde{a}^F=\tilde{a}^F_2&=-8\,,\\
f=\tilde f &= -8\,. \numberthis
\label{eq:nlo_ev_consts}
\end{align*}
The evanescent operators obtained in this manner agree with the literature, see for example Refs.~\cite{Herrlich:1994kh, Beneke:1998sy, Gorbahn:2009pp, Gerlach:2022wgb}. 

At NNLO the $\epsilon^2$ terms in \eqsto{eq:E1_def}{eq:e1qstF} matter. Furthermore, one encounters evanescent operators of the second
generation:
\begin{align*}
E^{(2)}_1 [Q]&= 4 (\overline{s}^c \gamma_{\mu_1} \gamma_{\mu_2} \gamma_{\mu_3} \gamma_{\mu_4} \gamma_{\mu_5} P_L b^d)\,(\overline{s}^d \gamma^{\mu_5} \gamma^{\mu_4} \gamma^{\mu_3} \gamma^{\mu_2} \gamma^{\mu_1} P_L b^c) - \left(16 + \tilde{h}\epsilon + \tilde{k}\epsilon^2\right) \widetilde{Q},\\
E^{(2)}_2 [Q]&= 4 (\overline{s}^c \gamma_{\mu_1} \gamma_{\mu_2} \gamma_{\mu_3} \gamma_{\mu_4} \gamma_{\mu_5} P_L b^c)\,(\overline{s}^d  \gamma^{\mu_5} \gamma^{\mu_4} \gamma^{\mu_3} \gamma^{\mu_2} \gamma^{\mu_1} P_L b^d) - \left(16 + {h}\epsilon + {k}\epsilon^2\right) Q\,. \numberthis\label{eq:E2_def}
\end{align*}
For $Q_S$ and $\widetilde{Q}_S$ the corresponding evanescent operators are
\begin{align*}
E^{(2)}[Q_S] =&\, 4 (\overline{s}^c \gamma_{\mu_1} \gamma_{\mu_2} \gamma_{\mu_3} \gamma_{\mu_4} P_R b^c)\,(\overline{s}^d \gamma^{\mu_4} \gamma^{\mu_3} \gamma^{\mu_2} \gamma^{\mu_1} P_R b^d)\\
&\,+ \left(128 + \tilde{c}\epsilon + \tilde{d}\epsilon^2\right) \widetilde{Q}_S - \left(c\epsilon + d \epsilon^2\right) Q_S\,,\\
E^{(2)}[\widetilde{Q}_S] =\, & 4 (\overline{s}^c \gamma_{\mu_1} \gamma_{\mu_2} \gamma_{\mu_3} \gamma_{\mu_4} P_R b^d)\,(\overline{s}^d \gamma^{\mu_4} \gamma^{\mu_3} \gamma^{\mu_2} \gamma^{\mu_1} P_R b^c)\\
&\, + \left(128 + \tilde{c}_2\epsilon + \tilde{d}_2\epsilon^2\right) Q_S - \left(c_2\epsilon + d_2\epsilon^2\right) \widetilde{Q}_S\,.\numberthis\label{eq:E2QS_def}
\end{align*}
For the Fierz-transformed basis we again choose the same evanescent operators for $Q$ and $Q^F=\widetilde{Q}$, 
\begin{align*}
      E^{(2)}_1[Q^F] &= E^{(2)}_1[Q]\,, \\
      E^{(2)}_2[Q^F] &= E^{(2)}_2[Q]\,,\numberthis
\end{align*}
while for $Q_S^F$ and $\widetilde{Q}_S^F$ we define the second generation evanescent operators as
\begin{align*}
E^{(2)}[Q_S^F] \equiv &- \frac23 (\overline{s}^c \gamma_{\mu_1} \gamma_{\mu_2}\gamma_{\mu_3} \gamma_{\mu_4} \sigma^{\mu\nu} \gamma^{\mu_4} \gamma^{\mu_3} \gamma^{\mu_2} \gamma^{\mu_1}  P_R b^d)\,(\overline{s}^d \sigma_{\mu\nu} P_R b^c) \\
&+ \frac13 (\overline{s}^c \gamma_{\mu_1} \gamma_{\mu_2} \gamma_{\mu_3} \gamma_{\mu_4} \sigma^{\mu\nu} P_R b^c)\,(\overline{s}^d \sigma_{\mu\nu} P_R \gamma^{\mu_4} \gamma^{\mu_3} \gamma^{\mu_2} \gamma^{\mu_1} b^d) \\
&+ \left(128 + \tilde{c}^F \epsilon + \tilde{d}^F \epsilon^2\right) \widetilde{Q}_S^F - \left(c^F\epsilon + d^F\epsilon^2\right) Q_S^F\,,\\
E^{(2)}[\widetilde{Q}_S^F] \equiv &- \frac23 (\overline{s}^c \gamma_{\mu_1} \gamma_{\mu_2}\gamma_{\mu_3} \gamma_{\mu_4}  \sigma^{\mu\nu} \gamma^{\mu_4} \gamma^{\mu_3} \gamma^{\mu_2} \gamma^{\mu_1} P_R b^c)\,(\overline{s}^d \sigma_{\mu\nu} P_R b^d) \\
&+ \frac13 (\overline{s}^c \gamma_{\mu_1} \gamma_{\mu_2}\gamma_{\mu_3} \gamma_{\mu_4}  \sigma^{\mu\nu} P_R b^d)\,(\overline{s}^d \sigma_{\mu\nu} P_R  \gamma^{\mu_4} \gamma^{\mu_3} \gamma^{\mu_2} \gamma^{\mu_1}  b^c) \\
&+ \left(128 + \tilde{c}_2^F \epsilon + \tilde{d}_2^F \epsilon^2\right) Q_S^F - \left(c_2^F\epsilon + d_2^F\epsilon^2\right) \widetilde{Q}_S^F\,.\numberthis
\end{align*}
The comparison of the renormalised physical matrix elements is completely analogous to the NLO calculation. Renormalising the $\langle Q \rangle^{(2), \text{ren}}$, $\langle Q_S \rangle^{(2), \text{ren}}$ and $\langle \widetilde{Q}_S \rangle^{(2), \text{ren}}$ amplitudes in the standard basis and the $\langle \widetilde{Q} \rangle^{(2), \text{ren}}$, $\langle Q_S^F \rangle^{(2), \text{ren}}$ and $\langle \widetilde{Q}_S^F \rangle^{(2), \text{ren}}$ amplitudes in the Fierz-transformed basis, we can then impose condition 1. Comparing just the $N_f$ terms and enforcing condition~2, we obtain the remaining coefficients of order $\epsilon^2$ in the first generation evanescent operators:
\begin{align*}
b=b_2&=4\,, & \tilde{b}=\tilde{b}_2&=0\,, \label{eq:first_ev_const}\\
b^F=b^F_2&=4\,, & \tilde{b}^F=\tilde{b}^F_2&=0\,,\\
g=\tilde{g}&=4\,.\numberthis
\end{align*}
This is consistent with the literature, see for example Ref.~\cite{Asatrian:2017qaz}.

From the remaining equations we get a solution space for the $c$, $d$, $h$ and $k$ coefficients of the second generation evanescent operators. We then solve the equations arising from condition 1 for $\{c, \tilde{c}, d, \tilde{d}, h, \tilde{h}\}$, leaving the rest of the constants undetermined, which marks a particular choice of a set of solutions. To further impose condition~3, we require the coefficients of $N_c$ and $N_c^2$ in the solutions of the evanescent constants to vanish. This chosen solution set consistent with all physical conditions is given in Appendix \ref{app:full_ev_consts}.

For concreteness, we can further pick a particular, convenient solution. For example, we can choose as many of the constants on the regular operator side to be equal to zero,
\begin{equation}
c = c_2 = \tilde c =  d = d_2 = \tilde d = \tilde d_2 = 0\,,
\end{equation}
and drop the $\epsilon^2$ terms of the second generation evanescent operators of $Q$ and $\widetilde Q$,
\begin{equation}
k = \tilde k = 0\,.
\end{equation}
With these additional constraints, the remaining evanescent constants are fixed uniquely:
\begin{align*}
\tilde c_2 &= -1024\,,\\
c^F &= \frac{-256(534 - 344\,N_c - 119\,N_c^2 - 366\,N_c^3 - 116\,N_c^4 + 163\,N_c^5)}{15(8 - 16\,N_c^2 + 2\,N_c^3 + 2\,N_c^4 + N_c^5)}\,,\\
c^F_2 &= \frac{256(92 + 46 \,N_c + 164\,N_c^2 + 104\,N_c^3 + 141\,N_c^4 + 17\,N_c^5)}{15(8 - 16\,N_c^2 + 2\,N_c^3 + 2\,N_c^4 + N_c^5)}\,,\\
\tilde{c}^F &= \frac{-128(196 - 220\,N_c + 26\,N_c^2 - 445\,N_c^3 + 256\,N_c^4 + 226\,N_c^5)}{15(8 - 16\,N_c^2 + 2\,N_c^3 + 2\,N_c^4 + N_c^5)}\,,\\
\tilde{c}^F_2 &= \frac{128(172 - 608\,N_c - 138\,N_c^2 - 730\,N_c^3 + 120\,N_c^4 + 25\,N_c^5)}{15(8 - 16\,N_c^2 + 2\,N_c^3 + 2\,N_c^4 + N_c^5)}\,,\\
d^F &= \frac{-32(1958 - 2608\,N_c + 6957\,N_c^2 - 3572\,N_c^3 - 2697\,N_c^4 + 1391\,N_c^5)}{15(8 - 16\,N_c^2 + 2\,N_c^3 + 2\,N_c^4 + N_c^5)}\,,\\
d^F_2 &= \frac{64(92 +46\,N_c + 164\,N_c^2 + 104\,N_c^3 + 141\,N_c^4 + 17\,N_c^5)}{15(8 - 16\,N_c^2 + 2\,N_c^3 + 2\,N_c^4 + N_c^5)}\,,\\
\tilde d^F &= \frac{-32(-964 - 1550\,N_c + 2696\,N_c^2 - 3610\,N_c^3 + 411\,N_c^4 + 1061\,N_c^5)}{15(8 - 16\,N_c^2 + 2\,N_c^3 + 2\,N_c^4 + N_c^5)}\,,\\
\tilde d^F_2 &= \frac{32(1132 - 608\,N_c - 2058\,N_c^2 - 490\,N_c^3 + 360\,N_c^4 + 145\,N_c^5)}{15(8 - 16\,N_c^2 + 2\,N_c^3 + 2\,N_c^4 + N_c^5)}\,,\\
h &=\frac{-64(-98  -158\,N_c + 23\,N_c^2 + 30\,N_c^3)}{-14 - 14\,N_c - 7\,N_c^2 + 6\,N_c^3}\,,\\
\tilde h &= -448\,.\numberthis\label{eq:last_ev_const}
\end{align*}
This is a particularly nice solution because the second generation evanescent operators on the regular operator side for $Q_S$ and $\widetilde{Q}_S$ have only one non-vanishing constant, $\tilde c_2$. Moreover, this constant turns out to be $N_c$-independent. This is at the expense of the Fierz-transformed operators, which are significantly more complicated but for practical calculations not relevant since the Fierz-transformed basis is not used. The second generation evanescent operators for $Q$ and $\widetilde{Q}$ have also been simplified by removing all $\epsilon^2$ terms. 
A computer-readable file is provided for 
the specific choice given here, see Appendix~\ref{app:full_ev_consts}.

It is worth noting that the evanescent operators are now QCD-specific, i.e.~they contain an explicit $N_c$ dependence. This feature arises because we want the \emph{renormalised} matrix elements to be invariant under a Fierz transformation, see condition 1.

For the matching calculation with the current-current operators of the $\Delta B = 1$ theory described in Section \ref{sub::match}, evanescent operators of the second generation  
must be included in the matching to cancel IR poles 
and are renormalised at NNLO. For this purpose we do not need to specify the $\mathcal{O}(\epsilon)$ terms of evanescent operators of higher generations, but we need to define them with the correct vanishing $\epsilon$-finite part. Therefore, we define the third generation as 
\begin{align*}
E^{(3)}_1 =&\, 4 (\overline{s}^c \gamma_{\mu_1} \gamma_{\mu_2} \gamma_{\mu_3} \gamma_{\mu_4} \gamma_{\mu_5} \gamma_{\mu_6} \gamma_{\mu_7} P_L b^d)\,(\overline{s}^d \gamma^{\mu_7} \gamma^{\mu_6} \gamma^{\mu_5} \gamma^{\mu_4} \gamma^{\mu_3} \gamma^{\mu_2} \gamma^{\mu_1} P_L b^c)\\
&\, - \left(64 + \mathcal{O}(\epsilon) \right) \widetilde{Q}\,,\\
E^{(3)}_2=&\, 4 (\overline{s}^c \gamma_{\mu_1} \gamma_{\mu_2} \gamma_{\mu_3} \gamma_{\mu_4} \gamma_{\mu_5} \gamma_{\mu_6} \gamma_{\mu_7} P_L b^c)\,(\overline{s}^d  \gamma^{\mu_7} \gamma^{\mu_6} \gamma^{\mu_5} \gamma^{\mu_4} \gamma^{\mu_3} \gamma^{\mu_2} \gamma^{\mu_1} P_L b^d)\\
&\, - \left(64 + \mathcal{O}(\epsilon) \right) Q\,,\\
E^{(3)}[Q_S] =&\, 4 (\overline{s}^c \gamma_{\mu_1} \gamma_{\mu_2} \gamma_{\mu_3} \gamma_{\mu_4} \gamma_{\mu_5} \gamma_{\mu_6} P_R b^c)\,(\overline{s}^d \gamma^{\mu_6} \gamma^{\mu_5} \gamma^{\mu_4} \gamma^{\mu_3} \gamma^{\mu_2} \gamma^{\mu_1} P_R b^d)\\
&\, + \left(2048 + \mathcal{O}(\epsilon) \right) \widetilde{Q}_S + \mathcal{O}(\epsilon)  Q_S\,,\\
E^{(3)}[\widetilde{Q}_S] =&\, 4 (\overline{s}^c \gamma_{\mu_1} \gamma_{\mu_2} \gamma_{\mu_3} \gamma_{\mu_4} \gamma_{\mu_5} \gamma_{\mu_6} P_R b^d)\,(\overline{s}^d \gamma^{\mu_6} \gamma^{\mu_5} \gamma^{\mu_4} \gamma^{\mu_3} \gamma^{\mu_2} \gamma^{\mu_1} P_R b^c)\\
&\, + \left(2048 + \mathcal{O}(\epsilon) \right) Q_S + \mathcal{O}(\epsilon)  \widetilde{Q}_S\,,\numberthis\label{eq:E3_def}
\end{align*}
and the fourth generation as
\begin{align*}
E^{(4)}_1 =&\, 4 (\overline{s}^c \gamma_{\mu_1} \gamma_{\mu_2} \gamma_{\mu_3} \gamma_{\mu_4} \gamma_{\mu_5} \gamma_{\mu_6} \gamma_{\mu_7} \gamma_{\mu_8} \gamma_{\mu_9} P_L b^d)\,(\overline{s}^d \gamma^{\mu_9} \gamma^{\mu_8} \gamma^{\mu_7} \gamma^{\mu_6} \gamma^{\mu_5} \gamma^{\mu_4} \gamma^{\mu_3} \gamma^{\mu_2} \gamma^{\mu_1} P_L b^c)\\
&\, - \left(256 + \mathcal{O}(\epsilon) \right) \widetilde{Q}\,,\\
E^{(4)}_2 =&\, 4 (\overline{s}^c \gamma_{\mu_1} \gamma_{\mu_2} \gamma_{\mu_3} \gamma_{\mu_4} \gamma_{\mu_5} \gamma_{\mu_6} \gamma_{\mu_7} \gamma_{\mu_8} \gamma_{\mu_9}P_L b^c)\,(\overline{s}^d \gamma^{\mu_9} \gamma^{\mu_8} \gamma^{\mu_7} \gamma^{\mu_6} \gamma^{\mu_5} \gamma^{\mu_4} \gamma^{\mu_3} \gamma^{\mu_2} \gamma^{\mu_1} P_L b^d)\\
&\, - \left(256 + \mathcal{O}(\epsilon) \right) Q\,,\\
E^{(4)}[Q_S] =&\, 4 (\overline{s}^c \gamma_{\mu_1} \gamma_{\mu_2} \gamma_{\mu_3} \gamma_{\mu_4} \gamma_{\mu_5} \gamma_{\mu_6} \gamma_{\mu_7} \gamma_{\mu_8} P_R b^c)\,(\overline{s}^d \gamma^{\mu_8} \gamma^{\mu_7} \gamma^{\mu_6} \gamma^{\mu_5} \gamma^{\mu_4} \gamma^{\mu_3} \gamma^{\mu_2} \gamma^{\mu_1} P_R b^d)\\
&\,+ \left(32768 + \mathcal{O}(\epsilon) \right) \widetilde{Q}_S + \mathcal{O}(\epsilon)  Q_S\,,\\
E^{(4)}[\widetilde{Q}_S] =&\, 4 (\overline{s}^c \gamma_{\mu_1} \gamma_{\mu_2} \gamma_{\mu_3} \gamma_{\mu_4} \gamma_{\mu_5} \gamma_{\mu_6} \gamma_{\mu_7} \gamma_{\mu_8} P_R b^d)\,(\overline{s}^d \gamma^{\mu_8} \gamma^{\mu_7} \gamma^{\mu_6} \gamma^{\mu_5} \gamma^{\mu_4} \gamma^{\mu_3} \gamma^{\mu_2} \gamma^{\mu_1} P_R b^c)\\
&\,+ \left(32768 + \mathcal{O}(\epsilon) \right) Q_S + \mathcal{O}(\epsilon)  \widetilde{Q}_S\,.\numberthis\label{eq:E4_def}
\end{align*}
The $\mathcal{O}(\epsilon)$ terms can be kept as arbitrary constants, and it can be checked that they drop out of the physical matching coefficients at the end.

\subsection{Renormalisation of the effective {$\mathbf{\Delta B = 2}$} theory}

In the following we describe the generic renormalisation procedure for the effective $\Delta B=2$ theory. We choose to renormalise the matching coefficients, 
\begin{equation}
  (\vec{C}^{\textrm{bare}}, \vec{C}_E^{\textrm{bare}}) =
  (\vec{C}^{\textrm{ren}}, \vec{C}_E^{\textrm{ren}}) \, Z \equiv
  (\vec{C}^{\textrm{ren}}, \vec{C}_E^{\textrm{ren}})  \begin{pmatrix} Z_{QQ} &
    Z_{QE} \\ Z_{EQ} & Z_{EE} \end{pmatrix}\,, \label{eq:zsub}
\end{equation}
where the sub-matrices $Z_{XY}$ with $XY \in \{QQ,EE\}$ of the matrix $Z$ are
understood to be an expansion in $\alpha_s$, \ie
\begin{equation}
  (Z_{XY})_{ij} = \delta_{ij} + \frac{\alpha_s}{4\pi} \frac{1}{\epsilon} \left(Z_{XY}^{(1,1)}\right)_{ij} + \mathcal{O}(\alpha_s^2)\,,
\end{equation}
and for the $QE$ part
\begin{equation}
  (Z_{QE})_{ij} = \frac{\alpha_s}{4\pi}\frac{1}{\epsilon} \left(Z_{QE}^{(1,1)}\right)_{ij}  + \mathcal{O}(\alpha_s^2)\,.
\end{equation}
The sub-matrix
$Z_{EQ}$ is special because it introduces \emph{finite} counterterms \cite{Buras:1989xd,Herrlich:1994kh}:
\begin{equation}
  (Z_{EQ})_{ij} = \frac{\alpha_s}{4\pi}   \left(Z_{EQ}^{(1,0)}\right)_{ij}  + \mathcal{O}(\alpha_s^2)\,.
\end{equation}
Note, however, that the $\alpha_s^2$ terms may contain both $\epsilon^0$ and $1/\epsilon$ terms.

%Let us also briefly explain the meaning of each sub-matrix of $Z$:
%\begin{itemize}
%\item $Z_{QQ}$ describes the mixing of the physical operators among
%  themselves under renormalisation. That is, beyond leading order a $C^i$
%  can generate a contribution proportional to $\frac{\alpha_s}{\epsilon} C^j$ with
%  $i \neq j$.
%\item $Z_{EE}$ describes the mixing of the evanescent operators among
%  themselves under renormalisation. That is, beyond leading order a
%  $C_E^i$ can generate a contribution proportional to
%  $ \frac{\alpha_s}{\epsilon} C_E^j$ with $i \neq j$.
%\item $Z_{QE}$ describes the mixing of the physical operators into evanescent
%  operators under renormalisation. That is, beyond leading order a $C_E^i$
%  can generate a contribution proportional to $\frac{\alpha_s}{\epsilon} C^j$.
%\item $Z_{EQ}$ describes the mixing of the evanescent operators into the
%  physical operators under renormalisation. That is, beyond leading order
%  a $C^i$ can generate a contribution proportional to $\alpha_s C_E^j$.
%\end{itemize}
To illustrate the meaning of the submatrices in \eq{eq:zsub} we consider 
$Z_{QE}$, which describes those counterterms to physical operators which are proportional 
to evanescent operators. When one assigns the counterterms to the Wilson coefficients instead of the operators,
$Z_{QE}$ involves  a contribution to $C_E^i$ which is proportional to $\frac{\alpha_s}{\epsilon} C^j$. 
The other submatrices correspond to the mixings among other subsets of operators. 
For our calculation we need $Z_{XY}$ to two-loop order. 
One-loop results can be found in Ref.~\cite{Gerlach:2022wgb} and
two-loop terms are presented in Appendix~\ref{app::Z}. The renormalisation of the operator $R_0$ is also 
needed and is discussed in Section \ref{sec:R0_ren}.

%- }}}
%- {{{ Matching of $\Delta B=1$ and $\Delta B=2$ theory:

%- }}}
%- {{{ R0:

\subsection{Renormalisation of the physical operators with a power-suppressed $\mathbf{R_0}$ }
\label{sec:R0_ren}

A proper treatment of the $\mathcal{O}(1/m_b)$ suppression of $\langle B_s | R_0 | \bar B_s\rangle$ requires a discussion of its renormalisation. The operator renormalisation involves three physical operators, which one can choose as $\{Q,\widetilde Q_S, Q_S\}$ or $\{Q,\widetilde Q_S, R_0\}$ because the renormalisation procedure is process-independent and the structure of the UV divergences is insensitive to the power suppression of $\langle B_s | R_0 | \bar B_s\rangle$. This power-suppression is a feature of the particular physical processes studied by us and calls for a finite counterterm.

The starting point of our discussion is the basis of physical operators $\{ Q, \widetilde{Q}_S, Q_S\}$, which is convenient because $\widetilde Q_S$ and $Q_S$ do not mix with $Q$ under renormalisation, meaning that their renormalisation constant is a $2\times 2$ matrix. As described in Section \ref{sec:physical_db2} we will later trade $Q_S$ for $R_0$, which results in a $3\times 3$ renormalisation matrix for $Q$, $\widetilde Q_S$ and $R_0$. However, this does not simplify our calculation since we cannot discard $R_0$ at this point. This is because renormalisation belongs to the \emph{off-shell}\ structure of our effective field theory while the power-suppression of $\langle B_s | R_0 | \bar B_s\rangle$ is an \emph{on-shell} phenomenon.

The $\ov{\rm MS}$ renormalised physical operators $\{ Q, \widetilde{Q}_S, Q_S\}$ are
\begin{align}
   Q&= Z_Q Q^{\rm bare}+\textrm{ev.}\, , &  
   %\begin{pmatrix} Q_S & \widetilde Q_S \end{pmatrix} &=
    %                     Z_{Q_S\widetilde Q_S}  \begin{pmatrix} Q_S^{\rm
     %                        bare}
      %                     & \widetilde Q_S^{\rm bare} \end{pmatrix}
 \begin{pmatrix} Q_S \\[7pt] \widetilde Q_S \end{pmatrix}
 &= Z_{Q_S\widetilde Q_S}  \begin{pmatrix} Q_S^{\rm
                             bare} \\[7pt] \widetilde Q_S^{\rm bare} \end{pmatrix}
                             +\textrm{ev.}\,,
                             \label{eq:zqq}
 \end{align}  
 where we have suppressed the mixing with evanescent operators, which are discussed in Section \ref{sec:eva}.
 We implement the $\ov{\rm MS}$ scheme by choosing
 $g^{\mathrm{bare}}= \bar\mu^\epsilon Z_g g$ with
 $\bar\mu^2=\mu^2e^{-\gamma_E}/(4\pi)$,
 such that the renormalisation constants only involve poles. All fields in the operators are understood as bare, so the quark fields are renormalised separately. 

 We expand all renormalisation constants  as
 \begin{align*}
   Z&= 1 + \frac{\alpha_s}{4\pi} \lt[ \frac{Z^{(11)}}{\epsilon} +
        Z^{(10)}\rt] + \lt( \frac{\alpha_s}{4\pi}\rt)^2
       \lt[ \frac{Z^{(22)}}{\epsilon^2} + \frac{Z^{(21)}}{\epsilon} +
        Z^{(20)}\rt] .
 \end{align*} 
For the  $\ov{\rm MS}$ constants in \eq{eq:zqq} one finds 
\begin{align*}
 Z_Q^{(11)}  &= 2\,, \qquad\qquad  Z_Q^{(22)} = -9 + \frac23 N_f\,, \qquad\qquad Z_Q^{(21)} = -\frac{45815}{516} + \frac19 N_f\,,\\[7pt]
  Z_{Q_S\widetilde Q_S}^{(11)} &= \begin{pmatrix} -\frac{14}{3} & \frac23 \\[7pt] \frac83 & \frac{16}{3} \end{pmatrix}\,, \qquad\qquad Z_{Q_S\widetilde Q_S}^{(22)} = \begin{pmatrix} \frac{337}{9} - \frac{14  N_f}{9} & -\frac{31}{9} + \frac{2 N_f}{9}  \\[7pt] -\frac{124}{9} + \frac{8 N_f}{9} & -\frac{128}{9}+ \frac{16 N_f}{9} \end{pmatrix},\\[7pt]
 Z_{Q_S\widetilde Q_S}^{(21)}
  &= \begin{pmatrix} \frac{547}{9} + \frac{22 N_f}{27} & -\frac{227}{9} + \frac{2 N_f}{27} \\[7pt] \frac{1235}{6}-\frac{19 N_f}{27} & \frac{641}{18} - \frac{83 N_f}{27} \end{pmatrix}\,, \numberthis
\end{align*}  
with the finite constants $ Z^{(j0)}$ vanishing. Note that the $Z^{(21)}$ terms depend on the choice of the evanescent operators. Here we show the results in the basis as specified in Eqs.~\eqref{eq:first_ev_const} to \eqref{eq:last_ev_const} and for $N_c=3$.

It is possible to switch to the basis $\{Q,\widetilde Q_S, R_0, E\}$, where $E$ stands for the evanescent operators. This basis leads to the renormalisation matrix $Z$:
\begin{equation}
   \begin{pmatrix} Q \\[7pt] \widetilde Q_S \\[7pt] R_0 \\[7pt] E \end{pmatrix} = Z  \begin{pmatrix}
     Q^{\rm bare} \\[7pt] \widetilde Q_S^{\rm bare} \\[7pt] R_0^{\rm bare} \\[7pt] E^{\rm bare} \end{pmatrix}\,. 
     \label{eq:ZR0_def}
\end{equation}
The results from the $\ov{\rm MS}$ scheme for the  $\{ Q, \widetilde{Q}_S, Q_S, E\}$ basis can be naively transformed to the $\{ Q, \widetilde{Q}_S, R_0, E\}$ basis by equating the renormalised matrix elements, 
\begin{align*}
\begin{pmatrix}
    Q \\[7pt]
    \widetilde{Q}_S  \\[7pt]
     \frac12 Q + \widetilde{Q}_S +  Q_S  \\[7pt]
     E
    \end{pmatrix}_{\{Q,Q_S,\widetilde{Q}_S\}} &=\quad \begin{pmatrix}
    Q \\[7pt]
    \widetilde{Q}_S  \\[7pt]
    R_0\\[7pt]
    E
    \end{pmatrix}_{\{Q,\widetilde{Q}_S,R_0\}}^\text{naive}\,, \numberthis
    \label{eq:naive_trafo}
\end{align*}
where Eqs.~\eqref{eq:zqq} and \eqref{eq:ZR0_def} are used to renormalise the left-hand side and right-hand side, respectively. The bare matrix elements of the two bases are related by Eq.~\eqref{eq::R0}. This transformation yields the ``naive'' renormalisation matrix of the $\{ Q, \widetilde{Q}_S, R_0, E\}$ basis,
\begin{equation}
    Z^{\text{naive}} = (Z_1^\text{naive}, Z_2^\text{naive}, Z_3^\text{naive}, Z_4^\text{naive}),
\end{equation}
where
\begin{align*}
    Z_1^{\text{naive}}  & = \begin{pmatrix}
        Z_Q\\[7pt] -\frac12 \left(Z_{Q_S\widetilde Q_S}\right)_{21} \\[7pt] \frac12 \lt( Z_Q \!- \left(Z_{Q_S\widetilde Q_S}\right)_{11}\!-  \left(Z_{Q_S\widetilde Q_S}\right)_{21}\rt)\\[7pt] Z_{EQ} - \frac12 Z_{E Q_S}
    \end{pmatrix}\,,\\[7pt]
    Z_2^{\text{naive}}  &= \begin{pmatrix}
        0 \\[7pt] - \left(Z_{Q_S\widetilde Q_S}\right)_{21} \! + \left(Z_{Q_S\widetilde Q_S}\right)_{22} \\ -\left(Z_{Q_S\widetilde Q_S}\right)_{11} \!+ \left(Z_{Q_S\widetilde Q_S}\right)_{12} \! -  \left(Z_{Q_S\widetilde Q_S}\right)_{21}\!+ \left(Z_{Q_S\widetilde Q_S}\right)_{22}  \\[7pt] Z_{E \widetilde{Q}_S} - Z_{E Q_S}
        \end{pmatrix}\,,\\[7pt]
    Z_3^{\text{naive}}  & = \begin{pmatrix}
        0 \\[7pt] \left(Z_{Q_S\widetilde Q_S}\right)_{21} \\[7pt] \left(Z_{Q_S\widetilde Q_S}\right)_{11} \! +  \left(Z_{Q_S\widetilde Q_S}\right)_{21} \\[7pt] Z_{E Q_S}
    \end{pmatrix}\,,\\[7pt]
    Z_4^{\text{naive}}  &= \begin{pmatrix}
        Z_{Q E} \\[7pt] Z_{\widetilde{Q}_S E} \\[7pt] \frac12  Z_{Q E} + Z_{Q_S E} +  Z_{\widetilde{Q}_S E}\\[7pt] Z_{EE}
    \end{pmatrix}\,.\numberthis\label{eq:Z_naive}
\end{align*}
Beyond tree-level, however, the matrix element of $R_0$ is not suppressed by $1/m_b$ and one must add a finite renormalisation to $R_0$, which is not captured by Eq.~\eqref{eq:Z_naive} \cite{Beneke:1998sy}. 
This feature stems from the fact that the $\overline{\rm MS}$ renormalised $\langle R_0 \rangle$ 
involves terms proportional to $\langle Q\rangle^{(0)}$ and $\langle \widetilde Q_S\rangle^{(0)}$ in the operator basis $\{Q,\widetilde{Q}_S,R_0\}$. Moreover, $R_0$ has an unsuppressed evanescent part which enters the calculation whenever IR singularities are regularised dimensionally, i.e.~when matching calculations are performed for $D\neq 4$. This requires us to renormalise the evanescent part of $R_0$ in the same way as other evanescent operators~\cite{Gerlach:2022wgb}.

To visualise how the finite renormalisation constants arise, it is instructive to consider the renormalised matrix element of $R_0$ as a linear combination of the renormalised $Q$, $\widetilde{Q}_S$ and $Q_S$ matrix elements,
\begin{equation}
  \braket{R_0}
  = \frac{1}{2} \alpha_1 \braket{Q}
  + \alpha_2 \braket{\widetilde{Q}_S}
      + \braket{Q_S}\,,
      \label{eq::R0_ME}
\end{equation}
which we require to be $1/m_b$-suppressed. Here and in the following we use the shorthand notation $\langle O \rangle$ to denote $\langle B_s| O |\bar{B}_s \rangle$. The constants $\alpha_1$ and $\alpha_2$ have an expansion
in $\alpha_s$,
\begin{equation}
  \alpha_i = 1 + \frac{\alpha_s}{4\pi} \alpha_i^{(1)}
               + \left(\frac{\alpha_s}{4\pi}\right)^2 \alpha_i^{(2)}
               + \mathcal{O}(\alpha_s^3)\,. \label{eq:alpha12_def}
\end{equation}
The coefficient of one of the operators, which we choose to be $Q_S$, can be set to one. This corresponds to a perturbative redefinition of the evanescent part of $R_0$. 
We discuss the calculation of $\alpha_{1,2}$ in detail in Section \ref{sec:alpha12_finite}.

The $\alpha_i^{(j)}$ constants are UV-finite quantities and will appear as finite pieces in the renormalisation constants for $R_0$. The true renormalisation matrix in the $\{Q,\widetilde{Q}_S,R_0\}$ basis is obtained in the same way as the naive renormalisation matrix except that Eq.~\eqref{eq:naive_trafo} is modified to read
\begin{equation}
    \begin{pmatrix}
    Q \\[7pt]
    \widetilde{Q}_S  \\[7pt]
     \frac12 \alpha_1 Q + \alpha_2 \widetilde{Q}_S + Q_S   \\[7pt]
     E
    \end{pmatrix}_{\{Q,Q_S,\widetilde{Q}_S\}} =\quad \begin{pmatrix}
    Q \\[7pt]
    \widetilde{Q}_S  \\[7pt]
    R_0  \\[7pt]
    E
    \end{pmatrix}_{\{Q,\widetilde{Q}_S,R_0\}}\,, \numberthis
    \label{eq:true_trafo}    
\end{equation}
This leads to the correct renormalisation matrix 
\begin{equation}
    Z = (Z_1, Z_2, Z_3, Z_4),
\end{equation}
where
\begin{align*}
    Z_1 & = \begin{pmatrix}
        Z_Q\\[7pt] -\frac12 \left(Z_{Q_S\widetilde Q_S}\right)_{21} \\[7pt] \frac12 \lt( \alpha_1 Z_Q \!- \left(Z_{Q_S\widetilde Q_S}\right)_{11}\!- \alpha_2 \left(Z_{Q_S\widetilde Q_S}\right)_{21}\rt)\\[7pt] Z_{EQ} - \frac12 Z_{E Q_S}
    \end{pmatrix}\,,\\[7pt]
    Z_2 &= \begin{pmatrix}
        0 \\[7pt] - \left(Z_{Q_S\widetilde Q_S}\right)_{21} \! + \left(Z_{Q_S\widetilde Q_S}\right)_{22} \\[7pt] -\left(Z_{Q_S\widetilde Q_S}\right)_{11} \!+ \left(Z_{Q_S\widetilde Q_S}\right)_{12} \! - \alpha_2 \left(Z_{Q_S\widetilde Q_S}\right)_{21}\!+ \alpha_2 \left(Z_{Q_S\widetilde Q_S}\right)_{22}  \\[7pt] Z_{E \widetilde{Q}_S} - Z_{E Q_S}
        \end{pmatrix}\,,\\[7pt]
    Z_3 & = \begin{pmatrix}
        0 \\[7pt] 
        \left(Z_{Q_S\widetilde Q_S}\right)_{21} \\[7pt] 
        \left(Z_{Q_S\widetilde Q_S}\right)_{11} \! + \alpha_2 \left(Z_{Q_S\widetilde Q_S}\right)_{21} 
        \\[7pt] Z_{E Q_S}
    \end{pmatrix}\,,\\[7pt]
    Z_4 &= \begin{pmatrix}
        Z_{Q E} \\[7pt]
        Z_{\widetilde{Q}_S E} \\[7pt]
        \frac12 \alpha_1 Z_{Q E} + Z_{Q_S E} + \alpha_2 Z_{\widetilde{Q}_S E}\\[7pt] 
        Z_{EE}
    \end{pmatrix}\,.\numberthis\label{eq:Z_true}
\end{align*}
Here it becomes apparent that the coefficients $\alpha_{1}$ and $\alpha_2$ induce $\epsilon$-finite renormalisation constants because they multiply the diagonal elements $Z_Q$ and $(Z_{Q_S \widetilde{Q}_S})_{22}$ whose leading term is one.

\subsection{Determination of the finite renormalisation constants for $R_0$}
\label{sec:alpha12_finite}

The one-loop corrections $\alpha_1^{(1)}$ and $\alpha_2^{(1)}$ in Eq.~\eqref{eq:alpha12_def} have been
computed in Ref.~\cite{Beneke:1998sy} and the two-loop fermionic corrections
in~\cite{Asatrian:2017qaz}. In the following we describe our calculation of
the two-loop terms $\alpha_1^{(2)}$ and $\alpha_2^{(2)}$.

\begin{figure}
    \centering
    \includegraphics[scale=0.55]{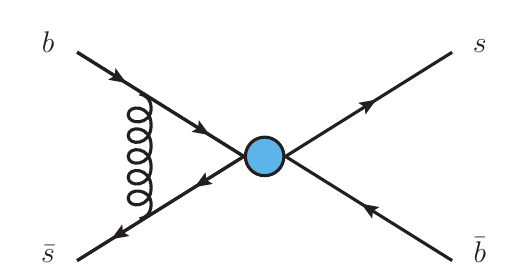} \quad
    \includegraphics[scale=0.55]{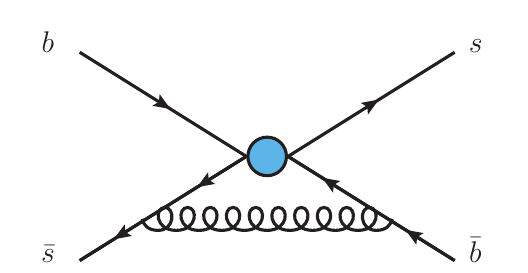} \quad
    \includegraphics[scale=0.55]{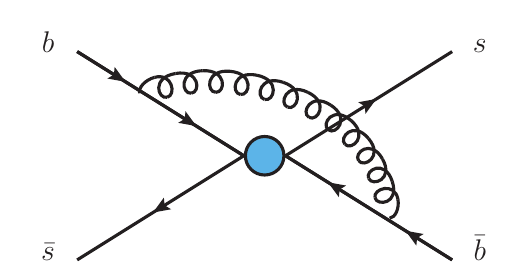}\\
    \centering
    \includegraphics[scale=0.55]{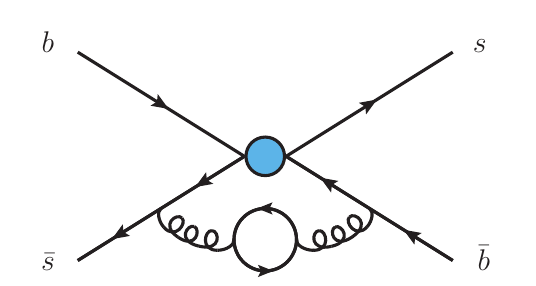}\quad
    \includegraphics[scale=0.55]{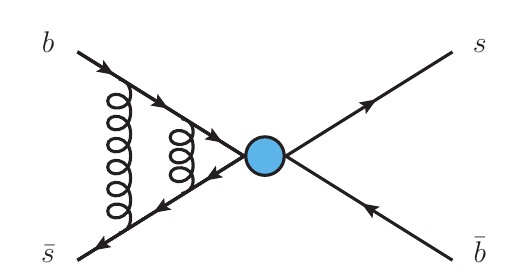}\quad
    \includegraphics[scale=0.55]{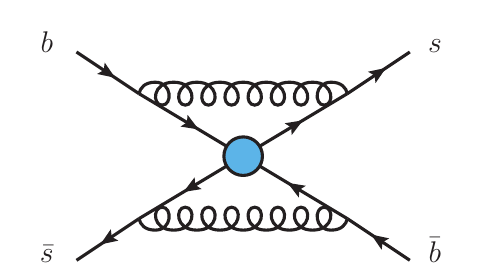}
    \caption{\label{fig::DB2mg}One- and two-loop Feynman diagrams needed for the calculation of
      $\alpha_1$ and $\alpha_2$. The blob stands for the insertion of one of the
      operators $Q$, $Q_S$ or $\widetilde{Q}_S$.
      }%~\\[-5mm]
%\hrule
\end{figure}

For the calculation of $\alpha_1$ and $\alpha_2$ it is important to
distinguish UV and IR divergences since otherwise $R_0$ is accompanied by an
unphysical evanescent operator
%, $E_{R_0}$, 
that is not suppressed by
$1/m_b$~\cite{Gerlach:2022wgb}. In the matching calculation we want to use
$R_0$ from Eq.~(\ref{eq::R0}) where $\alpha_1$ and $\alpha_2$ encode a finite UV
renormalisation which guarantees the $1/m_b$ suppression of $\braket{R_0}$. For
this purpose we have to compute the operator matrix elements $\braket{Q}$,
$\braket{Q_S}$ and $\braket{\widetilde{Q}_S}$ up to two loops. Sample Feynman
diagrams are shown in Fig.~\ref{fig::DB2mg}. 

For our calculation we use Feynman gauge and
we isolate the UV divergences by introducing a gluon mass $m_g$ in a minimal way 
into the gluon and ghost propagators via
\begin{eqnarray}
  \frac{i \delta^{ab} }{-p^2} 
  \left(g^{\mu \nu} + \xi \frac{p^\mu p^\nu}{-p^2}\right)
  &\to&  
        \frac{i \delta^{ab} }{m_g^2 - p^2}
        \left(g^{\mu \nu} + \xi \frac{p^\mu p^\nu}{m_g^2-p^2}\right)
        \,,
        \nonumber\\
	\frac{i \delta^{ab}}{-p^2}  &\to& \frac{i \delta^{ab}}{m_g^2 - p^2}\,,
                                          \label{eq::mg}
\end{eqnarray}
where $a$ and $b$ are colour indices.  
Note that we have to consider the gluon propagator with a general
gauge parameter $\xi$ since we have to renormalise $\xi$. 
Our final results for $\alpha_1$ and $\alpha_2$ do not depend on the gauge parameter.

The calculation of the diagrams
proceeds as described in Section~\ref{sec::technicalities}. 
The charm quark appears in diagrams with a closed quark loop insertion into the gluon propagator, cf.~Fig.~\ref{fig::DB2mg}. The charm-dependent contributions are taken from Ref.~\cite{Asatrian:2017qaz}.
When
preparing the amplitude, we expand in $m_g$ in those contributions where no IR
divergences are introduced. In the remaining parts we keep the gluon mass,
also during reduction to the master integrals.

\begin{figure}[t]
    \centering
    \begin{tabular}{ccccc}
    \includegraphics[width=0.8cm]{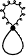}&
    \includegraphics[width=0.8cm]{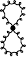}&
    \includegraphics[width=1.5cm]{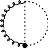}&
    \includegraphics[width=1.5cm]{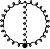}&
    \raisebox{0.25cm}{\includegraphics[width=2.5cm]{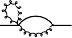}}\\
    \raisebox{0.25cm}{\includegraphics[width=2.5cm]{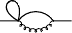}}&
    \includegraphics[width=2.5cm]{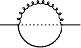}&
    \includegraphics[width=2.5cm]{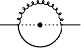}&
    \includegraphics[width=2.5cm]{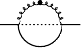}&
    \includegraphics[width=2.5cm]{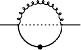}\\
    \includegraphics[width=2.5cm]{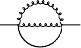}&
    \includegraphics[width=2.5cm]{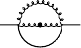}&
    \includegraphics[width=2.5cm]{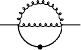}&
    \raisebox{0.35cm}{\includegraphics[width=2.5cm]{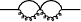}}&
    \includegraphics[width=2.5cm]{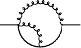}\\
    \includegraphics[width=2.5cm]{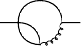}&
    \includegraphics[width=2.5cm]{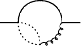}&
    \includegraphics[width=2.5cm]{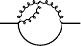}&
    \includegraphics[width=2.5cm]{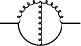}&
    \includegraphics[width=2.5cm]{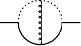}\\
    \includegraphics[width=2.5cm]{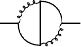}&
    \includegraphics[width=2.5cm]{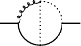}&
    \includegraphics[width=2.5cm]{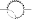}
%    \centerhfill 
    \end{tabular}
    \caption{\label{fig::DB2mg_MI}Two-loop master integrals where all lines
      represent scalar propagators.  Solid and curly lines carry mass $m_b$
      and $m_g$, respectively.  Dotted lines are massless. A dot on a line
      means that the corresponding propagator is squared.}
%      ~\\[-5mm]
%\hrule
\end{figure}

The matrix elements for $Q$, $Q_S$ or $\widetilde{Q}_S$ are expressed in terms
of the 23 master integrals shown in Fig.~\ref{fig::DB2mg_MI}. Note that
some of them have positive powers of $m_g$; for example, the second diagram in the top row scales as
$m_g^4$. However, they may be multiplied with $1/m_g$ poles up to order six, which requires a careful expansion
in $m_g$ such that only in the final expression for the amplitude
positive $m_g$ terms are set to zero.

For the expansion of the master integrals we apply an asymptotic expansion
(see, e.g.\ Ref.~\cite{Smirnov:2012gma}) for $m_g\ll m_b$.  In our case this
means that all components of each loop momentum can either be soft (i.e., of
order $m_g$) or hard (i.e., of order $m_b$), which splits each two-loop
integral into several regions. Depending on the routing there is also the possibility
that the sum or difference of the loop momenta is soft whereas the individual
momenta are hard. For illustration we describe the procedure in detail for the second
master integral in the bottom row of Fig.~\ref{fig::DB2mg_MI}, which has the following integrand
\begin{equation}
  \frac{1}{(m_g^2-p_1^2)(-p_1^2-2q\cdot p_1)(-p_2^2)(-p_2^2-2q\cdot p_2)(-(p_1-p_2)^2)}
  \,,
\end{equation}
with $q^2=m_b^2$.  Here we have to consider four regions. In case both loop
momenta are hard we can perform a Taylor expansion of the first propagator in
$m_g$ and thus $m_g$ disappears from the integrand. The resulting on-shell
integral is well-known in the literature (see, e.g.\ Ref.~\cite{Fleischer:1991xp}).
Similarly, if both loop momenta are soft, the second and fourth propagators
are expanded in $p_1^2$ and $p_2^2$, respectively. With the proper choice of
reference system a factor $1/m_b^2$ can be factored out and only a one-scale
integral remains. In case one of the momenta is soft and the other is hard
the two-loop integral factorises into two one-loop integrals which can be
solved in a straightforward way.

For the operator matrix elements we have to perform the standard UV
renormalisation for $\alpha_s$ and the external quark fields in the
$\overline{\rm MS}$ scheme and for the bottom quark mass in the on-shell scheme. Note
that in principle a non-zero gluon mass can also be introduced in these constants. However,
it only provides positive powers of $m_g$ to $\alpha_1$ and $\alpha_2$.
Furthermore, the usual renormalisation constants for operator mixing have to
be taken into account. As a consequence also evanescent operators enter the
tree-level and one-loop matrix elements.  As a non-standard counterterm we
have to introduce a renormalisation constant for the gluon mass. It is
determined from the requirement that the renormalised
gluon propagator is transverse. 
Note that in intermediate steps the renormalisation constants
for the gluon wave function and the QCD gauge parameter are 
needed, too.
Here it is possible to use the usual one-loop expressions in the
$\overline{\rm MS}$ scheme.

After renormalisation we obtain finite results for the operator matrix
elements $Q$, $Q_S$ and $\widetilde{Q}_S$
up to two loops. Note, however, that they still have a logarithmic
dependence on $m_g$. After extracting $\alpha_1$ and $\alpha_2$ from the
requirement that $\braket{R_0}$ in Eq.~(\ref{eq::R0_ME}) is ${\cal O}(1/m_b)$
the $\log(m_g)$ terms cancel since $\braket{R_0}$ is IR
finite. Of course, also $\alpha_1$ and $\alpha_2$ are IR finite.
They are given by
\begin{align*}
  \alpha_1^{(1)} &= \frac{26}{3} + 8 l_b \,,\\
  \alpha_1^{(2)} &= \frac{904309}{3096} + \frac{17 \pi^2}{3} +
 \frac{16 \pi^2 \log(2)}{3} - 8 \zeta(3) + \frac{649}{3} l_b + \frac{232}{3} l_b^2\\
  &\phantom{=} + N_L \left[ -\frac{218}{27} - \frac{8 \pi^2}{9} - \frac{104}{9} l_b - \frac{16}{3} l_b^2\right]\\
  &\phantom{=} + N_H\left[ -\frac{434}{27} + \frac{16 \pi^2}{9} - \frac{104}{9} l_b - \frac{16}{3} l_b^2 \right]  \\
  &\phantom{=} + N_V \left[-\frac{218}{27} - \frac{8 \pi^2}{9} + \frac{8 \pi^2 \sqrt{z}}{3} - 16 z + \frac{8 \pi^2 z^{3/2}}{3} - \frac{302 z^2}{27} - \frac{8 \pi^2 z^2}{9} + \frac{304 z^3}{225} \right.\\
  &\phantom{=} \left. + \frac{463 z^4}{4900} + \frac{7976 z^5}{297675} + \frac{1229 z^6}{117612} - \frac{104}{9}l_b - \frac{16}{3} l_b^2 \right.\\
 &\phantom{=}\left. + \left( \frac{52 z^2}{9} - \frac{32 z^3}{45} -
 \frac{3 z^4}{35} - \frac{32 z^5}{945} - \frac{5 z^6}{297}\right)  \log(z) - \frac{4 z^2}{3} \log^2(z) +{\cal O}(z^7) \right]\,,\\
  \alpha_2^{(1)} &= 8 + 16 l_b \,,\\
  \alpha_2^{(2)} &= \frac{3944}{9} + \frac{320 \pi^2}{27} + 440 l_b + \frac{752}{3} l_b^2 + \frac{32 \pi^2 \log(2)}{3} - 16 \zeta(3) \\
  &\phantom{=} + N_L \left[- \frac{422}{27} - \frac{16 \pi^2}{9} - \frac{208 l_b}{9} - \frac{32}{3} l_b^2 \right]\\
  &\phantom{=} + N_H \left[- \frac{854}{27} + \frac{32 \pi^2}{9} - \frac{208}{9} l_b - \frac{32}{3} l_b^2) \right] \\
  &\phantom{=} + N_V \left[ -\frac{422}{27} - \frac{16 \pi^2}{9} + \frac{16 \pi^2 \sqrt{z}}{3} - 32 z + \frac{16 \pi^2 z^{3/2}}{3} - \frac{604 z^2}{27} - \frac{16 \pi^2 z^2}{9} + \frac{608 z^3}{225} \right.\\
  &\phantom{=} + \frac{463 z^4}{1225} + \frac{31904 z^5}{297675} + \frac{1229 z^6}{29403} - \frac{208}{9} l_b - \frac{32}{3} l_b^2\\
 &\phantom{=} \left. + \left( \frac{104 z^2}{9} - \frac{64 z^3}{45} -
 \frac{12 z^4}{35} - \frac{128 z^5}{945} - \frac{20 z^6}{297} \right) \log(z) -
 \frac{8 z^2}{3} \log^2(z) +{\cal O}(z^7)\right]\,,
  \numberthis
\end{align*}
where $l_b=\log(\mu/m_b)$, $N_L$, $N_H$ and $N_V$ are the number of light (up, down, strange), bottom and charm quark flavours. $N_c$ has been set to three. Note that the specific choice of $\mathcal{O}(\epsilon)$ coefficients given in Eq.~\eqref{eq:nlo_ev_consts} and Eqs.~\eqref{eq:first_ev_const} to \eqref{eq:last_ev_const} has been used to simplify the expressions. 

The terms of order $\alpha_s$ and the massless fermionic $\alpha_s^2$ terms agree with Refs.~\cite{Beneke:1998sy} and~\cite{Asatrian:2017qaz}, respectively. All other terms are new. We also quote the higher order $z=m_c^2/m_b^2$ terms of the $\alpha_{1,2}^{(2)}$ constants, which stem from closed charm loops and have been calculated in Ref.~\cite{Asatrian:2017qaz}.\footnote{The original publication only contains terms up to $z^{3}$. The higher order terms up to $z^6$ quoted here have been kindly provided to us by one of the authors, Artyom Hovhannisyan.} For phenomenological applications the expansion to $z^6$ is sufficient as the relative contribution of the highest order term to $\alpha_{1,2}^{(2)}$ is smaller than $10^{-6}$.

Let us remark that there are multiple possibilities to introduce the gluon
mass. In our calculation we tested several approaches. For example,
it is possible to omit $m_g$ in the $\xi$-dependent term of the gluon
propagator in Eq.~(\ref{eq::mg}) or to leave the 
ghost propagator massless. The various options require different
expressions for $Z_{m_g}$ and also lead to different results for 
the renormalised matrix elements for
$Q$, $Q_S$ and $\widetilde{Q}_S$. However, in each case we obtain the same
results for $\alpha_1$ and $\alpha_2$.

%- }}}
%- {{{ Evanescent $\mathbf{\Delta B=2}$:

\subsection{\label{sub::match}Matching of $\Delta B=1$ and $\Delta B=2$ theory}
We now switch back to the general case of \bbmq\ with $q=d$ or $s$, because 
there are qualitative differences between the two systems, since the numerical suppression of
the CKM sub-leading terms involving $\lambda_u^q$ is much milder for $q=d$ than for $q=s$.

The off-diagonal element of the decay matrix can be written
as~\cite{Beneke:1998sy}
\begin{align*}
	\Gamma_{12}^q &= - (\lambda_c^q)^2\Gamma^{cc}_{12} 
	- 2\lambda_c^q\lambda_u^q \Gamma_{12}^{uc} 
	- (\lambda_u^q)^2\Gamma^{uu}_{12} \\
                      &= -(\lambda_t^q)^2 \left[
                          \Gamma_{12}^{cc} 
                          + 2 \frac{\lambda_u^q}{\lambda_t^q}\left(\Gamma_{12}^{cc}-\Gamma_{12}^{uc}\right)
                          + \left(\frac{\lambda_u^q}{\lambda_t^q}\right)^2 
                          \left(\Gamma_{12}^{uu}+\Gamma_{12}^{cc}-2\Gamma_{12}^{uc}\right)
                          \right]
	\,,\numberthis
	\label{eq::Gam12}
\end{align*}
where for the sake of clarity, we omit the superscript $q$ in the quantity $\Gamma_{12}^{ab}$. To obtain $\Gamma_{12}^q$, we calculate the absorptive part of
a bi-local matrix element containing a time-ordered product of two
$|\Delta B| = 1$ effective Hamiltonians. This yields
\begin{equation}
	\Gamma_{12}^{ab} 
	= \frac{G_F^2m_b^2}{24\pi M_{B_q}} \left[ 
	H^{ab}(z)   \langle B_q|Q|\bar{B}_q \rangle
	+ \widetilde{H}^{ab}_S(z)  \langle B_q|\widetilde{Q}_S|\bar{B}_q \rangle
	\right]
	+ \mathcal{O}(\Lambda_{\rm QCD}/m_b) \,,
	\label{eq::Gam^ab}
\end{equation}
with $z= m_c^2/m_b^2$. It is worth highlighting again that $R_0$ does not appear because its matrix element is $1/m_b$-suppressed in four dimensions thanks to the finite renormalisation outlined in Section \ref{sec:R0_ren}.

In our phenomenological analysis we include in the matching coefficients
$H^{ab}(z)$ and $\widetilde{H}^{ab}_S(z)$ the LO and NLO corrections from
current-current and penguin operators. 
Results for the former are exact in $z$ and taken from Ref.~\cite{Beneke:1998sy}. The penguin contributions are
expanded up to the linear term in $z$ and can be found in Refs.~\cite{Gerlach:2021xtb,Gerlach:2022wgb}. 
At NNLO our new contributions for
the current-current operators $Q_1$ and $Q_2$ are included.
For convenience of the reader we provide computer-readable results for the current-current contribution up to $z^{10}$ on the website~\cite{progdata}, where the specific choice of $\mathcal{O}(\epsilon)$ coefficients given in Eq.~\eqref{eq:nlo_ev_consts} and Eqs.~\eqref{eq:first_ev_const} to \eqref{eq:last_ev_const} has been used to simplify the expressions.
Our analysis also contains (in $z$ expanded) NNLO and N$^3$LO contributions
involving the chromomagnetic operator (usually denoted by $Q_8$)~\cite{Gerlach:2022wgb}. Note that they are numerically tiny.
For the $1/m_b$ suppressed terms see Refs.~\cite{Beneke:1996gn,Lenz:2006hd}. 

Writing the amplitude $\mathcal{M}$ of the $\bar{B}_q \rightarrow B_q$ transition calculated in the $\Delta B = 1$ theory as 
\begin{equation}
    \mathcal{M} = - (\lambda_c^q)^2 \mathcal{M}^{cc} 
	- 2\lambda_c^q\lambda_u^q \mathcal{M}^{uc} 
	- (\lambda_u^q)^2 \mathcal{M}^{uu} \,,
\end{equation}
we obtain the matching condition \cite{Nierste:2009wg}
\begin{equation}
    \Gamma_{12}^{ab} = \frac{1}{M_{B_q}} \,\text{Im} \,\mathcal{M}^{ab}\,.\label{eq:matching_condition}
\end{equation}

Since both the amplitude $\mathcal{M}$ and the matrix elements of the local $\Delta B = 2$ transition operators have IR poles beyond LO, lower-order matching calculations need to be carried out beyond the finite $\epsilon^0$ order. The two key issues here are the order in $\epsilon$ to which the amplitudes need to be evaluated as well as the generation of evanescent operators to be included in the local $\Delta B = 2$ matrix element. The order in $\epsilon$ is directly determined by the power of the IR poles, whereas the generations of evanescent operators on the $\Delta B = 2$ side follow the tree-level matrix elements of evanescent operators that appear on the $\Delta B = 1$ side. In the following we illustrate these two issues with a concrete matching calculation.

To obtain finite NNLO matching coefficients of the physical operators, the LO and NLO amplitudes need to be calculated to $\mathcal{O}(\epsilon^2)$ and $\mathcal{O}(\epsilon^1)$, respectively. This can be seen by expanding both sides of Eq.~\eqref{eq:matching_condition} in $\alpha_s$ and $\epsilon$. We can consider the matching equation order by order by expanding the matching coefficients in $\alpha_s$,
\begin{equation}
    H^{ab} =  H^{(0),ab} + \frac{\alpha_s}{4\pi}  H^{(1),ab} + \left(\frac{\alpha_s}{4\pi}\right)^2  H^{(2),ab} + \mathcal{O}(\alpha_s^3)\,,
    \label{eq:matching_coeff_alphas}
\end{equation}
and defining analogous expansions for $\Gamma_{12}^{ab}$, $\mathcal{M}^{ab}$ and the renormalised matrix elements $\langle O \rangle = \bra{B} O \ket{\bar{B}}$.
At LO we have
\begin{equation}
    (\Gamma_{12}^{ab} )^{(0)} = \frac{G_F^2m_b^2}{24\pi M_{B_q}} \left[H_P^{(0), ab}  \langle P \rangle^{(0)} + H_E^{(0), ab} \langle E \rangle^{(0)} \right]\,.\label{eq:LO_db2}
\end{equation}
The operators $P$ and $E$ schematically stand for any physical and evanescent operator, respectively. To be explicit, the physical operators we choose are $\{Q,\widetilde{Q}_S\}$ and the evanescent operators include $R_0$ in addition to the operators defined in Section \ref{sec:eva}. This is because after introducing finite renormalisation constants for $R_0$, its matrix element vanishes in four dimensions, while in $D$ dimensions $\langle R_0 \rangle$ possesses a power-unsuppressed evanescent piece~\cite{Gerlach:2022wgb}.

On the $\Delta B = 1$ side, the LO amplitude is
\begin{align*}
    \text{Im}(\mathcal{M}^{ab})^{(0)} = \frac{G_F^2m_b^2}{24\pi}  &\Big[  \left( a^{(0,0)} + \epsilon a^{(0,1)}  + \epsilon^2 a^{(0,2)} + \mathcal{O}(\epsilon^3) \right) \langle P \rangle^{(0)}\\
    &\phantom{\Big[} +  \left( b^{(0,0)} + \epsilon b^{(0,1)}  + \epsilon^2 b^{(0,2)} + \mathcal{O}(\epsilon^3) \right) \langle E \rangle^{(0)}  \Big]\,.\numberthis \label{eq:LO_db1}
\end{align*}
Equating Eqs.~\eqref{eq:LO_db2} and \eqref{eq:LO_db1}, the LO matching coefficients can be read off,
\begin{align*}
    H_P^{(0), ab} &= a^{(0,0)} + \epsilon a^{(0,1)}  + \epsilon^2 a^{(0,2)} + \mathcal{O}(\epsilon^3)  \,, \\
    H_E^{(0), ab} &= b^{(0,0)} + \epsilon b^{(0,1)}  + \epsilon^2 b^{(0,2)} + \mathcal{O}(\epsilon^3) \,. \numberthis \label{eq:LO_matching_coeffs}
\end{align*}

The procedure at NLO follows the same logic as the LO matching, but we encounter IR poles for the first time, which underlines the importance of keeping higher orders in $\epsilon$. Starting on the $\Delta B = 2$ side,  the first order correction is schematically
\begin{align*}
    (\Gamma_{12}^{ab} )^{(1)} = \frac{G_F^2m_b^2}{24\pi M_{B_q}} &\Big[H_P^{(0), ab}  \langle P \rangle^{(1)} + H_P^{(1), ab}  \langle P \rangle^{(0)}\\
    &\phantom{\Big[} + H_E^{(0), ab} \langle E \rangle^{(1)} +  H_E^{(1), ab} \langle E \rangle^{(0)}  \Big]\,.\numberthis\label{eq:NLO_db2}
\end{align*}
The matrix elements can be written as a series in $\epsilon$,
\begin{align*}
    \langle P \rangle^{(1)} &=  \left(\frac{ c^{(1,-1)}}{\epsilon} + c^{(1,0)} + \epsilon c^{(1,1)} \right) \langle P \rangle^{(0)} + \left( \frac{d^{(1,-1)}}{\epsilon}  + d^{(1,0)} + \epsilon d^{(1,1)} \right) \langle E \rangle^{(0)}\,,\\
    \langle E \rangle^{(1)} &=   \left( e^{(1,0)} + \epsilon e^{(1,1)} \right) \langle P \rangle^{(0)} + \left( \frac{f^{(1,-1)}}{\epsilon}  + f^{(1,0)} + \epsilon f^{(1,1)} \right) \langle E \rangle^{(0)} \,,\numberthis \label{eq:NLO_ome}
\end{align*}
where it is worth noting that the renormalised evanescent matrix elements are one order higher in $\epsilon$ than the renormalised physical operator matrix elements and hence do not have poles in front of $\langle P \rangle^{(0)}$. On the $\Delta B = 1 $ side, the NLO amplitude is parametrised as
\begin{align*}
    \text{Im}(\mathcal{M}^{ab} )^{(1)} = \frac{G_F^2m_b^2}{24\pi} &\Bigg[  \left( \frac{a^{(1,-1)}}{\epsilon} + a^{(1,0)}  + \epsilon a^{(1,1)} + \mathcal{O}(\epsilon^2) \right) \langle P \rangle^{(0)}\\
    &\phantom{\Big[} +  \left( \frac{b^{(1,-1)}}{\epsilon} +  b^{(1,0)}  + \epsilon b^{(1,1)} + \mathcal{O}(\epsilon^2) \right) \langle E \rangle^{(0)}  \Bigg]\,.\numberthis \label{eq:NLO_db1}
\end{align*}
Thus by inserting Eq.~\eqref{eq:NLO_ome} into Eq.~\eqref{eq:NLO_db2} and then equating with Eq.~\eqref{eq:NLO_db1}, the NLO matching coefficients can be extracted and read
\begin{align*}
    H_P^{(1), ab} &=  \frac{a^{(1,-1)}}{\epsilon} + a^{(1,0)}  + \epsilon a^{(1,1)} -  \left( \frac{ c^{(1,-1)}}{\epsilon} + c^{(1,0)} + \epsilon c^{(1,1)} \right) H_P^{(0), ab}\\
    &\phantom{=} -  \left( e^{(1,0)} + \epsilon e^{(1,1)} \right) H_E^{(0), ab}
    + \mathcal{O}(\epsilon^2)  \,, \\
    H_E^{(1), ab} &= \frac{b^{(1,-1)}}{\epsilon} +  b^{(1,0)}  + \epsilon b^{(1,1)} - \left( \frac{f^{(1,-1)}}{\epsilon}  + f^{(1,0)} + \epsilon f^{(1,1)} \right) H_E^{(0), ab}\\
    &\phantom{=} - \left( \frac{d^{(1,-1)}}{\epsilon}  + d^{(1,0)} + \epsilon d^{(1,1)} \right) H_P^{(0), ab} 
    + \mathcal{O}(\epsilon^2) \,. \numberthis \label{eq:NLO_matching_coeffs}
\end{align*}
We can now observe two important features: the cancellation of IR poles and the importance of $\epsilon$ parts of lower-order matching coefficients. The NLO matching coefficients reflect a physical quantity and need to be finite, meaning that
\begin{equation}
    a^{(1,-1)} - c^{(1,-1)} H_P^{(0), ab} = \mathcal{O}(\epsilon)\,,\label{eq:cross_check_nlo}
\end{equation}
which provides a strong cross-check for our calculation. Additionally, the $\epsilon$ parts of the LO matching coefficients enter the finite parts of the NLO matching coefficients. This can be seen from the pole $c^{(1,-1)}/\epsilon$ multiplying the LO matching coefficient $H_P^{(0),ab}$, resulting in a contribution of $a^{(0,1)}$ to the NLO matching coefficient $H_P^{(1),ab}$.

The matching procedure at NNLO is a natural extension of the NLO matching. The $\Delta B = 2$ side reads
\begin{align*}
    (\Gamma_{12}^{ab} )^{(2)} = \frac{G_F^2m_b^2}{24\pi M_{B_q}} &\Big[H_P^{(0), ab}  \langle P \rangle^{(2)} + H_P^{(1), ab}  \langle P \rangle^{(1)} +  H_P^{(2), ab}  \langle P \rangle^{(0)} \\
    &\phantom{\Big[} + H_E^{(0), ab} \langle E \rangle^{(2)} +  H_E^{(1), ab} \langle E \rangle^{(1)}  +  H_E^{(2), ab} \langle E \rangle^{(0)} \Big]\,,\numberthis\label{eq:NNLO_db2}
\end{align*}
where the renormalised NNLO part of the matrix element can be written in analogy to Eq.~\eqref{eq:NLO_ome} as
\begin{align*}
    \langle P \rangle^{(2)} &= \left(\frac{ c^{(2,-2)}}{\epsilon^2} + \frac{c^{(2,-1)}}{\epsilon} +  c^{(2,0)} \right) \langle P \rangle^{(0)} + \left( \frac{d^{(2,-2)}}{\epsilon^2}  + \frac{d^{(2,-1)}}{\epsilon} + d^{(2,0)} \right) \langle E \rangle^{(0)} \,,\\
    \langle E \rangle^{(2)} &=   \left( \frac{e^{(2,-1)}}{\epsilon} +  e^{(2,0)} \right) \langle P \rangle^{(0)} + \left( \frac{f^{(2,-2)}}{\epsilon^2}  + \frac{f^{(2,-1)}}{\epsilon} + f^{(2,0)} \right) \langle E \rangle^{(0)} \,.\numberthis \label{eq:NNLO_ome}
\end{align*}
Similar to Eq.~\eqref{eq:NLO_db1}, the NNLO amplitude on the $\Delta B = 1$ side is parametrised as
\begin{align*} 
    \text{Im}(\mathcal{M}^{ab} )^{(2)} = \frac{G_F^2m_b^2}{24\pi}  &\Bigg[  \left( \frac{a^{(2,-2)}}{\epsilon^2} + \frac{a^{(2,-1)}}{\epsilon}  + a^{(2,0)} + \mathcal{O}(\epsilon) \right) \langle P \rangle^{(0)}\\
    &\phantom{\Big[} +  \left( \frac{b^{(2,-2)}}{\epsilon^2} +  \frac{b^{(2,-1)}}{\epsilon}  +  b^{(2,0)} + \mathcal{O}(\epsilon) \right) \langle E \rangle^{(0)}  \Bigg]\,.\numberthis \label{eq:NNLO_db1}
\end{align*}
Hence by inserting Eq.~\eqref{eq:NNLO_ome} into Eq.~\eqref{eq:NNLO_db2} and then equating with Eq.~\eqref{eq:NNLO_db1}, the NNLO matching coefficients can be extracted and read
\begin{align*}
    H_P^{(2), ab} &= \frac{a^{(2, -2)}}{\epsilon^2} + \frac{a^{(2, -1)}}{\epsilon} + a^{(2, 0)} - e^{(1, 0)} H_E^{(1),ab} 
 -\left(\frac{c^{(1, -1)}}{\epsilon} + c^{(1, 0)}\right) H_P^{(1),ab}\\
    &\phantom{=} - \left(\frac{e^{(2, -1)}}{\epsilon} + e^{(2, 0)}\right) H_E^{(0),ab}  - \left(\frac{c^{(2, -2)}}{\epsilon^2} + \frac{c^{(2, -1)}}{\epsilon} + c^{(2, 0)}\right) H_P^{(0),ab} + \mathcal{O}(\epsilon)
    \,, \\
    H_E^{(2), ab} &= \frac{b^{(2, -2)}}{\epsilon^2} + \frac{b^{(2, -1)}}{\epsilon} +
 b^{(2, 0)}  - \left(\frac{f^{(1, -1)}}{\epsilon} + f^{(1, 0)}\right) H_E^{(1),ab} - \left(\frac{d^{(1, -1)}}{\epsilon} + d^{(1, 0)}\right) H_P^{(1),ab}\\
    &\phantom{=}- \left(\frac{f^{(2, -2)}}{\epsilon^2} + \frac{f^{(2, -1)}}{\epsilon} + f^{(2, 0)}\right) H_E^{(0),ab}    - \left(\frac{d^{(2, -2)}}{\epsilon^2} + \frac{d^{(2, -1)}}{\epsilon} + d^{(2, 0)}\right) H_P^{(0),ab} + \mathcal{O}(\epsilon). \numberthis \label{eq:NNLO_matching_coeffs}
\end{align*}
As the matching coefficients need to be finite, all $\epsilon$ poles need to cancel. This provides another cross-check analogous to Eq.~\eqref{eq:cross_check_nlo}. Since the lower order matching coefficients are multiplied by poles in $\epsilon$, it is now also apparent that the LO and NLO matching coefficients need to be extracted to $\mathcal{O}(\epsilon^2)$ and $\mathcal{O}(\epsilon)$, respectively.

For now we have discussed generic physical and evanescent operators, but it is also worth mentioning which evanescent operators need to be included in the matching. From Eq.~\eqref{eq:NNLO_matching_coeffs} it is evident that both the LO and NLO evanescent matching coefficients, $H_E^{(0)}$ and $H_E^{(1)}$, will contribute to the physical matching coefficient at NNLO. Furthermore, from Eq.~\eqref{eq:NNLO_db2} we can see that $H_E^{(0)}$ and $H_E^{(1)}$ enter through the renormalised NNLO and NLO evanescent matrix elements, respectively. This means that all evanescent operators which appear as tree-level matrix elements at LO on the $\Delta B = 1$ side, i.e.~in Eq.~\eqref{eq:LO_db1}, need to be renormalised to NNLO on the $\Delta B = 2$ side. Similarly all evanescent operators that appear at NLO on the $\Delta B = 1$ side, i.e.~in Eq.~\eqref{eq:NLO_db1}, need to be renormalised to NLO on the $\Delta B = 2$ side.

In our specific case where we consider two insertions of the current-current operators on the $\Delta B =1$ side, the most complicated evanescent operators appearing at LO are the second generation evanescent operators defined in Eqs.~\eqref{eq:E2_def} and \eqref{eq:E2QS_def}. At NLO we also get the third generation evanescent operators from Eq.~\eqref{eq:E3_def} and at NNLO we encounter the fourth generation evanescent operators from Eq.~\eqref{eq:E4_def} for the first time. This means we renormalise the second, third and fourth generation evanescent operators to NNLO, NLO and LO (i.e.~only tree-level), respectively. Incidentally this is exactly the hierarchy needed to renormalise a second generation evanescent operator at NNLO anyway.

%- }}}

%- }}}

%- {{{ Technicalities:

\section{\label{sec::technicalities}Technicalities}

For the calculation of the $\Delta B=1$ and $\Delta B=2$
amplitudes up to three and two loops, respectively, we have
used two independent codes. Both of them use
\texttt{qgraf}~\cite{Nogueira:2006pq} for the generation of the amplitudes for
the Feynman diagrams. 

The implementation of the Feynman rules for the four-fermion operators
requires special care. We find it convenient to 
split the four-particle vertices into two three-particle
vertices each.  The Feynman rules for the new vertices must be chosen such that one
can reproduce the correct colour structure of the original vertex. This
procedure ensures that relative signs between diagrams containing 
four-fermion operators are automatically correct.
This is also common to both setups.
All remaining steps are to a large extent different.

In the first setup 
we implemented the effective Hamiltonians in
\texttt{FeynRules}~\cite{Alloul:2013bka}\footnote{In our calculation it is
  important to disable the automatic simplification of chains with more than
  three Dirac matrices using the Chisholm identity. This is achieved by
  commenting the corresponding routine in the {\tt Processing} section of the
  file {\tt FeynArtsInterface.m}} and exporting the Feynman rules to
\texttt{FeynArts}~\cite{Hahn:2000kx}.  The relative signs of genuine
four-fermion operators are then automatically fixed by
\texttt{FeynCalc}~\cite{Mertig:1990an,Shtabovenko:2016sxi,Shtabovenko:2020gxv}
using the algorithm
adapted from \texttt{FormCalc}~\cite{Hahn:1998yk}. In this
way we were able to validate the approach based on the splitting of the four-particle vertices into two three-particle vertices each by
comparing amplitudes of selected diagrams.

The first setup further uses \texttt{q2e}~\cite{Harlander:1998cmq,Seidensticker:1999bb}
to rewrite the \texttt{qgraf} output to \texttt{FORM}~\cite{Kuipers:2012rf}.
Afterwards, each diagram is mapped to a
predefined integral family with the help of \texttt{exp}. The manipulations of
the amplitudes and the identification of scalar integrals of a given integral
family is done with the help of in-house \texttt{FORM} code.  For the
decomposition into the different operator structures, tensor integrals up to
rank ten have been implemented.  More details are provided in Ref.~\cite{Reeck:2024iwk}.
Note that the first setup to compute the three-loop corrections is restricted to the first two terms
in the expansion for $m_c/m_b\to 0$.  Thus, in the integral families the charm quark is massless.

We use version~6 of \texttt{FIRE}~\cite{Smirnov:2019qkx} in
combination with \texttt{LiteRed} \cite{Lee:2013mka} for the reduction to master
integrals. The calculation of the latter is described in detail in Section~\ref{app::masters} where also explicit results can be found.

The second setup uses \texttt{tapir}~\cite{Gerlach:2022qnc,Gerlach:2022fgs} 
which  automatically generates a number of
auxiliary files which in the first setup have to be generated manually. This
concerns in particular 
the definition of the integral families and the corresponding \texttt{FORM} code.
In the second approach we aim for a full dependence on $m_c$.
Thus, the integral families are significantly more involved as in the first approach.
Altogether we have 320 different integral families.

A major difference to the first approach is
also that here we use projectors instead of tensor integral reduction
in order to obtain the coefficients of the
various tensor structures. Details concerning the construction of the
projectors can again be found in Ref.~\cite{Reeck:2024iwk}.

We perform the reduction within each integral family to master integrals with the program {\tt Kira}~\cite{Maierhofer:2017gsa,Klappert:2020nbg}. 
At this step, the program \texttt{ImproveMasters.m}~\cite{Smirnov:2020quc}
is very useful in order to choose a good basis of master integrals such that the
coefficients only contain low-order polynomials in $\epsilon$ and $x$.
We use {\tt Kira} also for the minimisation of the master integrals across
all families. Keeping only those master integrals which have an imaginary part and completing the system of differential equations as outlined in Ref.~\cite{Reeck:2024iwk}, 
we obtain $342$ complex-valued master integrals.

For the computation of the master integrals we use the 
``expand and match'' approach ~\cite{Fael:2021kyg,Fael:2022rgm,Fael:2022miw,Fael:2023zqr}
and construct generalised series expansions around several
values for $z=m_c^2/m_b^2$ (see Ref.~\cite{Reeck:2024iwk}
for more details) such that we cover the range 
relevant for all renormalisation schemes. For the phenomenological
application it is sufficient to use the expansion around $z=0$ up to $z^{10}$ as the highest expansion term has a relative contribution of $10^{-6}$ to the coefficient
of $(\alpha_s/(4\pi))^2$ at the largest value of $z$ considered in the scale variation. 

%- }}}

%- {{{ Analytic results:

\section{\label{sec::analytic}Semi-analytic results}

For illustration purposes, we present the first $z^0$ term of the semi-analytic expansion of the matching coefficient contributions from current-current operators. We provide computer-readable results up to $z^{10}$ on the website~\cite{progdata}.

The matching coefficients are as defined in Eqs.~\eqref{eq::Gam^ab} and \eqref{eq:matching_coeff_alphas} and can be further decomposed by the respective contributions from the $\Delta B = 1$ Wilson coefficients,
\begin{equation}
    H^{(n),ab} = \sum_{i,j\in\{1,\dots,6,8\},j\geq i} C_i C_j p_{ij}^{(n),ab}\,,
\end{equation}
and analogously for $\widetilde{H}_S$. For the $z^0$ term the coefficients are identical for $ab\in \{uu,uc,cc\}$, so we omit these indices in the following. Our LO results read
\begin{align*}
    p_{11}^{(0)} &= 0.31944\,, & p_{11}^{S,(0)} &= - 0.55556\,,\\
    p_{12}^{(0)} &= 0.16667\,, & p_{12}^{S,(0)} &= - 1.33333\,,\\
    p_{22}^{(0)} &= 1.0000\,, & p_{22}^{S,(0)} &= 1.0000\,.
\end{align*}
At NLO we obtain
\begin{align*}
    p_{11}^{(1)} &= 3.2241 + 2.0802 L_1 + 2.7593 L_2\,, & p_{11}^{S,(1)} &= - 0.72285 - 0.93827 L_1 - 2.9630 L_2\,,\\
    p_{12}^{(1)} &= - 11.134 - 11.963 L_1 + 4.2222 L_2\,, & p_{12}^{S,(1)} &= 12.230 + 3.2593 L_1 - 7.1111 L_2\,,\\
    p_{22}^{(1)} &= - 14.764 - 3.1111 L_1 + 1.3333 L_2 \,,& p_{22}^{S,(1)} &= 0.64402 + 14.222 L_1 + 5.3333 L_2\,,
\end{align*}
and finally at NNLO we have
\begin{align*}
    p_{11}^{(2)} &= {68.566} + 126.22 L_1 + 39.642 L_1^2 + 44.635 L_2 + 53.132 L_1 L_2 - 5.7593 L_2^2 \,,\\
    p_{11}^{S,(2)} &= - 71.490 - 110.61 L_1 - 14.848 L_1^2 - 18.999 L_2 - 50.436 L_1 L_2 + 38.519 L_2^2\,,\\
    p_{12}^{(2)} &= {6.5554} - 153.65 L_1 - 129.70 L_1^2 - 2.7358 L_2 + 8.1975 L_1 L_2 - 0.22222 L_2^2 \,,\\
    p_{12}^{S,(2)} &= 36.006 + 238.79 L_1 + 2.1728 L_1^2 + 28.880 L_2 - 91.654 L_1 L_2 + 92.444 L_2^2\,,\\
    p_{22}^{(2)} &= {- 267.98} - 80.931 L_1 + 53.111 L_1^2 - 121.96 L_2 - 29.926 L_1 L_2 - 25.333 L_2^2\,, \\
    p_{22}^{S,(2)} &= - 396.55 - 146.85 L_1 + 89.481 L_1^2 + 30.694 L_2 + 157.63 L_1 L_2 - 69.333 L_2^2\,,
\end{align*}
where we have defined $L_1 = \ln (\mu_1/m_b)$ and $L_2 = \ln(\mu_2/m_b)$. An important cross-check of our results at three-loops is the calculation using analytical master integrals for $z=0$. The details of the computation are in  Appendix~\ref{app::masters}. Here and in the following $\mu_1$ is the scale at which the $\Delta B=1$ Wilson coefficients are evaluated, while $\mu_2$ is the scale at 
which the $\Delta B=2$ operators are defined. The $\mu_1$ dependence of the coefficients of the $\Delta B=2$ operators decreases with increasing orders of $\alpha_s$. The same holds true for the $\mu_2$ dependence only in combination with the $\mu_2$ dependence of the $\Delta B=2$ matrix elements, which requires a calculation of the lattice-continuum matching to the appropriate order in $\alpha_s$.

Let us briefly comment on the agreement of the matching coefficients obtained in this work with previous NNLO studies. The matching coefficients which form the basis of Ref.~\cite{Gerlach:2022hoj} have been reproduced as part of our calculation at intermediate steps. Note that the coefficients provided here and in the ancillary files are different because of the new operator basis used in this work, see Section \ref{sec:eva}. 

We also reproduced the part of the NNLO matching coefficients which stems from closed charm quark loops as calculated in Ref.~\cite{Asatrian:2017qaz}. Note that the latter contributions are independent of the choice of the second generation evanescent operators. Ref.~\cite{Gerlach:2022hoj} also incorporated the previously published closed charm loop contributions, but missed a numerically small correction, which has been resolved now.

%- }}}
%- {{{ Phenomenology: Delta\Gamma and afs:

\section{\label{sec::phen}Phenomenology}

In this Section we consider both the $B_s$ and $B_d$ system and
discuss numerical results for the width difference $\Delta\Gamma_q$,
the ratio $\Delta\Gamma_q/\Delta M_q$ and the CP
asymmetry in flavours-specific decays $a_{\rm fs}^q$.
We find good numerical agreement with previous calculations in Ref.~\cite{Gerlach:2022hoj}.

%- {{{ Input values and renormalisation schemes:

\subsection{\label{sub::input}Input values and renormalisation schemes}

\begin{table}[t]
  \begin{center}
  {\scalefont{0.8}
    \renewcommand{\arraystretch}{1.5}
    \begin{tabular}{rclc | rclc}
      \hline 
      $\alpha_s(M_Z)$ &=& $0.1180 \pm 0.0009$ & \cite{ParticleDataGroup:2024cfk}
      &
      $m_c(3~\mbox{GeV})$ &=& $0.993\pm 0.008$~GeV & \cite{Chetyrkin:2017lif}
      \\
      $m_t^{\rm pole}$ &=& $172.4\pm 0.7$~\mbox{GeV} & \cite{ParticleDataGroup:2024cfk} 
      &                                                 
      $m_b(m_b)$ &=& $4.163\pm 0.016$~GeV & \cite{Chetyrkin:2010ic} 
      \\
      $M_{B_s}$ &=& $5366.88 {\pm 0.14}$~\mbox{MeV} & \cite{ParticleDataGroup:2024cfk} 
      &
      $M_{B_d}$ &=& $5279.64 {\pm 0.12}$~\mbox{MeV} & \cite{ParticleDataGroup:2024cfk} 
      \\
      $B_{B_s}$ &=& $0.813\pm0.034$ & \cite{Dowdall:2019bea} 
      &
      $B_{B_d}$ &=& $0.806\pm0.041$ & \cite{Dowdall:2019bea}  
      \\
      $\widetilde{B}^\prime_{S,B_s}$ &=& $1.31\pm0.09$ & \cite{Dowdall:2019bea} 
      &
      $\widetilde{B}^\prime_{S,B_d}$ &=& $1.20\pm0.09$ & \cite{Dowdall:2019bea} 
      \\
      $f_{B_s}$ &=& $0.2303\pm0.0013$~GeV & \cite{Bazavov:2017lyh,Hughes:2017spc,ETM:2016nbo,Dowdall:2013tga} 
      &
      $f_{B_d}$ &=& $0.1905\pm0.0013$~\mbox{MeV} & \cite{Bazavov:2017lyh,Hughes:2017spc,ETM:2016nbo,Dowdall:2013tga} 
      \\
      \hline 
    \end{tabular}
    }
  \end{center}
  \caption{\label{tab::input}Input parameters for the numerical analysis.  The
    quoted $m_t^{\rm pole}$ corresponds to {
      $m_t(m_t)=(162.6 \pm 0.7)\,\gev$} in the $\overline{\rm MS}$ scheme. We
    use the values for $B_{B_q}=B_{B_q}(\mu_2)$ and
    $\widetilde{B}^\prime_{S,B_q}=\widetilde{B}^\prime_{S,B_q}(\mu_2)$ with
    $\mu_2=m_b^{\rm pole}$. }
%    ~\\[-5mm]
%\hrule
\end{table}

In Tab.~\ref{tab::input} we list the input values which we use for our
numerical analysis. Using $m_b(m_b)$ and the two-loop relation to the pole
mass we obtain {$m_b^{\rm pole}=4.758$~GeV}. 
The pole mass is well-defined at any finite order of QCD,
but suffers from a renormalon ambiguity of order $\lqcd$. While this feature is related to asymptotically large orders of perturbation theory, it is empirically known that perturbative series of physical quantities exhibit a poor convergence at low orders of $\alpha_s$. $\Gamma_{12}^q$ is proportional to two powers of $m_b$, and we 
will present results for different renormalisation schemes for this overall factor $m_b^2$. The results using the pole mass are for illustrative purposes only and are excluded from the final numerical values. For the bottom mass in the potential-subtracted (PS) scheme, we obtain {$m_b^{\rm{PS}}=4.480$~GeV} with the factorisation scale $\mu_f=\SI{2}{GeV}$ using the four-loop implementation in RunDec~\cite{Herren:2017osy}. Furthermore, we need
CKM parameters which are given by~\cite{CKMfitter}
\begin{align*}
%  \frac{\lambda^d_u}{\lambda^d_t} &=&  (0.01048 \pm 0.01065) - (0.425945 \pm 0.009055){\rm i}\,,\nonumber\\
    \frac{\lambda^d_u}{\lambda^d_t} &=  (0.0105 \pm 0.0107) - (0.4259 \pm 0.0091){\rm i}\,,\\
%  \frac{\lambda^s_u}{\lambda^s_t} &= -(0.00877 \pm 0.00043) +(0.01858 \pm 0.00038){\rm i}\,,\nonumber\\
  \frac{\lambda^s_u}{\lambda^s_t} &= -(0.00877 \pm 0.00043) +(0.01858 \pm 0.00038){\rm i}\,.
  \numberthis \label{eq::ckm_input}
\end{align*}
While only the ratio $\lambda^q_u/\lambda^q_t$ enters $\Gamma_{12}^q/M_{12}^q$, in a direct calculation of $\Delta \Gamma_q$ using Eq.~\eqref{eq:dgdmafsq}, the absolute values $|\lambda^d_t|$ and $|\lambda^s_t|$ are also needed, which depend on the value of $|V_{cb}|$. The current measurements from inclusive and exclusive decays are \cite{FlavourLatticeAveragingGroupFLAG:2024oxs}
\begin{align*}
    |V_{cb}^\text{incl}| & = (42.16 \pm 0.51) \times 10^{-3}\,\text{\cite{Bordone:2021oof}},\\
    |V_{cb}^\text{excl}| & = (39.45 \pm 0.56) \times 10^{-3}\,,\qquad [B \rightarrow (D,D^*)\ell\nu]\,,\,\, \textrm{FLAG average}, \,\,\text{\cite{MILC:2015uhg, Na:2015kha, FermilabLattice:2021cdg, Aoki:2023qpa, BaBar:2009zxk, Belle:2015pkj, Belle:2018ezy, Belle:2023bwv, Belle-II:2023okj, HFLAV:2022esi}},\numberthis \label{eq:vcbei}
\end{align*}
where for the inclusive $|V_{cb}|$ one can also use the result from Ref.~\cite{Bernlochner:2022ucr}, which leads to a similar result. The current SM fit which excludes measurements of $|V_{cb}|$ gives \cite{CKMfitter}
\begin{equation}
    |V_{cb}^\textrm{SM fit}| = (41.60^{+0.20}_{-0.58}) \times 10^{-3}\,. \label{eq:vcbfit}
\end{equation}
We obtain the following values for $\lambda_t^{{q}}$:
\begin{align*}
  |\lambda^{d,\text{SM fit}}_t| &= (8.56^{+0.08}_{-0.34})\times 10^{-3}\,, & |\lambda^{d,\text{excl}}_t| &= (8.12^{+0.15}_{-0.20})\times 10^{-3}\,, \\
  |\lambda^{s,\text{incl}}_t| &= (41.39 \pm 0.50) \times 10^{-3}\,, &|\lambda^{s,\text{excl}}_t| &= (38.73 \pm 0.55) \times 10^{-3}\,. \numberthis\label{eq:lambdat}
\end{align*}
Here $|\lambda^{d,\text{SM fit}}_t|$ is calculated from the $1\sigma$ best-fit result for 
$|V_{td}|$ and $|V_{tb}|$ found by the CKMFitter group from a global fit to the CKM parameters \cite{CKMfitter}. 
One cannot simply deduce which result one would find by fixing $|V_{cb}|$ to the value determined from exclusive semileptonic $B$ decays, which would come with a poor $p$-value, from the results of Ref.~\cite{CKMfitter}. Our $|\lambda^{d,\text{excl}}_t|$ is instead calculated by rescaling $|\lambda^{d,\text{SM fit}}_t|$ with the ratio of the two 
$|V_{cb}|^2$ values from \eqsand{eq:vcbei}{eq:vcbfit}. The $B_s$ system is easier because the involved CKM elements are fixed by CKM unitarity to $|V_{ts}| = (0.983\pm 0.001) |V_{cb}|$ and $|V_{tb}|=0.9991$ with essentially no sensitivity to other parameters. Thus we calculate $|\lambda^{s,\text{incl/excl}}_t|$ directly from $|V_{cb}^{\text{incl/excl}}|$ in this way.

The parameters $B_{B_s}$ and $\widetilde{B}^\prime_{S,B_s}$ in
Tab.~\ref{tab::input} are used to parametrise the matrix elements of the
leading operators $Q$ and $\widetilde{Q}_S$ as
\begin{align*}
 \bra{B_s} Q (\mu_2) \ket{\overline B_s}  &=
   \frac{8}{3} M^2_{B_s}\, f^2_{B_s} B_{B_s} (\mu_2), \\
\bra{B_s} \widetilde Q_S (\mu_2)\ket{\overline B_s} &= \frac{1}{3}  M^2_{B_s}\,
  f^2_{B_s} \widetilde  B_{S,B_s}^\prime (\mu_2)\,.\numberthis
      \label{eq:defb}
\end{align*}  
Analogous formulae hold for the $B_d$ system.
 Sum rule results for the matrix elements can be found in Refs.~\cite{Kirk:2017juj,DiLuzio:2019jyq}. They have, e.g.\ been used in Ref.~\cite{Lenz:2019lvd,Albrecht:2024oyn}.

For the $1/m_b$ suppressed correction to $\Gamma_{12}^q$ in Eq.~\eqref{eq::Gam^ab} % we 
%follow Ref.~\cite{Beneke:1996gn} and 
use for the overall factor $m_b^2$  the PS scheme.
The matrix elements of the $1/m_b$ suppressed corrections calculated with lattice QCD
are obtained from
Refs.~\cite{Davies:2019gnp,Dowdall:2019bea} and are given by
\begin{align}
  \braket{B_s|R_0|\bar{B}_s} &= - (0.43 \pm {0.18}) f_{B_s}^2 M_{B_s}^2\,,
  \nonumber\\
  \braket{B_s|R_1|\bar{B}_s} &=  (0.07 \pm 0.00) f_{B_s}^2 M_{B_s}^2\,,
  \nonumber\\
  \braket{B_s|\widetilde{R}_1|\bar{B}_s} &= (0.04 \pm 0.00)
  f_{B_s}^2 M_{B_s}^2\,, \nonumber\\
  \braket{B_s|R_2|\bar{B}_s} &= - (0.18 \pm 0.07) f_{B_s}^2 M_{B_s}^2\,,
  \nonumber\\
  \braket{B_s|\widetilde{R}_2|\bar{B}_s} &=  (0.18 \pm 0.07) f_{B_s}^2
  M_{B_s}^2\,, \nonumber\\ 
  \braket{B_s|R_3|\bar{B}_s} &=  (0.38 \pm 0.13) f_{B_s}^2 M_{B_s}^2\,,
  \nonumber\\
  \braket{B_s|\widetilde{R}_3|\bar{B}_s} &=  (0.29 \pm 0.10) f_{B_s}^2
                                       M_{B_s}^2\,,
  \label{eq::1/mb_ME}
\end{align}
where explicit expressions for the operators can be found in Refs.~\cite{FermilabLattice:2016ipl,Dowdall:2019bea}.
The results for the matrix elements of $R_2$, $\widetilde{R}_2$, $R_3$ and
$\widetilde{R}_3$ can be found in Ref.~\cite{Davies:2019gnp} and we extract the
remaining three matrix elements from~\cite{Dowdall:2019bea}. For
$\braket{B_s|R_1|\bar{B}_s}$ and $\braket{B_s|\widetilde{R}_1|\bar{B}_s}$ the
ratio of the bottom and strange quark masses is needed
${m_b(\mu)/m_s(\mu)} = 52.55 \pm 0.55$~\cite{Chakraborty:2014aca}.

The matrix elements of the $1/m_b$ suppressed corrections for the $B_d$ system
are given by
\begin{align}
  \braket{B_d|R_0|\bar{B}_d} &= - (0.35 \pm 0.19) f_{B_d}^2 M_{B_d}^2\,,
  \nonumber\\
  \braket{B_d|R_1|\bar{B}_d}         &=0\,, \nonumber\\
  \braket{B_d|\widetilde{R}_1|\bar{B}_d} &=0\,, \nonumber\\
  \braket{B_d|R_X|\bar{B}_d} &= {\frac{f_{B_d}^2 M_{B_d}^2}{f_{B_s}^2 M_{B_s}^2} }\braket{B_s|R_X|\bar{B}_s}
                               \,,
  \label{eq::1/mb_ME_d}
\end{align}
where $R_X \in \{R_2, \widetilde{R}_2, R_3, \widetilde{R}_3\}$~\cite{Davies:2019gnp}.
The matrix elements of $R_1$ and $\widetilde{R}_1$ are of order
$10^{-3}f_{B_d}^2 M_{B_d}^2$ which is due to the additional suppression factor
$m_d/m_s\approx 0.03$ as compared the corresponding quantities in the $B_s$
system. The  matrix element for $R_0$ is extracted from Ref.~\cite{Dowdall:2019bea}.
In the last expression in \eq{eq::1/mb_ME_d} flavour-SU(3) breaking effects 
beyond factorisation are neglected.

Following the matching procedure outlined in Section~\ref{sub::match}, we obtain an expression for $\Gamma_{12}^{ab}$ where
the $\alpha_s$ is renormalised in the $\overline{\rm MS}$ scheme and the charm
and bottom quark masses are in the on-shell scheme. In a first step we transform
the parameter $z$ in Eq.~(\ref{eq::Gam^ab}) to the $\overline{\rm MS}$ scheme
which introduces $\mu_c$ and $\mu_b$, the renormalisation scales of
$\overline{m}_{c}$ and $\overline{m}_{b}$.  For the overall factor
$m_b^2$ we adapt three different renormalisation schemes: we either leave it
in the pole scheme or transform it to the ${\overline{\rm MS}}$ or
PS mass~\cite{Beneke:1998rk}.  The latter is an example
for a so-called threshold mass. Alternatively we could have used the
kinetic~\cite{Bigi:1996si,Czarnecki:1997sz,Fael:2020njb} or
RS~\cite{Pineda:2001zq} schemes.  Note that the scheme transformation of
$m_b^2$ leads to a redefinition of $H^{ab}(z)$ and $\widetilde{H}^{ab}_S(z)$
such that the scheme dependence of $\Gamma_{12}^s$ is of higher order in $\alpha_s$.

Let us next discuss our choices for the various renormalisation and matching
scales which enter our analysis.  For the matching scale between the full
Standard Model to the \mbox{$|\Delta B|=1$} theory, $\mu_0$, we choose
$\mu_0=165~\mbox{GeV} \approx m_t(m_t)$.  We do not vary $\mu_0$ since it
barely affects the physical quantities. The matching scale $\mu_1$ and
the renormalisation scales $\mu_b$ and $\mu_c$ are all of the order of the
bottom quark mass. In our analysis we vary them simultaneously (i.e., $\mu_c=\mu_b=\mu_1$)
between $2.1$~GeV and $8.4$~GeV with the central scale $\mu_1=4.2$~GeV.
For reference, at the central scale $\mu_1=4.2$~GeV we have
{$\bar z= (m_c(\mu_1)/m_b(\mu_1))^2 \approx {0.04955}$}.

The $|\Delta B|=2$
operators are defined at the scale $\mu_2$ which has to be kept fixed, because
the $\mu_2$ dependence only cancels in the product of $H^{ab}(z)$ and
$\widetilde{H}^{ab}_S(z)$ with their respective matrix elements.  In our
analysis we set $\mu_2=m_b^{\rm pole}=4.758$~GeV
which is computed from $m_b(m_b)$ to two-loop accuracy.  The terms
of order $\lqcd/m_b$ in $\Gamma_{12}^q$ are only known to LO, so that the
$\mu_1$-dependence of these terms is non-negligible. 
In the numerical results, which we present below, the corresponding uncertainties
will be displayed separately.

%- }}}
%- {{{ $\Delta\Gamma_q$:

\subsection{$\Delta\Gamma_q$ and $\Delta\Gamma_q/\Delta M_q$}

\subsubsection{$B_s$ system}

We start with the discussion of the $B_s$ system and 
consider in a first step the ratio $\Delta\Gamma_s/\Delta M_s$. 

For the numerical evaluation we use the values for the input parameters
provided in Section~\ref{sub::input}. We vary each parameter within the given
uncertainty which leads to the central value for $\Delta\Gamma_s/\Delta M_s$
and the corresponding uncertainties.  The latter are added in quadrature and
taking into account the 100\% correlation of $\braket{B_s|R_2|\bar{B}_s}$ and
$\braket{B_s|\widetilde{R}_2|\bar{B}_s}$. In the following we provide separate
uncertainties for the scale variation (``scale''), the leading-power bag
parameters (``$B\widetilde{B}_S$''), matrix elements of the power-suppressed
corrections (``$1/m_b$''), and the variation of the remaining input parameters
(``input'').  In our three schemes we have 
\begin{align}
  \frac{\Delta \Gamma_s}{\Delta M_s}
  ~=&~ \left(
      {{3.84^{+0.53}_{-0.57}}_{\textrm{scale}}}
      {{{}^{+0.09}_{-0.19}}_{\textrm{scale, $1/m_b$}} } 
      {\pm 0.11_{B\widetilde{B}_S}} \rt. 
%\nonumber\\
%  &~
    \lt. \; {\pm 0.78_{1/m_b}} { \pm 0.06_{\textrm{input}}}\right) \times
    10^{-3}\ (\textrm{pole})\,, \nonumber\\
      \frac{\Delta \Gamma_s}{\Delta M_s} 
  ~=&~ \left(
      {{4.37^{+0.23}_{ -0.44}}_{\textrm{scale}}}
      {{{}^{+0.09}_{-0.19}}_{\textrm{scale, $1/m_b$}} } 
      {\pm  0.12_{B\widetilde{B}_S}} \rt. 
%\nonumber\\
%  &~
  \lt. \;     {\pm 0.78_{1/m_b} } {\pm 0.05_{\textrm{input}}}\right) \times 10^{-3}\ (\overline{\textrm{MS}})\,,
   \nonumber\\
  \frac{\Delta \Gamma_s}{\Delta M_s} ~=&~ 
 \left( {{ 4.27^{+0.36}_{ -0.37}}_{\textrm{scale}} }
      {{{}^{+0.09}_{-0.19}}_{\textrm{scale, $1/m_b$}} } 
      {\pm 0.12_{B\widetilde{B}_S} } \rt. 
%\nonumber\\
%  &~
  \lt. \;{\pm 0.78_{1/m_b}}
          {\pm 0.05_{\textrm{input}}}\right) \times 10^{-3}\ (\textrm{PS})\,.
          \label{eq::dGdM}
\end{align}
The dominant uncertainty comes from the matrix elements of the
power-suppressed corrections~\cite{Davies:2019gnp,Dowdall:2019bea}) followed
by the renormalisation scale uncertainty from the variation of $\mu_1$ in the
leading-power term.  The uncertainties from the leading-power bag parameters
and from the scale variation in the $1/m_b$ piece are much smaller.
The variation of the remaining input parameters is of minor relevance.

\begin{figure}[t]
  \begin{center}
      \includegraphics[width=0.8\textwidth]{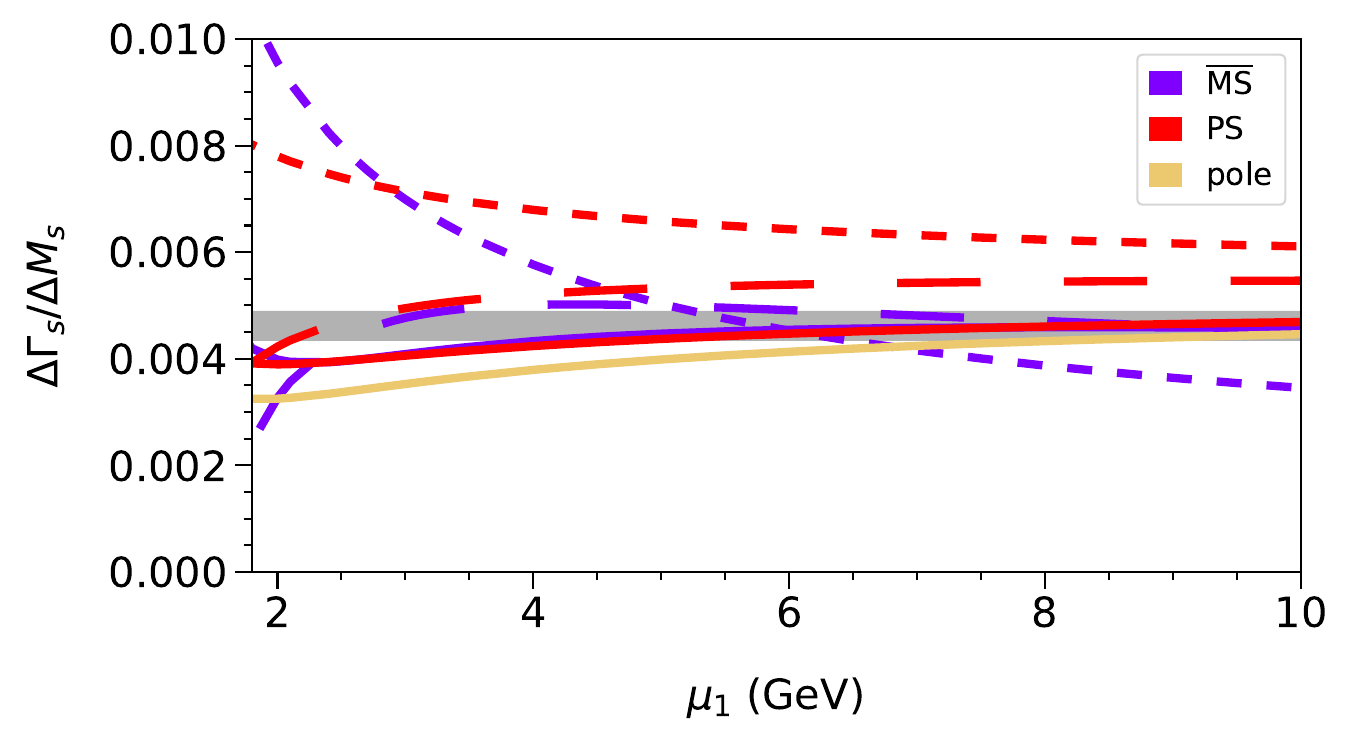}
  \end{center}
  \caption{\label{fig::DelGam_mu1} 
      Renormalisation scale dependence at LO (short dashes),
      NLO (long dashes) and NNLO (solid) for $\Delta\Gamma_s/\Delta M_s$.  The
      scale in the power-suppressed terms is kept fixed (so the displayed scale variation does not contain the uncertainties of the latter) while the scales $\mu_c=\mu_b=\mu_1$ are varied simultaneously.  The grey band
      represents the experimental results.
      }
%~\\[-5mm]
%\hrule
\end{figure}

In Fig.~\ref{fig::DelGam_mu1} we show the $\mu_1$ dependence of
$\Delta\Gamma_s/\Delta M_s$ with
$\mu_1=\mu_c=\mu_b$.  The experimental result is shown as a
grey band.  
To illustrate the effect of the NNLO
calculation we vary the renormalisation scale only in the leading $1/m_b$
corrections; in the power-suppressed terms it is kept fixed.   For the
$\overline{\rm MS}$ and PS schemes LO, NLO and NNLO results are shown. In both
schemes we observe that the dependence on $\mu_1$ decreases with increasing
order in perturbation theory. For example, in the interval $[2.1,8.4]$~GeV one
observes a variation of $\Delta\Gamma_s/\Delta M_s$ in the $\overline{\text{MS}}$ scheme of about {${123}$\%} at
LO which reduces to {${33}$\%} at NLO and {$15$\%} at NNLO, where we take the NNLO result as the reference point in each case and always include the $1/m_b$ corrections with the scale fixed at $\SI{4.2}{GeV}$. For the PS
scheme the corresponding numbers are {$\{35\%,26\%,17\%\}$}.
Let us stress that the NNLO results of both schemes are very close
in the whole $\mu_1$ interval.

It is interesting to consider the perturbative expansion at the central
scale $\mu=4.2$~GeV. In the PS scheme the NLO corrections are negative
and reduce the LO predictions by about { ${22}$\%}. Also the NNLO
corrections are negative and about a factor ${1.6}$ smaller
than at NLO. This is different for the $\overline{\rm MS}$ scheme.
Here the NNLO corrections amount to about { ${10}$\%} and 
are about a factor  ${1.1}$ larger than at NLO.
This is because the central scale is close to $\mu_1\approx 5$~GeV
where the NLO corrections vanish. The situation is completely
different at other values for $\mu_1$. For example, for 
$\mu_1\approx 2.5$~GeV or $\mu_1\approx {9}$~GeV we observe 
NNLO corrections which are much smaller than the NLO ones,
indicating a perfect behaviour of perturbation theory.

Both NNLO predictions lie inside the uncertainty band of
experimental result for $\mu_1 \gtrsim 4$~GeV in the $\overline{\rm MS}$ and
for $\mu_1 \gtrsim 5$~GeV in the PS scheme. For smaller values
the NLO $\overline{\rm MS}$ curve lies within the uncertainty band
for most of the $\mu_1$ values. However, in the PS scheme
the prediction is outside the band for $\mu_1 \gtrsim 3$~GeV.

For comparison we show in Fig.~\ref{fig::DelGam_mu1} also the NNLO result
in the pole scheme. We observe a stronger dependence on $\mu_1$.
Furthermore, the agreement with the experimental
result is worse.

One of the benefits of the ratio $\Delta\Gamma_q/\Delta M_q$ is the reduced
dependence on the hadronic matrix elements.  $\Delta M_q$ only depends on the
numerically dominant matrix element $\bra{B_q} Q (\mu_2) \ket{\overline B_q}$
and thus the leading term of $\Delta\Gamma_q/\Delta M_q$ is independent of
hadronic matrix elements. The numerically sub-leading terms are either
proportional to
$ \bra{B_q}\widetilde Q_S(\mu_2)\ket{\overline B_q}/ \bra{B_q}
Q(\mu_2)\ket{\overline B_q}$ (from the leading power in the $1/m_b$ expansion)
or involve matrix elements from the $1/m_b$ contributions divided by 
$\bra{B_q} Q(\mu_2)\ket{\overline B_q}$.
To illustrate the relative size of the
contributions, we show the decomposition of the central value from Eq.~\eqref{eq::dGdM} in the $\overline{\rm MS}$
scheme for the $B_s$ system,
\begin{equation}
    \frac{\Delta \Gamma_s}{\Delta M_s} = 4.37 \times 10^{-3} \approx \left(4.20 + 1.69_{\widetilde{Q}_S/Q} - 1.53_{1/m_b}\right) \times 10^{-3}\ (\overline{\textrm{MS}})\,,
\end{equation}
where the first number in the round brackets denotes the term independent of hadronic matrix elements, $\widetilde{Q}_S/Q$ the term proportional to the ratio of the leading order hadronic matrix elements and $1/m_b$ the contribution from the sub-leading terms in the HQE.
Note that
the sum of the latter two makes up less than {4}\% of the leading contribution and is
of the order of the hadronic uncertainty of the leading $1/m_b$ contribution.

The most precise prediction for $\Delta\Gamma_s$ is obtained from the results
in Eq.~(\ref{eq::dGdM}) combined with the experimental
result~\cite{LHCb:2021moh}
\begin{equation}
  \Delta M_s^{\rm exp} = 17.7656 \pm 0.0057~\mbox{ps}^{-1}\,.
\end{equation}
Using Eq.~(\ref{eq::dGdM}) we obtain for the three renormalisation schemes
\begin{align}
  \Delta \Gamma_s
  ~=&~ \left(
      {{6.82^{+0.94}_{-1.02}}_{\textrm{scale}}}
      {{}^{+0.16}_{-0.34}}_{\textrm{scale, $1/m_b$}}
      \pm 0.19_{B\widetilde{B}_S}
    \; \pm 1.39_{1/m_b} \pm 0.10_{\textrm{input}}
    \right) \times
    10^{-2}\mbox{ps}^{-1}\ 
      (\textrm{pole})\,, \nonumber\\
      \Delta \Gamma_s
  ~=&~ \left(
      {{7.76^{+0.40}_{-0.79}}_{\textrm{scale}}}
      {{}^{+0.16}_{-0.34}}_{\textrm{scale, $1/m_b$}}
      \pm 0.21_{B\widetilde{B}_S}
    \; \pm 1.39_{1/m_b} \pm 0.09_{\textrm{input}}
    \right) \times
    10^{-2}\mbox{ps}^{-1}\ 
     (\overline{\textrm{MS}})\,,
   \nonumber\\
  \Delta \Gamma_s 
  ~=&~ \left(
      {{7.58^{+0.63}_{-0.66}}_{\textrm{scale}}}
      {{}^{+0.16}_{-0.34}}_{\textrm{scale, $1/m_b$}}
      \pm 0.20_{B\widetilde{B}_S}
    \; \pm 1.39_{1/m_b} \pm 0.09_{\textrm{input}}
    \right) \times
    10^{-2}\mbox{ps}^{-1}\ 
                      (\textrm{PS})\,.
          \label{eq::dGamma_s}
\end{align}
For the final result for $\Delta\Gamma_s$ we only consider the
$\overline{\rm MS}$ and PS schemes. The total uncertainty in each scheme is obtained by adding the upper and lower bounds in quadrature, which are then symmetrised. Averaging the central values and the symmetrised uncertainties of the two
%First we symmetrise the scale dependence and add the various uncertainties in quadrature. Afterwards we average the two
schemes and obtain
\begin{equation}
  \Delta\Gamma_s = {{(0.077 \pm 0.016)}} ~\mbox{ps}^{-1}\,. \label{eq:dGs_average}
\end{equation}
The comparison to Eq.~(\ref{eq:expdgs}) shows that the uncertainty is only bigger by about
a factor of three  than the experimental uncertainty and dominated by the $1/m_b$
corrections. Equation~(\ref{eq::dGamma_s}) shows that thanks to the NNLO calculation the perturbative uncertainty of the leading-power part as estimated from the scale dependence 
is only marginally larger
than the current experimental error in \eq{eq:expdgs}. Clearly, a better calculation of the $1/m_b$ contribution is needed now.

In the analysis performed in
Refs.~\cite{Lenz:2019lvd,Albrecht:2024oyn} the NLO result
$\Delta\Gamma_s = {{(0.091 \pm 0.015)}} ~\mbox{ps}^{-1}$
has been obtained, which we can compare to our NLO prediction of $\Delta\Gamma_s = {(0.091 \pm 0.020)} ~\mbox{ps}^{-1}$, analogous to Eq.~\eqref{eq:dGs_average}. As can be seen from Fig.~\ref{fig::DelGam_mu1} and as  discussed above, the NNLO corrections shift the central value down. It is noteworthy that the central values for the two NLO predictions are quite close together, even though in Refs.~\cite{Lenz:2019lvd,Albrecht:2024oyn} only the leading term from the penguin operators is taken into account. For the operator matrix elements of the leading dimension-6
operators sum rule results
from Refs.~\cite{Kirk:2017juj,DiLuzio:2019jyq} are used. Furthermore, the estimate of the uncertainties from renormalisation scale variation and the one from sub-leading matrix elements are smaller than in our analysis.

%\clearpage

\subsubsection{$B_d$ system}

Results for $\Delta \Gamma_d$ can be obtained
by repeating the analysis performed for the $B_s$ system
after adjusting the relevant parameters.
For the three renormalisation schemes we get
\begin{align}
  \frac{\Delta \Gamma_d}{\Delta M_d}
  ~=&~ \left(
      {{3.69^{+0.52}_{-0.57}}_{\textrm{scale}}  }
      {{}^{+0.12}_{-0.20}}_{\textrm{scale, $1/m_b$}}
      \pm 0.11_{B\widetilde{B}_S}
    \pm 0.80_{1/m_b} \pm 0.06_{\textrm{input}}
    \right) \times
    10^{-3}\ (\textrm{pole})\,, \nonumber\\
      \frac{\Delta \Gamma_d}{\Delta M_d} 
  ~=&~ \left(
      {{4.21^{+0.23}_{ -0.44}}_{\textrm{scale}}   }
      {{}^{+0.12}_{-0.20}}_{\textrm{scale, $1/m_b$}}
      \pm 0.12_{B\widetilde{B}_S}
    \pm 0.80_{1/m_b} \pm 0.05_{\textrm{input}}
   \right) \times 10^{-3}\ (\overline{\textrm{MS}})\,,
   \nonumber\\
  \frac{\Delta \Gamma_d}{\Delta M_d} ~=&~ 
    \left( 
    {{ 4.11^{+0.35}_{ -0.37}}_{\textrm{scale}} } 
      {{}^{+0.12}_{-0.20}}_{\textrm{scale, $1/m_b$}}
      \pm 0.12_{B\widetilde{B}_S}
      \pm 0.80_{1/m_b}
          \pm 0.05_{\textrm{input}}
    \right) \times 10^{-3}\ (\textrm{PS})\,.
          \label{eq::dGdM_d}
\end{align}

From the results in Eq.~(\ref{eq::dGdM_d})
we obtain after multiplication with $\Delta M_d^{\rm exp}$
\begin{align}
  \Delta \Gamma_d
  ~=&~ \left(
      {{1.87^{+0.26}_{-0.29}}_{\textrm{scale}}    }
      {{}^{+0.06}_{-0.10}}_{\textrm{scale, $1/m_b$}}
      \pm 0.06_{B\widetilde{B}_S}
    \; \pm 0.40_{1/m_b} \pm 0.03_{\textrm{input}}  
    \right) \times
    10^{-3}\mbox{ps}^{-1}\ 
      (\textrm{pole})\,, \nonumber\\
      \Delta \Gamma_d
  ~=&~ \left(
      {{2.13^{+0.11}_{-0.23}}_{\textrm{scale}}  }
      {{}^{+0.06}_{-0.10}}_{\textrm{scale, $1/m_b$}}
      \pm 0.06_{B\widetilde{B}_S}
    \; \pm 0.40_{1/m_b} \pm 0.03_{\textrm{input}}  
    \right) \times
    10^{-3}\mbox{ps}^{-1}\ 
     (\overline{\textrm{MS}})\,,
   \nonumber\\
  \Delta \Gamma_d
  ~=&~ \left(
      {{2.08^{+0.18}_{-0.19}}_{\textrm{scale}}   }
      {{}^{+0.06}_{-0.10}}_{\textrm{scale, $1/m_b$}}
      \pm 0.06_{B\widetilde{B}_S}
    \; \pm 0.40_{1/m_b} \pm 0.03_{\textrm{input}}  
    \right) \times
    10^{-3} \mbox{ps}^{-1}\ 
                      (\textrm{PS})\,,
          \label{eq::dGamma_d_2}
\end{align}
where~\cite{HFLAV:2022esi}
\begin{equation}
  \Delta M_d^{\rm exp} = (0.5065 \pm 0.0019)\, \mbox{ps}^{-1}
\end{equation}
has been used.

We add the various uncertainties in quadrature for the upper and lower bounds separately, symmetrise the total uncertainty in each scheme and average the results for the $\overline{\text{MS}}$ and PS schemes to obtain
\begin{equation}
    \Delta \Gamma_d = {{(0.00211 \pm 0.00045)}} ~\mbox{ps}^{-1}\,.
\end{equation}

For the $B_d$ system the pattern of the uncertainties is very similar
to the $B_s$ system: The dominant contribution comes from the
matrix elements of the power-suppressed corrections, followed
by the renormalisation scale uncertainty in the
leading-power term, the uncertainty in the leading-power bag parameters
and from the scale variation in the $1/m_b$ piece.
Also for $B_d$
the variation of the remaining input parameters is only a minor effect.

\begin{figure}[t]
  \begin{center}
      \includegraphics[width=0.8\textwidth]{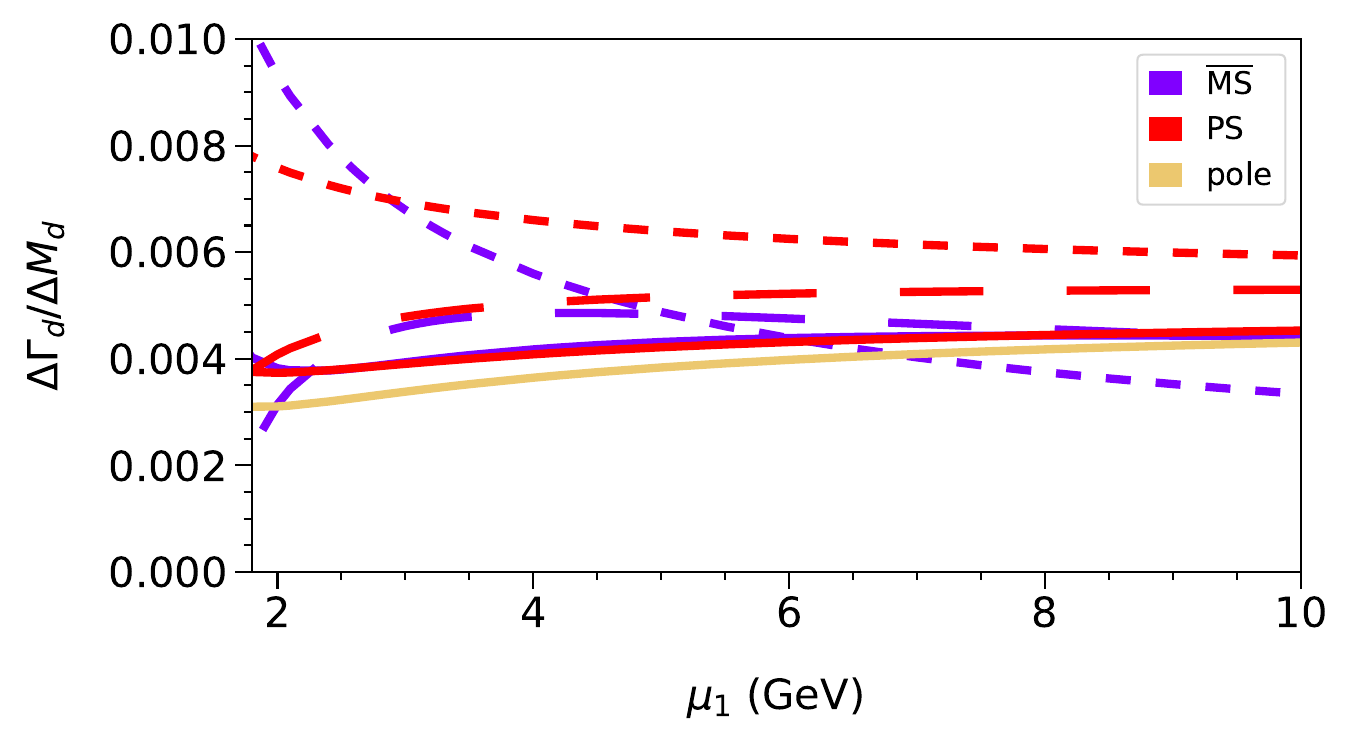}
  \end{center}    \caption{\label{fig::DelGam_mu1_Bd} 
      Renormalisation scale dependence at LO (short dashes),
      NLO (long dashes) and NNLO (solid) for $\Delta\Gamma_d/\Delta M_d$.  The
      scale in the power-suppressed terms is kept fixed while the scales $\mu_c=\mu_b=\mu_1$ are varied simultaneously. 
      }  
%  ~\\[-5mm]
%\hrule    
\end{figure}

In Fig.~\ref{fig::DelGam_mu1_Bd} we show the $\mu_1$ dependence of
$\Delta\Gamma_d/\Delta M_d$. We refrain
from showing the experimental results since the uncertainties are quite
large and would cover the whole plot range.
For the $\overline{\rm MS}$ and PS scheme we again show LO, NLO and NNLO results and observe the same convergence pattern as for $B_s$. Also for $B_d$ it is a welcome feature that the NNLO predictions for both renormalisation schemes
are very close together. The NNLO prediction for the pole
scheme is only shown for reference; it is not used for our
final prediction.

In an alternative approach it is possible to obtain
$\Delta \Gamma_d$ from
$\Delta\Gamma_s/\Delta M_s$ after multiplying with the experimental value for
$\Delta M_d$. This is possible since the CKM-suppressed contribution to
$\Delta\Gamma_d/\Delta M_d$ is only subleading. A priori it is  expected to be numerically
relevant due to $|\lambda^d_u/\lambda^d_t|\gg |\lambda^s_u/\lambda^s_t|$, but there are significant numerical cancellations in the sum of
$uc$ and $uu$ contributions, which has been observed for the first time in
Ref.~\cite{Beneke:2003az}. Furthermore, the non-perturbative hadronic matrix
elements for the $B_s$ and $B_d$ system agree well within the uncertainties as
can be seen in Tab.~\ref{tab::input}. Thus, the agreement of the central
values for $\Delta\Gamma_d/\Delta M_d$ and $\Delta\Gamma_s/\Delta M_s$ is well
below the uncertainties given in Eq.~(\ref{eq::dGdM}).
Our results are given by
\begin{align}
  \Delta \Gamma_d
  ~=&~ \left(
      {{1.94^{+0.27}_{-0.29}}_{\textrm{scale}}  }
      {{}^{+0.05}_{-0.10}}_{\textrm{scale, $1/m_b$}} 
      \pm 0.05_{B\widetilde{B}_S}
    \; \pm 0.40_{1/m_b} \pm 0.03_{\textrm{input}}  
    \right)  
    \times
    10^{-3}\mbox{ps}^{-1}\ 
      (\textrm{pole})\,, \nonumber\\
      \Delta \Gamma_d
  ~=&~ \left(
      {{2.21^{+0.11}_{-0.22}}_{\textrm{scale}} } 
      {{}^{+0.05}_{-0.10}}_{\textrm{scale, $1/m_b$}} 
      \pm 0.06_{B\widetilde{B}_S}
    \; \pm 0.40_{1/m_b} \pm 0.03_{\textrm{input}} 
    \right) \times
    10^{-3}\mbox{ps}^{-1}\ 
     (\overline{\textrm{MS}})\,,
   \nonumber\\
  \Delta \Gamma_d
  ~=&~ \left(
      {{2.16^{+0.18}_{-0.19}}_{\textrm{scale}}   }
      {{}^{+0.05}_{-0.10}}_{\textrm{scale, $1/m_b$}} 
      \pm 0.06_{B\widetilde{B}_S}
    \; \pm 0.40_{1/m_b} \pm 0.03_{\textrm{input}}
    \right) \times
    10^{-3} \mbox{ps}^{-1}\ 
                      (\textrm{PS})\,,
          \label{eq::dGamma_d}
\end{align}
Adding the various uncertainties in quadrature for the upper and lower bounds separately, symmetrising the total
uncertainty in each scheme, and averaging the results for the $\overline{\text{MS}}$ and PS schemes, we obtain
\begin{equation}
    \Delta \Gamma_d = {{(0.00219 \pm 0.00045)}} ~\mbox{ps}^{-1}\,.\label{eq:dGd_average}
\end{equation}

The comparison of Eqs.~(\ref{eq::dGamma_d_2}) and~(\ref{eq::dGamma_d}) 
shows good agreement in the central values well within the uncertainties. 
Also most of the individual uncertainties
agree well for the two approaches to compute $\Delta\Gamma_d$. 

The SM prediction of $\dg_s/\dm_s$ is almost completely independent  of any CKM element, while 
the individual predictions for $\dg_s$ and $\dm_s$  depend on $|V_{cb}|$ 
(which determines $|V_{ts}|\simeq (0.982 \pm 0.001) |V_{cb}|$ from the unitarity of the 
CKM matrix), so that the ratio $\dg_s/\dm_s$ (or, equivalently, 
$\dg_s$ in \eq{eq:dGs_average} calculated from this ratio) probes, e.g.\ new physics in $\dm_s$ without the need for a resolution of the $V_{cb}$ puzzle.

The situation is  different for 
$\dg_d/\dm_d$, which depends on the apex $(\bar\rho,\bar\eta)$ of the CKM unitarity triangle 
through $\lambda_u^d/\lambda_t^d$. Thus the SM predictions of $\dg_d/\dm_d$ in \eq{eq::dGdM_d} and 
of $\dg_d$ in \eq{eq:dGd_average} assume that not only the predicted quantities are free from new-physics contributions, but also the observables entering the global fit determining $(\bar\rho,\bar\eta)$. Coincidentally, if the fitted values for $(\bar\rho,\bar\eta)$ 
and thus the value for {$\lambda_u^d/\lambda_t^d$} in \eq{eq::ckm_input} are correct, the linear and quadratic terms in the square bracket of \eq{eq::Gam12} almost exactly cancel in $\dg_d$ (which is proportional to the real part of the square bracket) \cite{Beneke:2003az}. As a consequence, for the values in \eq{eq::ckm_input} we find $\dg_d/\dm_d \simeq \dg_s/\dm_s$.

%\clearpage

\subsubsection{$\Delta\Gamma_q/\Delta M_q$ versus $\Delta\Gamma_q$ and $\Delta M_q$}

\begin{figure}[t]
  \begin{center}
    \begin{tabular}{cc}
      \includegraphics[width=0.45\textwidth]{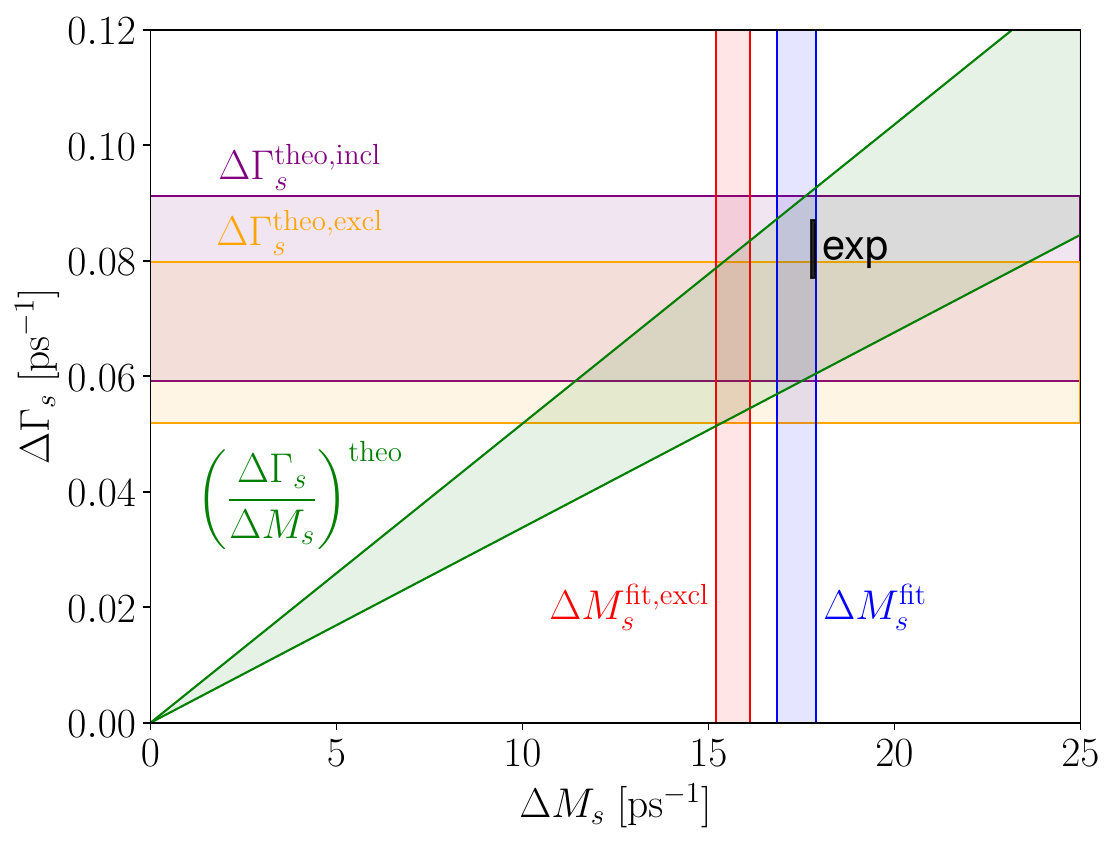}
      &
%      \\
      % \includegraphics[width=0.45\textwidth]{figs_phen/dGdM}
      \includegraphics[width=0.45\textwidth]{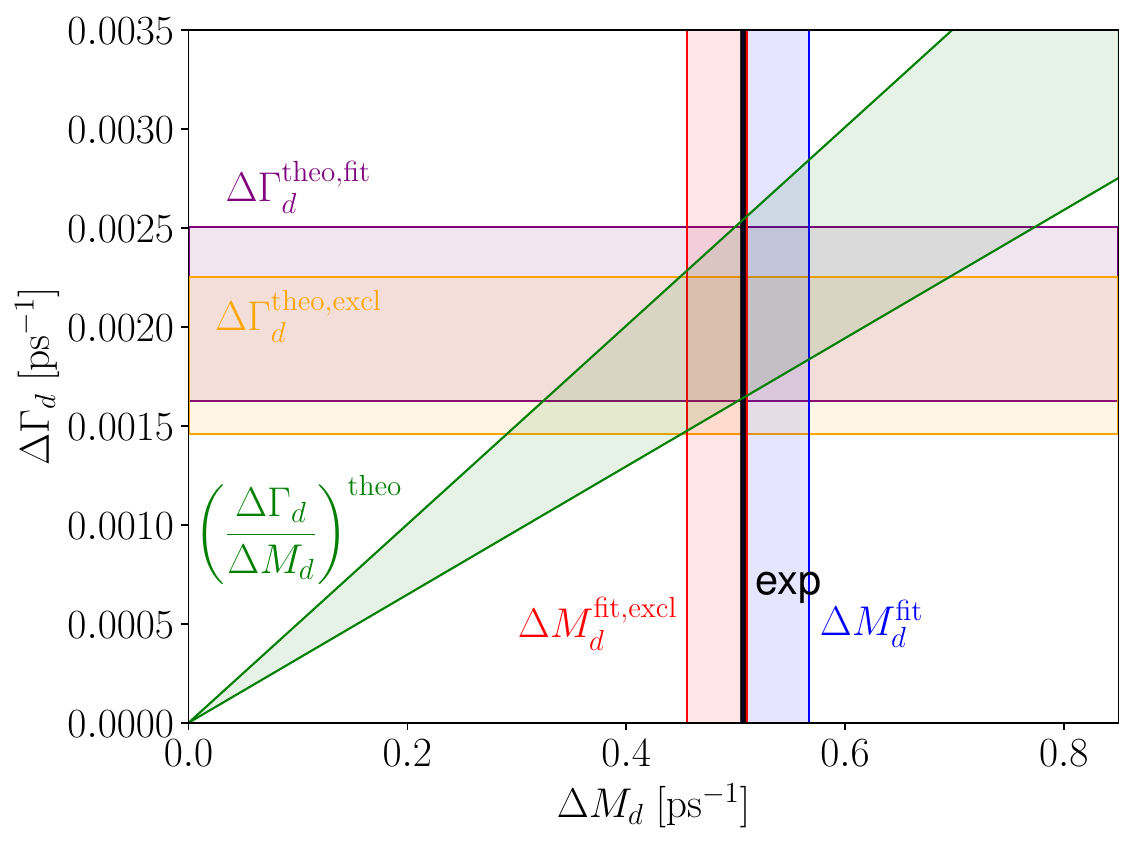}
%      \\
%      (a) & (b)
    \end{tabular}
     \end{center}
     \caption{\label{fig::dGdM}$\Delta\Gamma_q$ versus $\Delta M_q$ for the
      $B_s$ (left) and $B_d$ (right) schemes. The different horizontal and vertical
      coloured bands refer to different input values for $|V_{cb}|$. The bands for $\Delta \Gamma_s$ are obtained from a direct calculation according to Eq.~\eqref{eq:dgdmafsq} using the values for $\lambda_t^{s}$ as given in Eq.~\eqref{eq:lambdat}. The band for $\Delta M_s^\textrm{fit}$ is obtained from the value and  $1\sigma$ uncertainty given in Ref.~\cite{CKMfitter} where $\Delta M_s$ was excluded from the fit. $\Delta M_s^\textrm{fit,excl}$ is the corresponding value rescaled by $|V_{cb}^\textrm{excl}|^2/|V_{cb}^\textrm{SM fit}|^2$. The bands for $B_d$ are completely analogous with the exception of $\Delta \Gamma_d^\textrm{theo,fit}$, which was calculated using $|\lambda_t^{d, \textrm{SM fit}}|$.
      The experimental values shown are as given in Eqs.~\eqref{eq:expdms} to \eqref{eq:expdgd} with the uncertainty on $\Delta M_s$ scaled up by a factor of ten for visibility.}
%  ~\\[-5mm]
%\hrule
\end{figure}

We are now in the position to confront the ratio $\Delta\Gamma_q/\Delta M_q$
($q=d,s$)
with the individual predictions for $\Delta\Gamma_q$ and $\Delta M_q$ which is
done in Fig.~\ref{fig::dGdM}.  The $|V_{cb}|$ controversy in \eq{eq:vcbei} (red versus blue vertical and orange versus purple horizontal strips) prevents any conclusion on
possible new physics from $\Delta M_d$ or $\Delta M_s$ alone. 
As stated above, the predictions for $\dg_s$ and $\dm_s$ depend on no other CKM parameters than
$|V_{cb}|$ and we can label the corresponding predictions with ``excl'' and ``incl''. 

In the case of $\dg_d$ and $\dm_d$ we use the prediction using the CKM 
parameters from the global fit instead of the $|V_{cb}^{\rm incl}|$ scenario. To this end we use 
the CKMFitter result from a fit in which $\dm_d$ has not been used to delimit the allowed region for $\dm_d$. The fit result corresponds to $|\lambda_t^{d,\textrm{SM fit}}|$ in Eq.~\eqref{eq:lambdat} which is then used for the band for $\dg_d$. 
For the case that the true value is $|V_{cb}^{\rm excl}|$, one cannot easily determine the 
allowed regions in Fig.~\ref{fig::dGdM} because the small value of $|V_{cb}^{\rm excl}|$ gives a very bad fit to $(\bar\rho,\bar\eta)$. For simplicity, we choose the same 
$(\bar\rho,\bar\eta)$ as for $\dm_d^{\rm fit}$, i.e.\ we simply rescale $\dg_d$ and $\dm_d$ by
the corresponding values of $|V_{cb}|^2$ in \fig{fig::dGdM}, as explained after \eq{eq:lambdat}.

Clearly the experimental uncertainty of 
$\Delta \Gamma_d$ is much bigger than that of the theoretical prediction 
at present, calling for a better measurement.

In conclusion, a combined analysis of
$\Delta M_q$ and $\Delta\Gamma_q$ adds important information because the SM prediction of $\Delta\Gamma_q/\Delta M_q$, shown as green wedge, is independent of $|V_{cb}|$ 
and one can disentangle the probe of new physics 
from the determination of CKM parameters.

%- }}}

%\clearpage

%- {{{ $a_{\rm fs}^q$:

\subsection{$a_{\rm fs}^q$}

Before presenting our results for $a_{\rm fs}^q$, let us briefly discuss possible measurements of the observable. A flavour-specific $B_q \to f$ decay is defined by the property that $B_q\to f$ is allowed while 
$B_q \to \bar f$ and $\bar B_q \to f$ are forbidden. Here $\bar f$ is the CP conjugate final state to $f$. Traditionally one uses semileptonic decays with $f=X \ell^+\nu_\ell$ and $\bar f=\bar X \ell^-\bar \nu_\ell$. But to gain statistics, it is desirable to include as many flavour-specific decays as possible into the measurement of $a_{\rm fs}^q$. For instance, $a_{\rm fs}^d$ can be measured as well with $B_d\to J/\psi K^+ \pi^-$, $B_d\to D_s^+ D^-$, $B_d\to D^- K^+$, and many more decays. 

$a_{\rm fs}^q$ is related to the time-dependent decay rates as
\begin{equation}
 \frac{\gqbtf -  \gdtfb}{\gqbtf + \gdtfb} 
= a_{\rm fs}^q  + a_{\rm CP}^{\rm dir}, 
\label{afstc} 
\end{equation}
where $B_q(t)$ denotes a meson tagged as $B_q$ at time $t=0$ with an analogous definition of $\bar B_q(t)$. For $t>0$ a non-zero $a_{\rm fs}^q$ is generated, because an initially produced $\bar B_q$ oscillates into $B_q$ which then decays to $f$. 
\begin{equation}
  a_{\rm CP}^{\rm dir} =
  \frac{\Gamma(B_q\to f)-\Gamma(\bar B_q\to \bar f)}{\Gamma(B_q\to f)+\Gamma(\bar B_q\to \bar f)}
\end{equation}
is the direct CP asymmetry in the studied decay.
Usually $a_{\rm CP}^{\rm dir}$ is omitted when \eq{afstc} is quoted, because $a_{\rm CP}^{\rm dir}$
is zero in the SM for semileptonic decays. Yet this is not true 
for e.g.\ $B_d\to J/\psi K^+ \pi^-$ and $B_d\to D_s^+ D^-$.

It is well-known that $a_{\rm fs}^q$ can be measured without the need for flavour tagging. With the time-dependent untagged decay rate
\begin{equation}
\guntf = \gqtf + \gqbtf    .   \label{guntf}
\end{equation}
one finds 
\begin{equation}
a_{\rm fs, unt}^{q}(t) \equiv
   \frac{\guntf -  \guntfb}{\guntf +\guntfb}
=  a_{\rm CP}^{\rm dir} + 
        \frac{a_{\rm fs}^q}{2} \left(1 
        - \left(a_{\rm CP}^{\rm dir}\right)^2\right) \,
        \left(1- \frac{\cos (\dm_q\, t)}{\cosh (\dg_q t/2) } \right)
        . \,  \label{fsun}
\end{equation}
In all formulae terms of order $(a_{\rm fs}^q)^2$ are neglected. While the time dependence drops out of \eq{afstc}, this is not the case for \eq{fsun}. Tracking the oscillatory time dependence helps to 
disentangle $a_{\rm fs}^q$ from $a_{\rm CP}^{\rm dir}$ and possible experimental detection 
asymmetries \cite{Nierste:2004uz}.

Our results for the CP asymmetry in flavour-specific decays for the
$B_s$ system read
\begin{align*}
  a_{\rm fs}^s
  &= 
      \left(
      {{2.28^{{+0.01}}_{-0.04}}_{\textrm{scale}}} 
      { {{}^{+0.01}_{-0.00}}_{\textrm{scale, $1/m_b$}}}
      {\pm 0.01_{B\widetilde{B}_S}}
      {\pm 0.06_{1/m_b}}
      {\pm 0.07_{\textrm{input}}}
      \right) \times
      10^{-5}\ (\textrm{pole})\,,\\
  a_{\rm fs}^s
  &= 
      \left(
      {{2.25^{{+0.10}}_{-0.19}}_{\textrm{scale}}}
      { {{}^{+0.01}_{-0.00}}_{\textrm{scale, $1/m_b$}}}
      {\pm 0.01_{B\widetilde{B}_S}}
      {\pm 0.06_{1/m_b}}
      {\pm 0.07_{\textrm{input}}}
      \right) \times
      10^{-5}\ (\overline{\rm MS})\,,\\
  a_{\rm fs}^s
  &= 
      \left(
      {{2.31^{{+0.03}}_{-0.07}}_{\textrm{scale}}}
      { {{}^{+0.01}_{-0.00}}_{\textrm{scale, $1/m_b$}}}
      {\pm 0.01_{B\widetilde{B}_S}}
      {\pm 0.06_{1/m_b} }
      {\pm 0.07_{\textrm{input}}}
      \right) \times
      10^{-5}\ (\textrm{PS})\,,\numberthis
  \label{eq:afsnum_s}
\end{align*}
and for the $B_d$ system we have
\begin{align*}
  a_{\rm fs}^d
  &= -
      \left(
      {{5.21^{{+0.00}}_{{-0.09}}}_{\textrm{scale}}   }
      {{}^{{+0.03}}_{{-0.08}}}_{\textrm{scale, $1/m_b$}}
      \pm {0.03_{B\widetilde{B}_S}}
      \pm {0.14_{1/m_b} }
      \pm 0.16_{\textrm{input}}
      \right) \times
      10^{-4}\ (\textrm{pole})\,,\\
  a_{\rm fs}^d
  &= -
      \left(
      {{5.15^{{+0.21}}_{{-0.43}}}_{\textrm{scale}}  }
      {{}^{{+0.03}}_{{-0.08}}}_{\textrm{scale, $1/m_b$}}
      \pm {0.03_{B\widetilde{B}_S}  }
      \pm {0.14_{1/m_b}   }
      \pm {0.16_{\textrm{input}}}
      \right) \times
      10^{-4}\ (\overline{\rm MS})\,,\\
  a_{\rm fs}^d
  &= -
      \left(
      {{5.28^{{+0.08}}_{{-0.17}}}_{\textrm{scale}}   }
      {{}^{{+0.03}}_{{-0.08}}}_{\textrm{scale, $1/m_b$}}
      \pm {0.03_{B\widetilde{B}_S}}
      \pm {0.14_{1/m_b} }
      \pm 0.16_{\textrm{input}}
      \right) \times
      10^{-4}\ (\textrm{PS})\,.\numberthis
  \label{eq:afsnum_d}
\end{align*}
As before, we separately add the various uncertainties in quadrature for the upper and lower bounds, symmetrise the total uncertainty in each scheme, and average the results for the $\overline{\text{MS}}$ and PS schemes to obtain
\begin{align*}
    a_{\rm fs}^s &= \left({2.28 \pm 0.14} \right) \times 10^{-5}\,,\\
    a_{\rm fs}^d &= -\left({5.21 \pm 0.32} \right) \times 10^{-4}\,.\numberthis
\end{align*}

To a large extend we observe the same pattern in the uncertainties as for
$\Delta\Gamma_q/\Delta M_q$. However, there is an important difference
since the uncertainty from the parameter variation is of the about the same size
as the one from the power-suppressed operator matrix elements. This is due
to the parameters $\lambda^q_u/\lambda^q_t$ and $m_c$, which dominate the
``input'' uncertainties with contributions of comparable size. 

Note that in the limit $m_c\to0$ the CP asymmetries vanish.
It is interesting to note that the contribution beyond the leading $m_c$ contribution only amounts to {13\%} for both $a_{\rm fs}^s$ and $a_{\rm fs}^d$ in both the PS and $\overline{\textrm{MS}}$ scheme.
%both in PS and MSbar scheme

In the following, we discuss the dependence on the 
scale $\mu_1=\mu_b=\mu_c$ in analogy to $\Delta\Gamma_q/\Delta M_q$
in the previous subsection. In particular, we focus again on the
$\overline{\rm MS}$ and PS schemes and show results for the pole scheme only
for reference.

\begin{figure}[t]
  \begin{center}
    \begin{tabular}{c}
      \includegraphics[width=0.8\textwidth]{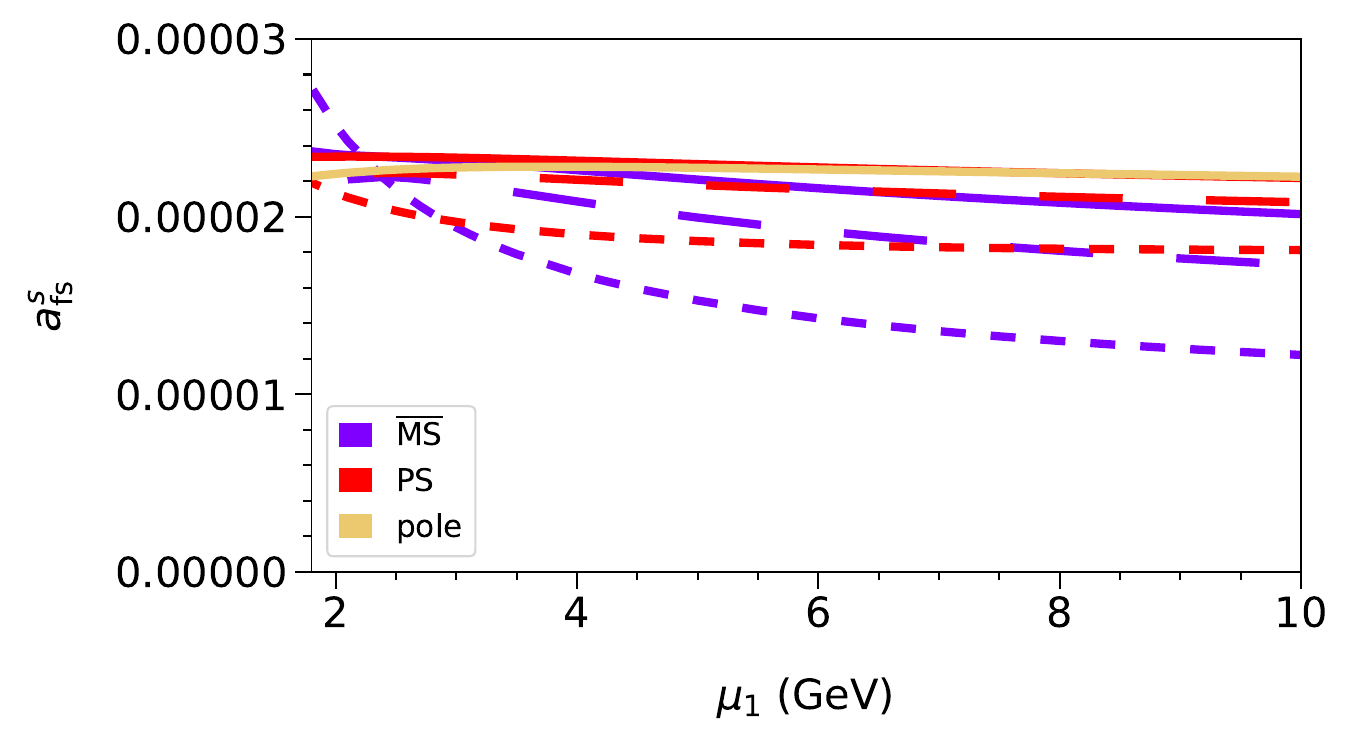}
    \\ 
    %(a) \\
      \includegraphics[width=0.8\textwidth]{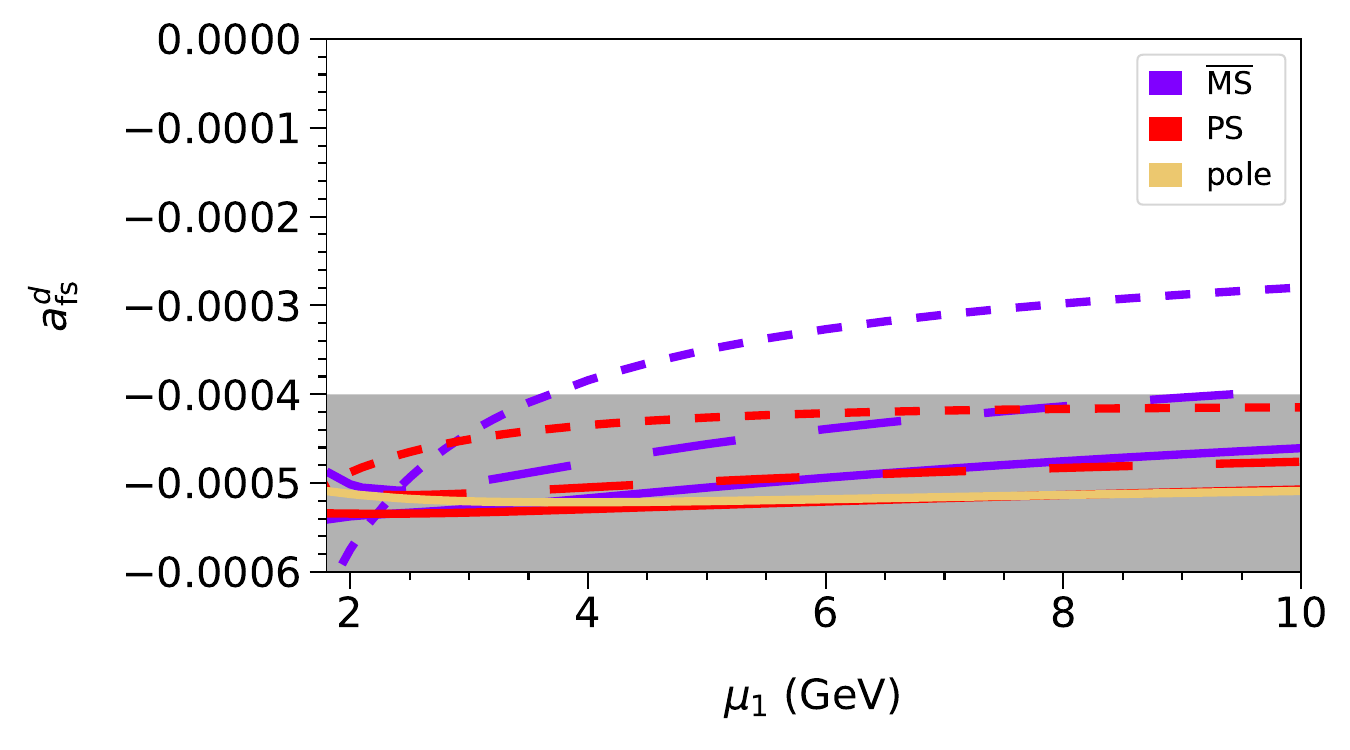}
    \\ 
    %(b)
    \end{tabular}
  \end{center}   \caption{\label{fig::afd_afs_mu1}
  Renormalisation scale dependence at LO (short dashes), NLO (long dashes) and
      NNLO (solid) for $a_{\rm fs}^s$ (top) and $a_{\rm fs}^d$ (bottom). The scale
      in the power-suppressed terms is kept fixed. 
      The grey band represents the $1\sigma$ 
      experimental error of $a_{\rm fs}^d$ from \eq{eq:expafsd}.}
%~\\[-5mm]
%\hrule
\end{figure}

In Fig.~\ref{fig::afd_afs_mu1} we show the $\mu_1$ dependence of the leading
$1/m_b$ contribution of $a_{\rm fs}^s$ and $a_{\rm fs}^d$ and keep the scale
in the power-suppressed terms fixed.  For the $\overline{\rm MS}$ and PS
scheme LO, NLO and NNLO predictions are shown. Similarly to
Fig.~\ref{fig::DelGam_mu1} we observe in both schemes a decreased dependence
on $\mu_1$ for the higher order perturbative corrections. For both
$a^s_{\rm fs}$ and $a^d_{\rm fs}$ the NNLO terms are about a factor of 3 smaller than the NLO
corrections at the central scale $\mu_1=4.2$~GeV in the PS scheme.
Both for $a^s_{\rm fs}$ and $a^d_{\rm fs}$ 
we observe a good convergence of the perturbative series
and an overlap of the bands
from the variation of $\mu_1$.

The NNLO prediction in the PS scheme is impressively flat and shows a
variation of only about {$5$\%} for both $a^s_{\rm fs}$ and $a^d_{\rm fs}$.  For small values of $\mu_1$ the
$\overline{\rm MS}$ and PS schemes agree.  However, in contrast to
Figs.~\ref{fig::DelGam_mu1} and \ref{fig::DelGam_mu1_Bd}, 
the $\mu_1$ dependence in the $\overline{\rm MS}$
is worse than the one in the PS scheme and a variation of about {$13$\%}
is observed for both $a^s_{\rm fs}$ and $a^d_{\rm fs}$.

The current experimental uncertainty for $a_{\rm fs}^d$ is too big
to draw conclusions about possible deviations from the
SM prediction. However, it is interesting to note that the NNLO
predictions are well inside the 
experimental band. 

For comparison we show in Fig.~\ref{fig::afd_afs_mu1} also the
NNLO prediction in the pole scheme. In contrast to
$\Delta\Gamma_q/\Delta M_q$ it shows a flat behaviour and
it is furthermore very close to the 
result in the PS scheme.

Finally we come back to the plot on the right panel of Fig.~\ref{fig:box} and discuss how a measurement of $a_{\rm fs}^d$ constrains the parameters $\bar{\rho}$ and $\bar{\eta}$ from the improved \cite{Buras:1994ec,Herrlich:1995hh} Wolfenstein \cite{Wolfenstein:1983yz} parametrisation of the CKM matrix.
We follow Ref.~\cite{Beneke:2003az}
and write $a_{\rm fs}^d$ as
\begin{equation}
  a_{\rm fs}^d = \left[a\, \text{Im} \frac{\lambda_u^d}{\lambda_t^d} + b\, \text{Im} \frac{(\lambda_u^d)^2}{(\lambda_t^d)^2} \right]\times 10^{-4}
  \,,
  \label{eq::afs_pred}
\end{equation}
with
\begin{equation}
\frac{\lambda_u^d}{\lambda_t^d} = \frac{1-\bar\rho - i \bar\eta}{(1-\bar\rho)^2+\bar\eta^2} - 1
\,.\label{eq:lambdUoT_param}
\end{equation}
From our calculation we can predict the quantities $a$ and $b$ in Eq.~(\ref{eq::afs_pred}) to LO, NLO and NNLO. We separate $a$ and $b$ into the leading $1/m_b$ and sub-leading $1/m_b$ contributions,
\begin{align*}
    a &= a_0 + a_1\,,\\
    b &= b_0 + b_1\,, \numberthis
\end{align*}
where the subscripts $0$ and $1$ indicate the leading and sub-leading terms, respectively. Our results for the leading term in the $\overline{\textrm{MS}}$ and PS schemes are given in Tab.~\ref{tab:afs_d}. The sub-leading contributions read
\begin{align*}
    a_1 &= {0.622^{+0.073}_{-0.020}}_\textrm{scale, $1/m_b$} \pm {0.43_{\textrm{para}}}\,,\\
    b_1 &= {0.091^{+0.011}_{-0.003}}_\textrm{scale, $1/m_b$} \pm {0.046_{\textrm{para}}}\,,
\end{align*}
which are identical for all schemes and orders as they are only known to LO. The uncertainties given here are the perturbative scale uncertainty from varying $\mu_1=\mu_b=\mu_c$ simultaneously between $\SI{2.1}{GeV}$ and $\SI{8.4}{GeV}$ and the combined uncertainties of all input parameters to our calculation. This means that all sources of uncertainties labelled with $B\widetilde{B}_S$, $1/m_b$ and input in Eq.~\eqref{eq:afsnum_d} have been added in quadrature. Therefore, our leading $1/m_b$ predictions $a_0$ and $b_0$ are free of input uncertainties.
\begin{table}[t]
    \centering
    \begin{tabular}{cc}
    \begin{tabular}{@{} l | l l l @{}}
         $\overline{\rm MS}$& LO & NLO & NNLO  \\
         \midrule
         $a_0$ & ${8.20^{+4.20}_{-1.94}}$ & ${10.47^{+0.76}_{-1.48}}$ & ${11.46^{+0.51}_{-1.00}}$ \\[5pt]
         $b_0$ & ${0.069^{+0.037}_{-0.020}}$ & ${0.112^{+0.043}_{-0.020}}$ & ${0.134^{+0.042}_{-0.022}}$
    \end{tabular}
    &
    \begin{tabular}{@{} l | l l l @{}}
         PS & LO & NLO & NNLO  \\
         \midrule
         $a_0$ & ${9.53^{+1.17}_{-0.39}}$ & ${11.18^{+0.26}_{-0.53}}$ & ${11.77^{+0.18}_{-0.40}}$ \\[5pt]
         $b_0$ & ${0.081^{+0.011}_{-0.009}}$ & ${0.122^{+0.027}_{-0.008}}$ & ${0.140^{+0.034}_{-0.015}}$
    \end{tabular}
    \end{tabular}
    \caption{Our results for the values $a_0$ and $b_0$ in the $\overline{\textrm{MS}}$ (left) and PS (right) schemes which form the basis of Fig.~\ref{fig::afd_CKM}. Note that the uncertainty shown here refers to the perturbative scale uncertainty from varying $\mu_1=\mu_b=\mu_c$ simultaneously between $\SI{2.1}{GeV}$ and $\SI{8.4}{GeV}$.}
    \label{tab:afs_d}
\end{table}

\begin{figure}[t]
    \centering
    \begin{tabular}{c}
    \mbox{
    \includegraphics[width=0.5\linewidth]{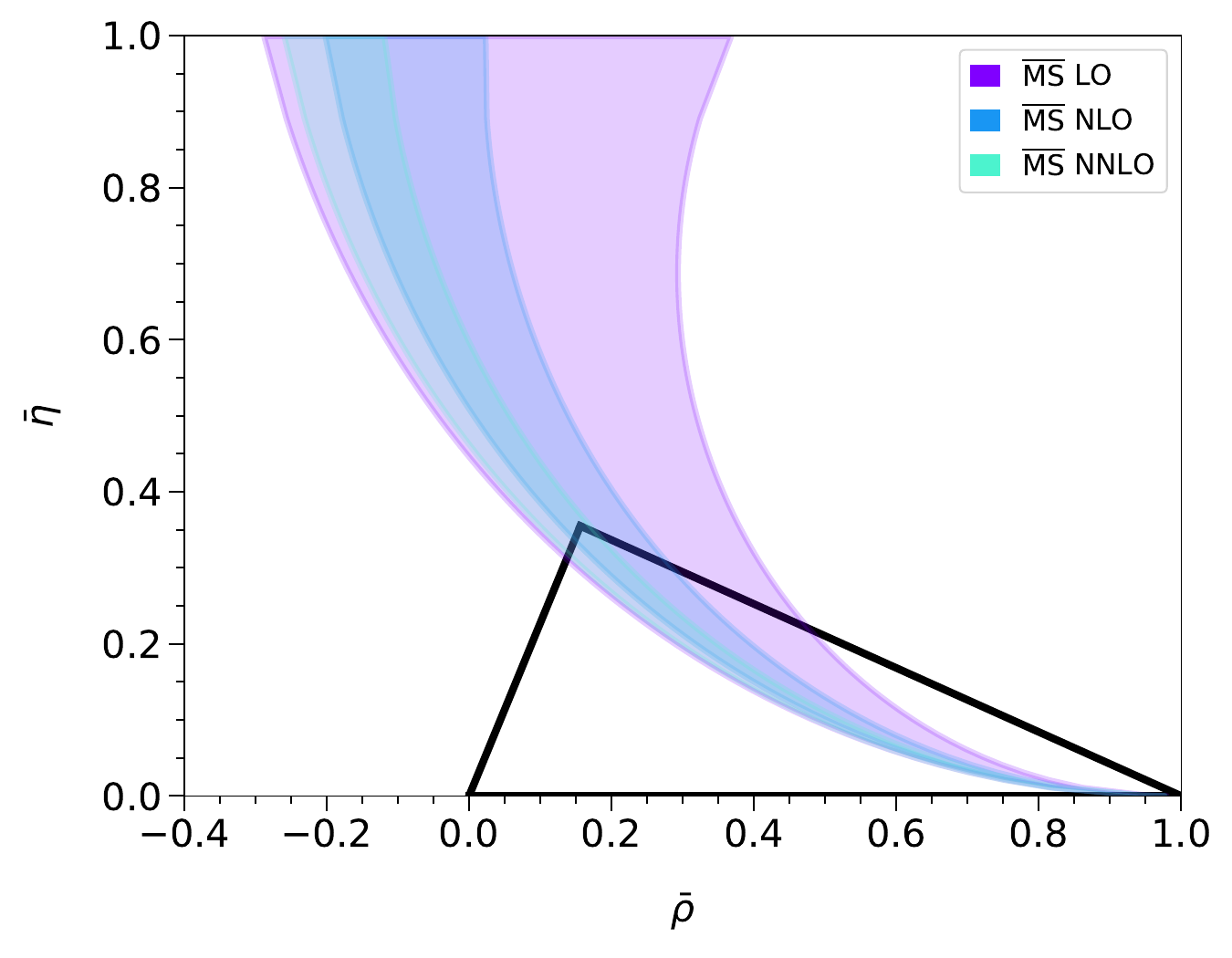}
    \includegraphics[width=0.5\linewidth]{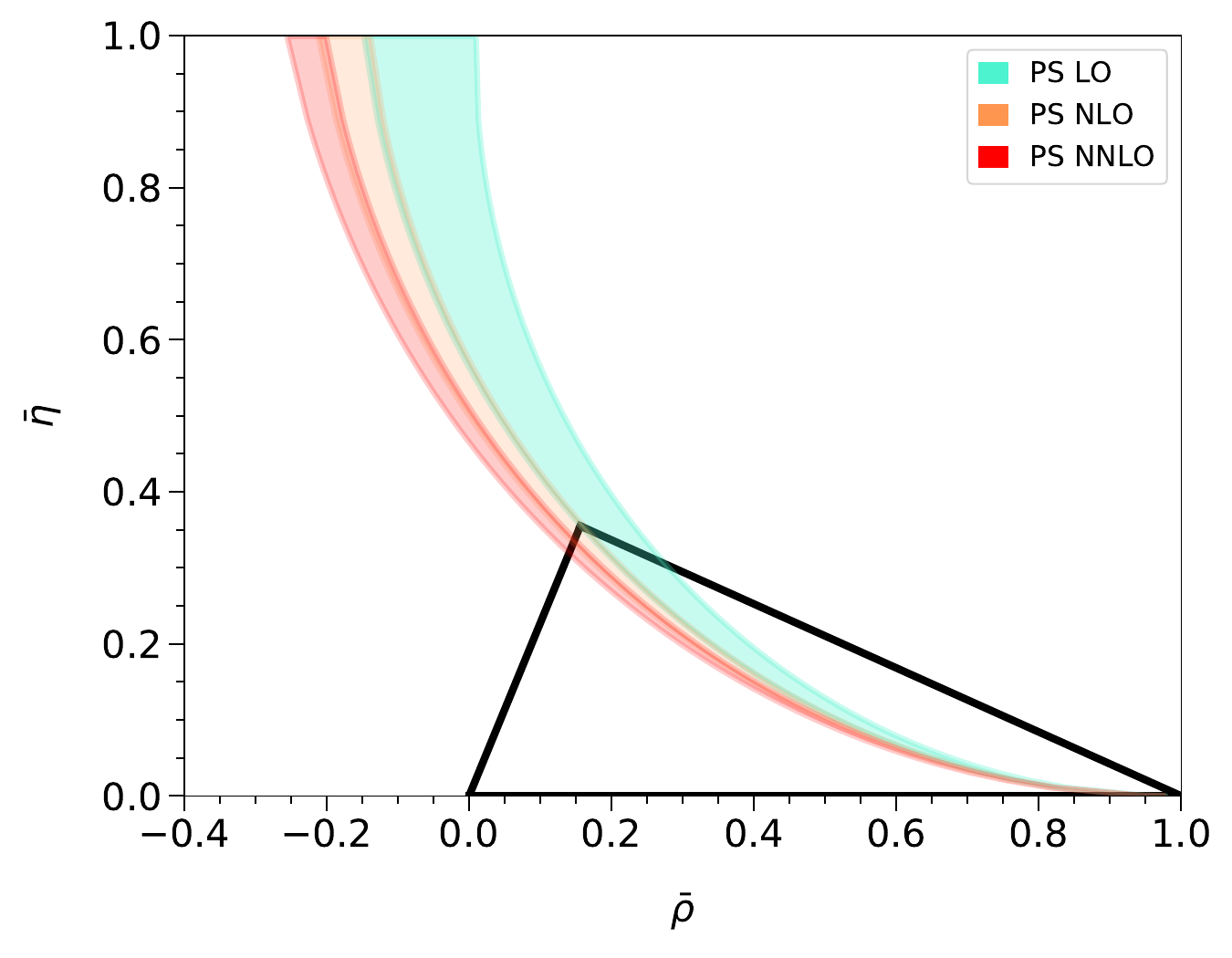}
     }
    \\
    \includegraphics[width=0.5\linewidth]{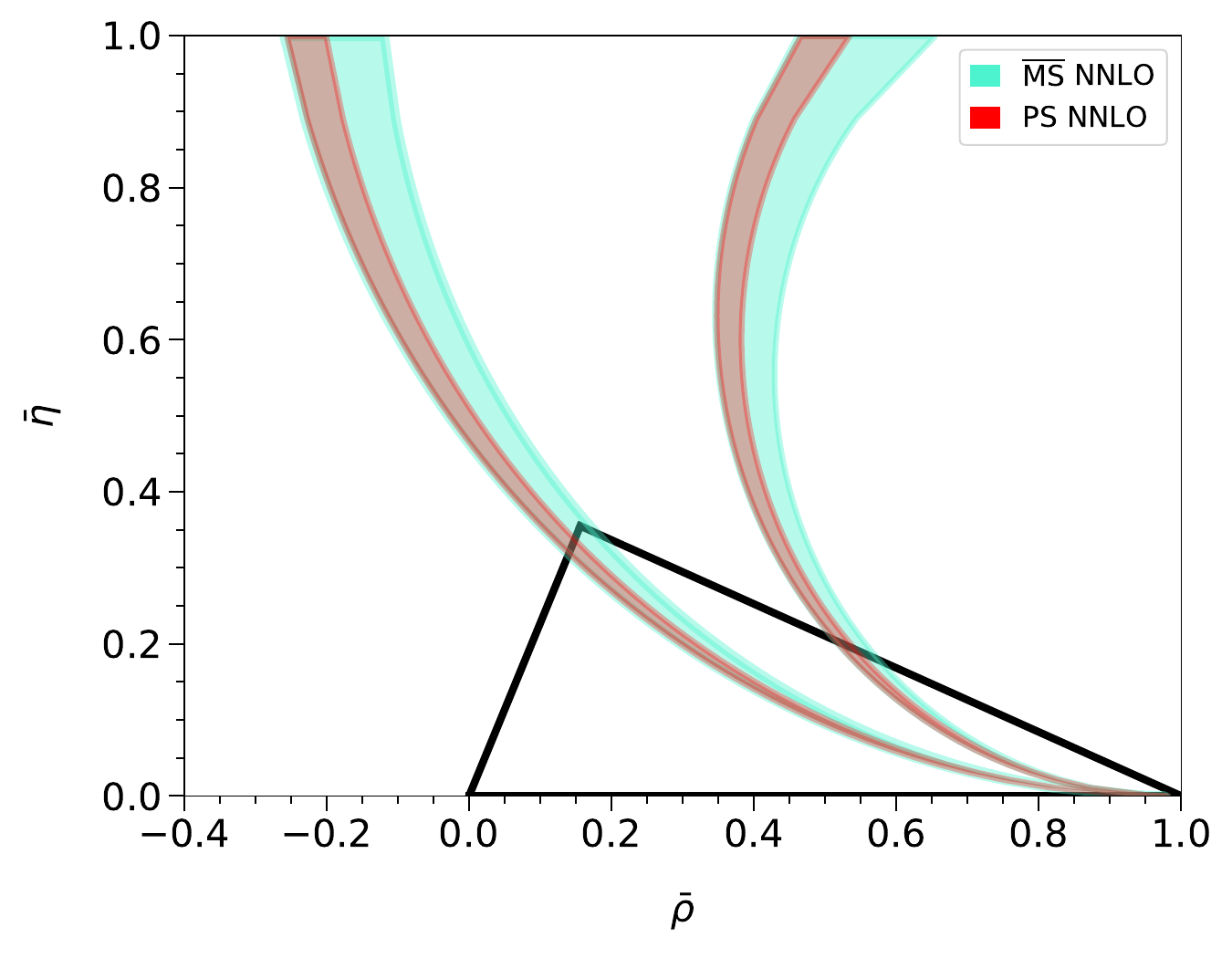}
    \end{tabular}
    \caption{Constraints on the apex of the CKM triangle in the $(\bar\rho, \bar\eta)$ 
    plane, assuming a measurement of $( a_{\rm fs}^d)^\text{exp} = -5 \times 10^{-4}$.
    In the top row LO, NLO and NNLO results are shown for the $\overline{\rm MS}$ and PS scheme.
    In the lower panel we compare the NNLO results in both schemes and also show solutions
    for $( a_{\rm fs}^d)^\text{exp} = -1 \times 10^{-3}$ (bands on the right). Note that the
    NNLO PS band is completely contained in the $\overline{\rm MS}$ band. The current 
    global average of the CKM triangle is shown in black~\cite{CKMfitter}.
    \label{fig::afd_CKM}}
%        ~\\[-1mm]
%\hrule
\end{figure}

If we use for the left-hand side of Eq.~(\ref{eq::afs_pred}) the experimentally measured result and insert Eq.~\eqref{eq:lambdUoT_param} into Eq.~\eqref{eq::afs_pred}, 
we obtain a constraint in the $\bar{\rho}$-$\bar{\eta}$ plane. 
In Fig.~\ref{fig::afd_CKM} we show the constraints that would arise from an experimental measurement of $( a_{\rm fs}^d)^\text{exp} = -5 \times 10^{-4}$. The plots on the top show LO, NLO and NNLO results for both the $\overline{\rm MS}$ and PS schemes.
The bands indicate the combined scale uncertainty of the leading and sub-leading
terms in the HQE, which are obtained by varying $\mu_1=\mu_b=\mu_c$ between $\SI{2.1}{GeV}$ and $\SI{8.4}{GeV}$. The lower panel compares the NNLO results for both schemes.
Furthermore, it contains results for 
$( a_{\rm fs}^d)^\text{exp} = -1 \times 10^{-3}$ (green and red bands on the right). 
For reference we also show the currently favoured CKM triangle~\cite{CKMfitter}.

One observes that with increasing order of perturbation theory the
widths of the bands decrease significantly. In the case of the
$\overline{\rm MS}$ scheme the NLO band is completely contained within the LO band. Furthermore, there is a significant overlap of the NNLO and NLO bands.
In general, the uncertainties in the PS scheme are smaller. Here we observe only a small overlap of the NNLO and NLO bands. However,
the NNLO band is completely contained in the NNLO band of the $\overline{\rm MS}$.

Fig.~\ref{fig::afd_CKM} is an extension of Fig.~3 in Ref.~\cite{Beneke:2003az} to NNLO, where it is worth noting that the authors of~\cite{Beneke:2003az} chose to plot the constraints arising from the averages of the pole and $\overline{\text{MS}}$ schemes.
The $\overline{\text{MS}}$ results that we provide, and which are the basis for our Fig.~\ref{fig::afd_CKM}, are consistent with Ref.~\cite{Beneke:2003az} up to NLO.

%- }}}

%- }}}

%\clearpage

%- {{{ Conclusions:

\section{\label{sec::concl}Conclusions}

In this paper we have computed the 
the NNLO corrections induced by current-current operators
to the decay matrix element
$\Gamma_{12}^q$ appearing in $B_q-\bar{B}_q$ mixing.
The result is expressed as a deep semi-analytic 
expansion in  $z=m_c^2/m_b^2$, which for
all relevant values of the charm and bottom quark masses
is equivalent to an exact result.

The new results are used to perform a phenomenological analysis
of the width difference $\Delta\Gamma_q$ and the 
CP asymmetry in flavour-specific $B_q$ decays, $a_{\rm fs}^q$,
both for the $B_s$ and the $B_d$ system.
The corrections of order $\alpha_s^2$
significantly stabilise the dependence on the
renormalisation scale and thus lead to a reduction
of the uncertainties.

We provide computer-readable results for the matching coefficients~\cite{progdata} which can be combined with the
non-perturbative matrix elements in a straightforward way.

For a further reduction of the uncertainty it would be important to obtain more precise results for the 
matrix elements of the dimension-7 operators.
Furthermore, it is desirable to compute
the corresponding NLO
and NNLO contributions from the penguin operators
to the leading term matching coefficient in the $1/m_b$ expansion.
At the central scale their numerical effect is expected to be small. However, their inclusion could lead to an even
more stable dependence on the renormalisation scale.

The calculated quantities permit to probe new physics in mass or decay matrix 
of the $B_s$ and $B_d$ systems from \bbm\ observables alone, without the information from a global 
fit to CKM parameters, which is sensitive to new physics as well. To this end more experimental effort 
on the determination of $\dg_d$ and $a_{\rm fs}^{d,s}$ is highly desirable.

%- }}}

%- {{{ Acknowledgements:

\section*{Acknowledgements}  

The work was supported by the Deutsche Forschungsgemeinschaft (DFG, German
Research Foundation) under grant 396021762 --- TRR 257 ``Particle Physics
Phenomenology after the Higgs Discovery''.
We would like to thank Artyom Hovhannisyan for providing to us unpublished
results from Ref.~\cite{Asatrian:2017qaz}. Pascal Reeck would like to thank the Studienstiftung des deutschen Volkes for supporting him.

%- }}}

%- {{{ Appendix:

\begin{appendix}

%- {{{ renormalisation constants:

\section{\label{app::Z} Renormalisation constants of the $\Delta B = 2$ theory}

Computer-readable files are provided for the renormalisation constants of the $\Delta B = 2$ theory up to $\alpha_s^2$ and up to third\footnote{The third generation is not required for the current-current calculation, but we provide the renormalisation constants for completeness.} generation evanescent operators in Ref.~\cite{progdata}.  We supply the renormalisation constants in the $\{Q, Q_S, \widetilde{Q}_S\}$ and $\{Q,\widetilde{Q}_S, R_0\}$ bases. For the latter we quote the ``naive'' renormalisation constants as well as the ones that include the $\epsilon$-finite renormalisation of $R_0$, see Section \ref{sec:R0_ren}. The specific choice of $\mathcal{O}(\epsilon)$ coefficients given in Eq.~\eqref{eq:nlo_ev_consts} and Eqs.~\eqref{eq:first_ev_const} to \eqref{eq:last_ev_const} has been used here to simplify the expressions. The notation of the files and their contents is explained in the \texttt{README} file provided with them.

%- }}}

%- {{{ VIA

\section{\label{app::VIA} Vacuum insertion approximation}

The leading term of the $1/N_c$ expansion of the $\Delta B=2$ matrix elements can be 
calculated analytically by factorising the matrix elements of the four-quark operators 
into two current matrix elements. At sub-leading order in $1/N_c$ the factorisation of 
the matrix element is an approximation, 
called vacuum insertion approximation (VIA).
This is illustrated for $\widetilde Q_S$ in the following, highlighting the step in 
which Fierz symmetry is crucial.

With $\Gamma \equiv (1+\gamma_5)$ we calculate this matrix element in VIA to find 
\begin{equation}
 \bra{B_s} \widetilde Q_S \ket{\bar B_s} 
 \stackrel{\rm VIA}{=} 2 \Gamma_{jk}\Gamma_{lm} \left[ \bra{B_s} \bar s_j^\alpha b_k^\beta \ket{0} \bra{0} 
  \bar s_l^\beta b_m^\alpha \ket{\bar B_s} \, -\, 
  \bra{B_s} \bar s_j^\alpha b_m^\alpha \ket{0} \bra{0} 
   \bar s_l^\beta b_k^\beta \ket{\bar B_s}\right]\,. 
\end{equation}
Here $j,k,l,m$ are Dirac indices and $\alpha,\beta$ are colour indices. While  in the first term we immediately recognise the matrix element of the pseudoscalar current, we need a Fierz transformation to contract the indices of the  Dirac matrices with those of the quark fields:
\begin{equation}
 \bra{B_s} \widetilde Q_S \ket{\bar B_s} 
 \stackrel{\rm VIA}{=} 2 \bra{B_s} \bar s^\alpha \Gamma b^\beta \ket{0} \bra{0} 
  \bar s^\beta \Gamma b^\alpha \ket{\bar B_s} \, -\, 
  \bra{B_s} \bar s^\alpha \Gamma^F b^\alpha \ket{0} \bra{0} 
   \bar s^\beta \Gamma^F b^\beta \ket{\bar B_s} \quad \label{eq:via}
\end{equation}
with 
\begin{equation}
 \Gamma^F_{jm} \Gamma^F_{lk}
 = \frac12 \Gamma_{jm} \Gamma_{lk} +\frac18 \left[\sigma_{\mu\nu} (1+\gamma_5) \right]_{jm} 
 \left[ \sigma^{\mu\nu} (1+\gamma_5)\right]_{lk} ,\quad \label{eq:fz}
\end{equation}
where only the first term gives a non-zero contribution to \eq{eq:via}. Thus 
\begin{align*}
 \bra{B_s} \widetilde Q_S (\mu_2) \ket{\bar B_s} 
 &\overset{\rm VIA}{=} \left( \frac{2}{N_c} - 1 \right)  \bra{B_s} \bar s \gamma_5 b \ket{0} \bra{0} 
  \bar s \gamma_5 b \ket{\bar B_s}\\
  &\overset{\phantom{\rm VIA}}{=} \left( 1- \frac{2}{N_c} \right) \frac{f_{B_s}^2 M_{B_s}^4}{\left[ m_b(\mu_2) +m_s(\mu_2)\right]^2}. \numberthis \label{eq:viaf}
\end{align*}
Here $M_{B_s}$ and $f_{B_s}$ are mass and decay constant of the $B_s$ meson, $N_c=3$ is the number of 
colours, and we show the dependence of the matrix element on the renormalisation scale $\mu_2$ at which 
our $\Delta B=2$ operator $\widetilde Q_S$ is defined. In VIA the $\mu_2$ dependence enters the result only 
through the quark masses $m_{b,s}(\mu_2)$. 

Importantly, the VIA result in \eq{eq:viaf} is exact in the limit $N_c\to \infty$. The VIA result also includes 
a colour-suppressed term at $1/N_c$, which is referred to as the factorisable $1/N_c$ correction. Comparing VIA with the lattice-QCD predicition of $\bra{B_s} \widetilde Q_S  \ket{\bar B_s}$ shows that VIA gives a 
good approximation, i.e.\ non-factorisable corrections are small. This is remarkable, because there is a 
large cancellation between 1 and $2/N_c$ in \eq{eq:viaf}. 

Since $\bra{B_s} \widetilde Q_S  \ket{\bar B_s}$ is uniquely fixed in terms of quantities like $f_{B_s}$ (which could be determined through other measurements),
the Wilson coefficient of $\widetilde Q_S$ must be also unique in the large-$N_c$ limit. This feature also means that the coefficient is Fierz-symmetric, because the  large-$N_c$ term in \eq{eq:viaf} stems from 
$\Gamma^F$. If one defined the evanescent operators in an arbitrary way, 
one would find arbitrary, scheme-dependent terms in the large-$N_c$ part of the coefficient in order $\alpha_s$ and beyond, and there is no term in \eq{eq:viaf} to cancel this scheme-dependence. 

In other words, $\widetilde Q_S$ and 
$\widetilde Q_S^F$ are identical in four dimensions and both evaluate to the same VIA result in \eq{eq:viaf}. 
In the perturbative calculation one is entitled to either work with $\widetilde Q_S$ 
or $\widetilde Q_S^F$ and, of course, the final prediction for physical observables must be the same,
which means that the coefficients $\widetilde C_S$ and $\widetilde C_S^F$ must be identical as well. 

Since the VIA results are good approximations and furthermore are the reference values to which the lattice QCD results are normalised, it is desirable to impose Fierz symmetry also on the sub-leading terms in the colour counting, i.e.\ exactly. Furthermore, the leading $N_c$ term is uniquely determined. 

%- }}}

%- {{{ Three-loop master integrals:

\section{\label{app::masters}Three-loop master integrals}

Our phenomenological analysis is based on the approach where
the master integrals are computed with the help of
a semi-analytic method. However, an important cross check is provided
by the analytic computation of the first expansion
terms of order\footnote{In this approach the contributions 
  from a closed charm quark loop in the gluon propagator are not considered.}
$(m_c^2/m_b^2)^0$ and $(m_c^2/m_b^2)^1$.
In this section we briefly describe the computation of the corresponding
three-loop master integrals. For the matching
coefficient we only need the imaginary part of the master integrals. However,
we find it more convenient to compute the complete diagrams and take the
imaginary parts afterwards. All integrals have an on-shell external momentum
$q^2=m_b^2$ and internal lines which carry the mass $m_b$ or are massless.
Some integrals reduce to vacuum integrals.

\begin{figure}[t]
    \centering
    \begin{tabular}{ccccc}
    \includegraphics[width=4cm, height=1.5cm, keepaspectratio]{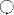}&
    \includegraphics[width=4cm, height=1.5cm, keepaspectratio]{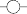}&
    \includegraphics[width=4cm, height=1.5cm, keepaspectratio]{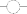}
    \\
    \includegraphics[width=4cm, height=1.5cm, keepaspectratio]{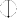}    &
    \includegraphics[width=4cm, height=1.5cm, keepaspectratio]{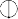}    &
    \includegraphics[width=4cm, height=1.5cm, keepaspectratio]{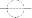}    &
    \includegraphics[width=4cm, height=1.5cm, keepaspectratio]{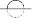}    &
    \includegraphics[width=4cm, height=1.5cm, keepaspectratio]{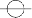}    \\
    \includegraphics[width=4cm, height=1.5cm, keepaspectratio]{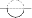}    &
    \includegraphics[width=4cm, height=1.5cm, keepaspectratio]{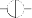}    &
    \includegraphics[width=4cm, height=1.5cm, keepaspectratio]{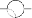}    &
    \includegraphics[width=4cm, height=1.5cm, keepaspectratio]{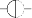}    &
    \includegraphics[width=4cm, height=1.5cm, keepaspectratio]{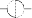}    \\
    \includegraphics[width=4cm, height=1.5cm, keepaspectratio]{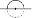}    &
    \includegraphics[width=4cm, height=1.5cm, keepaspectratio]{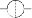}    &
    \includegraphics[width=4cm, height=1.5cm, keepaspectratio]{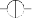}    &
    \includegraphics[width=4cm, height=1.5cm, keepaspectratio]{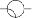}
      \end{tabular}
    \caption{One- and two-loop master integrals. 
      Solid and dotted lines denote massive (mass $m_b$) and massless propagators, respectively.
      A dot on a line denotes an additional power of the respective denominator.}
    \label{fig:2-loop-masters}
%   ~\\[-5mm]
%\hrule 
\end{figure}

For completeness we show in Fig.~\ref{fig:2-loop-masters} the one- and two-loop master integrals needed for
our calculation. Note that some of the three-loop master integrals factorise;
for this reason there are also lower-loop integrals which do not have an
imaginary part. The master integrals in Fig.~\ref{fig:2-loop-masters} are also needed for the two-loop $\Delta B=2$
calculation. Thus both real and imaginary parts have to be computed,
which is straightforward using standard techniques (see, e.g.\
Ref.~\cite{Smirnov:2012gma}). Analytic expressions for most of them can be
found in Ref.~\cite{Fleischer:1999hp}.

\begin{figure}[t]
	\centering
    \centerhfill
    \subfloat[]{\includegraphics[width=4cm, height=1.7cm, keepaspectratio]{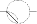}\label{fig:3-loop-masters-i}} % prop3L3topo010000000(0, 1, 1, 0, 0, 1, 1, 1, 0) X
    \hspace{0.2cm} \centerhfill \hspace{0.2cm}    
    \subfloat[]{\includegraphics[width=4cm, height=1.7cm, keepaspectratio]{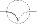}\label{fig:3-loop-masters-j}} % prop3L3topo010000000(0, 1, 1, 1, 0, 1, 1, 0, 0) X    
    \hspace{0.2cm} \centerhfill \hspace{0.2cm}
    \subfloat[]{\includegraphics[width=4cm, height=1.7cm, keepaspectratio]{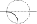}\label{fig:3-loop-masters-k}} % prop3L3topo010000000(0, 1, 1, 1, 1, 0, 1, 0, 0) X    
    \hspace{0.2cm} \centerhfill \hspace{0.2cm}
    \subfloat[]{\includegraphics[width=4cm, height=1.7cm, keepaspectratio]{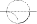}\label{fig:3-loop-masters-l}} % prop3L3topo010000000(0, 1, 1, 1, 1, 1, 0, 0, 0) X    
    \centerhfill \\ \centerhfill
    \subfloat[]{\includegraphics[width=4cm, height=1.7cm, keepaspectratio]{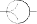}\label{fig:3-loop-masters-a}} % prop3L1topo010000100(0, 1, 1, 1, 1, 1, 1, 0, 0) X
    \hspace{0.2cm} \centerhfill \hspace{0.2cm}
    \subfloat[]{\includegraphics[width=4cm, height=1.7cm, keepaspectratio]{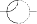}\label{fig:3-loop-masters-b}} % prop3L1topo100100100(1, 0, 1, 1, 1, 0, 1, 1, 0) X    
    \hspace{0.2cm} \centerhfill \hspace{0.2cm}
    \subfloat[]{\includegraphics[width=4cm, height=1.7cm, keepaspectratio]{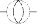}\label{fig:3-loop-masters-c}} % prop3L2topo000110000(1, 0, 1, 1, 1, 0, 1, 1, 0) X  
    \hspace{0.2cm} \centerhfill \hspace{0.2cm}
    \subfloat[]{\includegraphics[width=4cm, height=1.7cm, keepaspectratio]{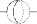}\label{fig:3-loop-masters-x}} % prop3L3topo011000000(1, 1, 1, 0, 0, 1, 1, 1, 0) X
    \centerhfill \\ \centerhfill
     \subfloat[]{\includegraphics[width=4cm, height=1.7cm, keepaspectratio]{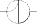}\label{fig:3-loop-masters-m}} % prop3L3topo011000000(1, 1, 1, 0, 1, 0, 1, 1, 0) X
    \hspace{0.2cm} \centerhfill \hspace{0.2cm}
     \subfloat[]{\includegraphics[width=4cm, height=1.7cm, keepaspectratio]{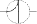}\label{fig:3-loop-masters-n}} % prop3L3topo011000000(1, 1, 1, 1, 0, 1, 1, 0, 0) X
     \hspace{0.2cm} \centerhfill \hspace{0.2cm}    
     \subfloat[]{\includegraphics[width=4cm, height=1.7cm, keepaspectratio]{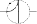}\label{fig:3-loop-masters-y}} % prop3L3topo011000000(1, 1, 1, 1, 0, 1, 2, 0, 0) X
      \hspace{0.2cm} \centerhfill \hspace{0.2cm}
     \subfloat[]{\includegraphics[width=4cm, height=1.7cm, keepaspectratio]{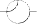}\label{fig:3-loop-masters-r}} % prop3L3topo101000000(1, 1, 1, 0, 1, 0, 1, 1, 0) X
     \centerhfill \\ \centerhfill
     \subfloat[]{\includegraphics[width=4cm, height=1.7cm, keepaspectratio]{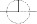}\label{fig:3-loop-masters-o}} % prop3L3topo011000000(1, 1, 1, 1, 1, 0, 1, 0, 0) X
   	 \hspace{0.2cm} \centerhfill \hspace{0.2cm}
     \subfloat[]{\includegraphics[width=4cm, height=1.7cm, keepaspectratio]{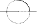}\label{fig:3-loop-masters-t}} % prop3L3topo101000000(1, 1, 1, 1, 1, 0, 1, 0, 0) X
     \hspace{0.2cm} \centerhfill \hspace{0.2cm}     
     \subfloat[]{\includegraphics[width=4cm, height=1.7cm, keepaspectratio]{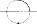}\label{fig:3-loop-masters-u}} % prop3L3topo101000000(1, 1, 1, 1, 1, 0, 2, 0, 0) X
     \hspace{0.2cm} \centerhfill \hspace{0.2cm}
     \subfloat[]{\includegraphics[width=4cm, height=1.7cm, keepaspectratio]{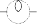}\label{fig:3-loop-masters-a1}} % prop3L2topo000110000(1, 0, 1, 1, 1, 1, 1, 1, 0) X
     \centerhfill \\ \centerhfill
     \subfloat[]{\includegraphics[width=4cm, height=1.7cm, keepaspectratio]{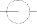}\label{fig:3-loop-masters-d}} % prop3L3topo001000000(1, 1, 1, 1, 1, 1, 1, 0, 0)  X
     \hspace{0.2cm} \centerhfill \hspace{0.2cm}
     \subfloat[]{\includegraphics[width=4cm, height=1.7cm, keepaspectratio]{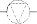}\label{fig:3-loop-masters-g}} % prop3L3topo001000000(1, 1, 1, 0, 1, 1, 1, 1, 0) X    
     \hspace{0.2cm} \centerhfill \hspace{0.2cm}
    \subfloat[]{\includegraphics[width=4cm, height=1.7cm, keepaspectratio]{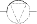}\label{fig:3-loop-masters-s}} % prop3L3topo011000000(1, 1, 1, 1, 1, 1, 0, 1, 0) X
    \hspace{0.2cm} \centerhfill \hspace{0.2cm}
    \subfloat[]{\includegraphics[width=4cm, height=1.7cm, keepaspectratio]{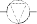}\label{fig:3-loop-masters-z}} % prop3L3topo101000000(1, 1, 1, 1, 0, 1, 1, 1, 0) X
    \centerhfill \\ \centerhfill
    \subfloat[]{\includegraphics[width=4cm, height=1.7cm, keepaspectratio]{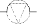}\label{fig:3-loop-masters-v}} % prop3L3topo101000000(1, 1, 1, 1, 1, 1, 0, 1, 0) X
    \hspace{0.2cm} \centerhfill \hspace{0.2cm}
    \subfloat[]{\includegraphics[width=4cm, height=1.7cm, keepaspectratio]{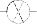}\label{fig:3-loop-masters-p}} % prop3L3topo011000000(1, 1, 1, 1, 1, 1, 1, 1, 0) X
   \hspace{0.2cm} \centerhfill \hspace{0.2cm}
    \subfloat[]{\includegraphics[width=4cm, height=1.7cm, keepaspectratio]{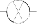}\label{fig:3-loop-masters-w}} % prop3L3topo101000000(1, 1, 1, 1, 1, 1, 1, 1, 0) X
    \hspace{0.2cm} \centerhfill \hspace{0.2cm}
    \subfloat[]{\includegraphics[width=4cm, height=1.7cm, keepaspectratio]{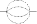}\label{fig:3-loop-masters-h}} % 
    \centerhfill \\ \centerhfill
    \subfloat[]{\includegraphics[width=4cm, height=1.7cm, keepaspectratio]{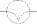}\label{fig:3-loop-masters-q}} % 
    % ???   
    \hspace{0.2cm} \centerhfill \hspace{0.2cm}
    \subfloat[]{\includegraphics[width=4cm, height=1.7cm, keepaspectratio]{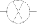}\label{fig:3-loop-masters-f}} %  
	% ????   
    \hspace{0.2cm} \centerhfill \hspace{0.2cm}
	\subfloat[]{\includegraphics[width=4cm, height=1.7cm, keepaspectratio]{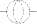}\label{fig:3-loop-masters-e}} % 
	% ???   
    \centerhfill \\
	\caption{Three-loop master integrals arising from the $\Delta B=1$
          NNLO calculation.
          Solid and dotted lines denote massive (mass $m_b$) and massless propagators, respectively.
          A dot on a line denotes an additional power of the respective denominator.}
	\label{fig:3-loop-mi-massive}
%    ~\\[-5mm]
%\hrule
\end{figure}

The 27 three-loop master integrals which do not factorise into lower-loop
integrals are shown in Fig.~\ref{fig:3-loop-mi-massive}.  Most of the
integrals are not available in the literature. The four integrals with only
massless propagators are can be obtained from
\textsc{MINCER}~\cite{Larin:1991fz}
(see, e.g.\  also Refs.~\cite{Bekavac:2005xs,Baikov:2010hf}).
Furthermore, some of them have been calculated in the
context of the fermionic NNLO corrections and can be found in
Ref.~\cite{Asatrian:2017qaz,Hovhannisyan:2022miy}.  Among multiple techniques
available for the analytic evaluation of Feynman integrals (see, e.g.\
Ref.~\cite{Weinzierl:2022eaz} for a recent overview), we have chosen the
method of direct integration of the Feynman parametric representation using
\textsc{HyperInt}~\cite{Panzer:2014caa}. A detailed description of how
to use \textsc{HyperInt} for tackling this class of integrals can be found in~\cite{Reeck:2024iwk}. Upon processing the results with \textsc{HyperLogProcedures}~\cite{Schnetz:2021ebf}
and \textsc{PolyLogTools}~\cite{Duhr:2019tlz} we can rewrite the imaginary part of each master integral in
terms of simple constants given by
\begin{equation}
	\pi,\, \log(2),\, \zeta_2,\, \zeta_3,\, \zeta_4,\, \textrm{Cl}_2 
\left(\frac{\pi }{3}\right),\, \sqrt{3},\,
	\textrm{Li}_{4} \left(\frac{1}{2}\right),\, \log \left ( \frac{1+\sqrt{5}}{2} \right ).
\end{equation}
The real parts of the master integrals are admittedly much more complicated
and still involve multiple constant GPLs, but since those are irrelevant for
$\Delta \Gamma_s$, we do not discuss them here.

In the following we present the analytic results for the imaginary parts of
those master integrals which have at least one massive line. We use the normalisation with
$(e^{\gamma_E { \epsilon}}/ (i \pi )^{D/2})^3$ and introduce
\begin{equation}
	\textrm{Im} (I^{3}_i) = \pi M_i^{(3)}\,.
\end{equation}

Our results for the non-trivial three-loop integrals read\footnote{The results for purely massless integrals were taken from \textsc{MINCER}~\cite{Larin:1991fz}, cf. also Refs.~\cite{Bekavac:2005xs,Baikov:2010hf}.}
\begin{align}
		M_{1}^{(3)} &= -\frac{1}{2 \epsilon ^2} -\frac{5}{2 \epsilon } -\frac{17}{2} -\frac{3 \zeta_2}{4} +\epsilon  \left(-\frac{49}{2} -\frac{15 \zeta_2}{4}+\frac{5 \zeta_3}{2}\right) \nonumber \\
		& +	\epsilon ^2 \left(-\frac{129}{2} -\frac{51 \zeta_2}{4}+\frac{25 \zeta_3}{2}-\frac{57 \zeta_4}{32}\right), \nonumber \\
		% prop3L3topo010000000(0, 1, 1, 0, 0, 1, 1, 1, 0) 
		M_{2}^{(3)} &=	\frac{1}{2 \epsilon } +\frac{11}{2} -\zeta_2 +\epsilon  \left(\frac{77}{2} -\frac{33 \zeta_2}{4}-8 \zeta_3\right) +\epsilon ^2 \left(\frac{439}{2} -\frac{195 \zeta_2}{4}-\frac{107 \zeta_3}{2}-\frac{123 \zeta_4}{4}\right), \nonumber \\
		% prop3L3topo010000000(0, 1, 1, 1, 0, 1, 1, 0, 0) X   
		M_{3}^{(3)} &= \frac{1}{2 \epsilon }+\frac{15}{4} +\epsilon \left(\frac{145}{8}-\frac{9 \zeta_2}{4}\right)  + \epsilon ^2 \left(\frac{1155}{16} -\frac{135 \zeta_2}{8}-\frac{11 \zeta_3}{2}\right) , \nonumber \\
		% prop3L3topo010000000(0, 1, 1, 1, 1, 0, 1, 0, 0) X    
		M_{4}^{(3)} &= \frac{7}{4} -\zeta_2 +\epsilon  \left(\frac{175}{8} -\frac{9 \zeta_2}{2}-11 \zeta_3\right) + \epsilon ^2 \left(\frac{2681}{16}-\frac{241 \zeta_2}{8}-\frac{99 \zeta_3}{2}-\frac{189 \zeta_4}{4}\right), \nonumber \\
		% prop3L3topo010000000(0, 1, 1, 1, 1, 1, 0, 0, 0) X 
		M_{5}^{(3)} &= 2 -2 \sqrt{3} \text{Cl}_2\left(\frac{\pi }{3}\right)+\zeta_2, \nonumber \\		
		% prop3L1topo010000100(0, 1, 1, 1, 1, 1, 1, 0, 0) X
		M_{6}^{(3)} &= \frac{1}{2 \epsilon^2 } + \frac{5}{2 \epsilon} + \frac{15}{2} -\frac{\zeta_2}{4} -2 \zeta_3 +\epsilon  \left(\frac{29}{2} + \frac{11 \zeta_2}{4}-\frac{19 \zeta_3}{2} +\frac{3 \zeta_4}{2}\right),   \nonumber \\		
		% prop3L1topo100100100(1, 0, 1, 1, 1, 0, 1, 1, 0) X    
		M_{7}^{(3)} &= \frac{1}{\epsilon ^2}+\frac{7}{\epsilon } +33 -\frac{17 \zeta_2}{2} -\frac{12 \zeta_2}{\sqrt{5}}-\frac{48}{5} \zeta_2 \log \left(\frac{1+\sqrt{5}}{2} \right)+\frac{8 \zeta_3}{5}, \nonumber \\
		% prop3L2topo000110000(1, 0, 1, 1, 1, 0, 1, 1, 0) X  
		M_{8}^{(3)} &=\frac{1}{2 \epsilon ^2}+\frac{7}{2 \epsilon } +\frac{33}{2} -\frac{5 \zeta_2}{4}-2 \zeta_3 +\epsilon  \left(\frac{131}{2} -\frac{35 \zeta_2}{4}-\frac{21 \zeta_3}{2}-\frac{7 \zeta_4}{2}\right),		\nonumber \\
		% prop3L3topo011000000(1, 1, 1, 0, 0, 1, 1, 1, 0) X
		M_{9}^{(3)} &= 1 + \epsilon (14-4 \zeta_3)  +\epsilon ^2 \left(119-\frac{9 \zeta_2}{2}-28 \zeta_3-30 \zeta_4\right), \nonumber \\	
		% prop3L3topo011000000(1, 1, 1, 0, 1, 0, 1, 1, 0) X		
		M_{10}^{(3)} &= \frac{1}{2 \epsilon ^2}+\frac{5}{2 \epsilon } +\frac{15}{2} +\frac{3 \zeta_2}{4}  -3 \zeta_2 \log (2)+\frac{5 \zeta_3}{4} +\epsilon  \biggl(\frac{29}{2} +\frac{11 \zeta_2}{4} \nonumber \\
		& -6 \zeta_2 \log (2)+4 \zeta_3 + 8 \text{Li}_4\left(\frac{1}{2}\right) -\frac{33 \zeta_4}{8}-2 \zeta_2 \log ^2(2)+\frac{\log ^4(2)}{3}\biggr),\nonumber \\
		% prop3L3topo011000000(1, 1, 1, 1, 0, 1, 1, 0, 0) X						
		M_{11}^{(3)} &= -\frac{\zeta_2}{\epsilon } -2 \zeta_2-3 \zeta_2 \log (2)-\frac{7 \zeta_3}{4} +\epsilon  \biggl(-4 \zeta_2 -6 \zeta_2 \log (2) -\frac{7 \zeta_3}{2} \nonumber \\
		& + 8 \text{Li}_4\left(\frac{1}{2}\right) -\frac{217 \zeta_4}{8} -2 \zeta_2 \log ^2(2)+\frac{\log ^4(2)}{3}		
		\biggr), \nonumber \\
		% prop3L3topo011000000(1, 1, 1, 1, 0, 1, 2, 0, 0) X				
		M_{12}^{(3)} &=\frac{1}{\epsilon } +11-\zeta_2-\zeta_3 +\epsilon  \left(77 -\frac{23 \zeta_2}{2} -13 \zeta_3 -7 \zeta_4\right), \nonumber \\
		% prop3L3topo101000000(1, 1, 1, 0, 1, 0, 1, 1, 0) X
		M_{13}^{(3)} &= 1 -\zeta_2+\zeta_3 +\epsilon  (14-7 \zeta_2-3 \zeta_3+5 \zeta_4),\nonumber \\
		% prop3L3topo011000000(1, 1, 1, 1, 1, 0, 1, 0, 0) X			
		M_{14}^{(3)} &= -2 +\zeta_2-3 \zeta_2 \log (2)+\frac{13 \zeta_3}{4} +\epsilon  \biggl(-26 +4 \zeta_2 -10 \zeta_2 \log ^2(2) \nonumber \\
		& -6 \zeta_2 \log (2)+\frac{39 \zeta_3}{2}+ 40 \text{Li}_4\left(\frac{1}{2}\right) -\frac{65 \zeta_4}{8} +\frac{5 \log ^4(2)}{3}\biggr),\nonumber \\
		% prop3L3topo101000000(1, 1, 1, 1, 1, 0, 1, 0, 0) X
		M_{15}^{(3)} &= -\frac{\zeta_2}{\epsilon }-2 \zeta_2-3 \zeta_2 \log (2)-\frac{39 \zeta_3}{4} +\epsilon  \biggl(-4 \zeta_2  -6 \zeta_2 \log (2) \nonumber \\
		 &-10 \zeta_2 \log ^2(2) -\frac{39 \zeta_3}{2} + 40 \text{Li}_4\left(\frac{1}{2}\right)-\frac{629 \zeta_4}{8}+\frac{5 \log ^4(2)}{3}\biggr), \nonumber \\		
		% prop3L3topo101000000(1, 1, 1, 1, 1, 0, 2, 0, 0) X	
		M_{16}^{(3)} &= -16 +\frac{24 \zeta_2}{\sqrt{5}}-\frac{24}{5} \zeta_2 \log \left(\frac{1+\sqrt{5}}{2}\right)-\frac{56 \zeta_3}{5}, \nonumber \\
		% prop3L2topo000110000(1, 0, 1, 1, 1, 1, 1, 1, 0) X			
		M_{17}^{(3)} &= 4   \zeta_4, \nonumber \\
		% prop3L3topo001000000(1, 1, 1, 1, 1, 1, 1, 0, 0)  X		
		M_{18}^{(3)} &= 6   \zeta_4, \nonumber \\
		% prop3L3topo001000000(1, 1, 1, 0, 1, 1, 1, 1, 0) X 		
		M_{19}^{(3)} &= 5   \zeta_4, \nonumber \\
		 % prop3L3topo011000000(1, 1, 1, 1, 1, 1, 0, 1, 0) X	
		M_{20}^{(3)} &= \frac{27}{4}  \zeta_4, \nonumber \\
	    % prop3L3topo101000000(1, 1, 1, 1, 0, 1, 1, 1, 0) X	
		M_{21}^{(3)} &= 6   \zeta_4, \nonumber \\
		% prop3L3topo101000000(1, 1, 1, 1, 1, 1, 0, 1, 0) X		
		M_{22}^{(3)} &= \frac{27}{4}   \zeta_4, \nonumber \\
	    % prop3L3topo011000000(1, 1, 1, 1, 1, 1, 1, 1, 0) X		
		M_{23}^{(3)} &= \frac{61}{4}   \zeta_4, \nonumber \\
		% prop3L3topo101000000(1, 1, 1, 1, 1, 1, 1, 1, 0) X         
		M_{24}^{(3)} &= \frac{1}{12} +\frac{71 \epsilon }{72} + \epsilon^2 \left(\frac{3115}{432}-\frac{7 \zeta_2}{8}\right)  +\epsilon ^3 \left(\frac{109403}{2592}-\frac{497 \zeta_2}{48}-\frac{29 \zeta_3}{12}\right) \nonumber \\ 
        & + \epsilon ^4 \left(\frac{3386467}{15552}-\frac{21805 \zeta_2}{288}-\frac{2059 \zeta_3}{72}+\frac{291 \zeta_4}{64}\right), \nonumber \\
		% prop3L3topo010000000X001111000 X
        M_{25}^{(3)} &= \frac{1}{\epsilon }+10 + \epsilon \left(64-\frac{21 \zeta_2}{2}\right)  \nonumber\\
        & +\epsilon ^2 (336-105 \zeta_2-23 \zeta_3) + \epsilon ^3 \left(1584-672 \zeta_2-230 \zeta_3+\frac{1017 \zeta_4}{16}\right), \nonumber \\         
        % prop3L3topo100000000(0,1,0,1,1,1,0,1,0)        
        M_{26}^{(3)} &= 0, \nonumber \\        
        % prop3L3topo000000000(1,1,1,1,1,1,1,1,0)
        M_{27}^{(3)} &=  \frac{1}{\epsilon ^2}+\frac{7}{\epsilon } +31 -\frac{21 \zeta_2}{2} +\epsilon  \left(103-\frac{147 \zeta_2}{2}+7 \zeta_3\right)  \nonumber \\        
        & +\epsilon ^2 \left(235-\frac{651 \zeta_2}{2}+49 \zeta_3+\frac{1737 \zeta_4}{16}\right), 
        \label{eq::MI3l}
        %prop3L3topo000000000(1,0,1,1,1,1,0,1,0)
\end{align}
with the Clausen function
\begin{equation}
\text{Cl}_2\left(\frac{\pi }{3}\right) = \textrm{Im} \left ( \textrm{Li}_2(e^{\frac{i \pi}{3}})\right )	\approx 1.014941\,.
\end{equation}
Notice that while the massless 8-line integral $I_{26}^{(3)}$ does have a cut, its imaginary part vanishes at $\mathcal{O}(\epsilon^0)$ since the integral itself is finite and proportional to $1/m_b^6$. The imaginary part together with $\log(m_b)$ terms start contributing $\mathcal{O}(\epsilon)$.

In Eqs.~\ref{eq::MI3l} we set $m_b$ to unity, since the corresponding prefactor can be 
trivially
recovered using dimensional analysis
\begin{equation}
	M_i^{(3)} \to  m_b^{12-2 n_p - 6 \epsilon} \mu^{6 \epsilon} 
M_i^{(3)},
\end{equation}
where $n_p$ denotes the number of propagators (including dots) present 
in the given integral.

All these results have been checked numerically with the help of
{\tt pySecDec}~\cite{Borowka:2017idc,Borowka:2018goh,Heinrich:2021dbf} 
at different values of $m_b$ using the default precision of the QMC integrator.

%- }}}

\section{Full results for the $\mathbf{\Delta B = 2}$ evanescent operator definitions}
\label{app:full_ev_consts}

Imposing condition~3 on the equations arising from condition 1 for $\{c, \tilde{c}, d, \tilde{d}, h, \tilde{h}\}$, we obtain a particular solution for the constants appearing in the second generation evanescent operators consistent with all physical conditions from Section \ref{sec:eva}. It is given by:
\begin{align*}
c &= \frac{1}{15\,N_c^3\,(2 + N_c^2)}\Big( 45568 - 54272\,N_c  - 151808\,N_c^2 - 55808\,N_c^3 - 6656\,N_c^4 \\
&\qquad + 22528\,N_c^5 + c^F \left( -48 - 576\,N_c^2 + 30\,N_c^3 - 96\,N_c^4 + 15\,N_c^5 \right)\\
&\qquad + \tilde{c}^F_2 \left(  - 240\,N_c - 300\,N_c^3 + 90\,N_c^5 \right) + d^F  \left( 192 + 384\,N_c^2 + 144\,N_c^4 \right) \Big)\\
\tilde{c} &= \frac{1}{15\,N_c^3\,(2 + N_c^2)}\Big( -59392 - 20992\,N_c  - 177664\,N_c^2 - 63616\,N_c^3 - 48640\,N_c^4\\
&\qquad  + 2816\,N_c^5 + \tilde{c}^F \left( -48 - 576\,N_c^2 + 30\,N_c^3 - 96\,N_c^4 + 15\,N_c^5 \right)\\
&\qquad + c^F_2 \left(  - 240\,N_c - 300\,N_c^3 + 90\,N_c^5 \right) + \tilde{d}^F  \left( 192 + 384\,N_c^2 + 144\,N_c^4 \right) \Big)\\
d &= \frac{1}{30\,N_c^3\,(2 + N_c^2)}\Big( 79744 - 94976\,N_c  - 265664\,N_c^2 - 112384\,N_c^3 + 21952\,N_c^4 \\
&\qquad + 47424\,N_c^5 + c^F  \left( -84 -1008\,N_c^2 - 93\,N_c^4 \right) + \tilde{c}^F_2 \left(  - 420\,N_c - 600\,N_c^3 + 195\,N_c^5 \right)\\
&\qquad + d^F \left( 336 + 672\,N_c^2 + 60\,N_c^3 + 252\,N_c^4 + 30\,N_c^5 \right)  \Big)\\
\tilde{d} &= \frac{1}{30\,N_c^3\,(2 + N_c^2)}\Big( -103936 - 36736\,N_c  - 310912\,N_c^2 - 123008\,N_c^3 - 74560\,N_c^4 \\
&\qquad + 11328\,N_c^5 + \tilde{c}^F  \left( -84 -1008\,N_c^2 - 93\,N_c^4 \right) + c^F_2 \left(  - 420\,N_c - 600\,N_c^3 + 195\,N_c^5 \right)\\
&\qquad + \tilde{d}^F \left( 336 + 672\,N_c^2 + 60\,N_c^3 + 252\,N_c^4 + 30\,N_c^5 \right)  \Big)\\
h &= \frac{4}{-14 - 14\,N_c - 7\,N_c^2 + 6\,N_c^3} \Big(1568 + 2528\,N_c - 368\,N_c^2 - 480\,N_c^3 \\
&\qquad + k \left(- 4  - 4\,N_c  - 2\,N_c^2 + N_c^3 \right)  + \tilde{k}\left( - 10 - 20\,N_c + 5\,N_c^2 + 5\,N_c^3 \right) \Big)\\
c_2 &= \frac{3584}{5} + c^F_2 +\frac{12}{5}\,\tilde{c}^F - \frac{8}{5}\,\tilde{d}^F \\
\tilde{c}_2 &= \frac{3456}{5} + \frac{12}{5}\,c^F +\tilde{c}^F_2 - \frac{8}{5}\,d^F\\
d_2 &= \frac{896}{5} + \frac14\,c^F_2 +\frac{3}{5}\,\tilde{c}^F - \frac{2}{5}\,\tilde{d}^F\\
\tilde{d}_2 &= \frac{2144}{5} + \frac{3}{5}\,c^F +\frac{1}{4}\,\tilde{c}^F_2 - \frac{2}{5}\,d^F\\
d^F_2 &= \frac14\,c^F_2\\\
\tilde{d}^F_2 &= 256 + \frac14\,\tilde{c}^F_2\\
\tilde h &= -448 + 4\,\tilde{k} \numberthis
\end{align*}
It can be checked that the leading $\mathcal{O}(N_c^2)$ term of the renormalised physical matrix elements does not depend on the remaining undetermined constant in the evanescent operator definitions.

We provide the generic solution quoted above as well as the specific choice given in Eq.~\eqref{eq:nlo_ev_consts} and Eqs.~\eqref{eq:first_ev_const} to \eqref{eq:last_ev_const} in computer-readable format which can be downloaded from Ref.~\cite{progdata}. The notation of the files and their contents are described in a separate \texttt{README} file.

%- }}}

\end{appendix}

%- {{{ Bibl.:

\bibliographystyle{JHEP}
\bibliography{inspire_mod.bib,extra.bib}

%- }}}

\end{document}